# Nearby M, L, and T Dwarfs Discovered by the Wide-field Infrared Survey Explorer (WISE)

Maggie A. Thompson[a], J. Davy Kirkpatrick[b], Gregory N. Mace[b,c], Michael C. Cushing[d], Christopher R. Gelino[b], Roger L. Griffith[b], Michael F. Skrutskie[i], Peter R. M. Eisenhardt[k], Edward L. Wright[c], Kenneth A. Marsh[e], Katholeen J. Mix[b], Charles A. Beichman[b], Jacqueline K. Faherty[f], Odette Toloza[g], Jocelyn Ferrara[h], Brian Apodaca[j], Ian S. McLean[c], and Joshua S. Bloom[l]

[a] Department of Astrophysical Sciences, Princeton University, Peyton Hall, 4 Ivy Lane, Princeton, NJ, 08544-1001; mat3@princeton.edu

[b] Infrared Processing and Analysis Center, MS 100-22, California Institute of Technology, Pasadena, CA 91125

[c] Department of Physics and Astronomy, UCLA, Los Angeles, CA 90095-1547

[d] Department of Physics and Astronomy, MS 111, University of Toledo, 2801 W. Bancroft St., Toledo, OH 43606-3328

[e] School of Physics and Astronomy, Cardiff University, Cardiff CF24 3AA, UK

[f] Department of Astronomy, Universidad de Chile Cerro Calan, Las Condes, Chile

[g] Departamento de Física y Astronomía, Universidad de Valparaiso, Av. Gran Bretaña 1111, Valparaiso, Chile

[h] Department of Physics and Astronomy, Barnard College, 504A Altschul, 3009 Broadway, New York, New York 10027

[i] Department of Astronomy, University of Virginia, Charlottesville, VA, 22904

[j] California State Polytechnic University, Pomona, CA 91768

[k] Jet Propulsion Laboratory, California Institute of Technology, MS 169-327, 4800 Oak Grove Drive, Pasadena, CA 91109

[l] Department of Astronomy, University of California, Berkeley, CA 94720-3411

*Abstract*

In our effort to complete the census of low-mass stars and brown dwarfs in the immediate Solar Neighborhood, we present spectra, photometry, proper motions, and distance estimates for forty-two low-mass star and brown dwarf candidates discovered by the Wide-field Infrared Survey Explorer (WISE). We also present additional follow-up information on twelve candidates selected using WISE data but previously published elsewhere. The new discoveries include fifteen M dwarfs, seventeen L dwarfs, five T dwarfs, and five objects of other type. Among these discoveries is a newly identified "unusually red L dwarf" (WISE J223527.07+451140.9), four peculiar L dwarfs whose spectra are most readily explained as unresolved L+T binary systems, and a T9 dwarf (WISE J124309.61+844547.8). We also show that the recently discovered red L dwarf WISEP J004701.06+680352.1 (Gizis et al. 2012) may be a low-gravity object and hence young and potentially low mass (< 25 $M_{Jup}$).

Key Words: brown dwarfs – solar neighborhood – proper motion – spectroscopy – binaries

## *1. Introduction*

Data from the Wide-Field Infrared Survey Explorer (WISE) offer a rare opportunity to probe the entire sky at wavelengths where cold stars and brown dwarfs are relatively bright and at a recent epoch that can be used to reveal motions of the nearest examples when compared to earlier surveys. Beginning in early 2010 and continuing through early 2011, WISE mapped the entire sky in four mid-infrared bandpasses centered at wavelengths of 3.4, 4.6, 12, and 22 $\mu$m, which are referred to as W1, W2, W3, and W4, respectively (Wright et al. 2010). WISE's two shortest wavebands, W1 and W2, are the most valuable in searching for cold stars and brown dwarfs because these objects emit a substantial amount of their flux at these wavelengths.

Many papers (Mainzer et al. 2011; Burgasser et al. 2011; Kirkpatrick et al. 2011, 2012; Cushing et al. 2011; Mace et al. 2013) have exploited WISE color-based selections to uncover brown dwarfs that were either not recognized in earlier surveys or colder than those previously identified. Other papers (Scholz et al. 2011; Gizis et al. 2011; Gizis, Troup, & Burgasser 2011; Gizis et al. 2012; Castro et al. 2012; Luhman et al. 2012) have leveraged WISE data to search for high proper motion objects via comparison to large-area data sets such as the Two Micron All Sky Survey (2MASS; Skrutskie et al. 2006) or the Sloan Digital Sky Survey (SDSS; York et al. 2000). Sometimes, an object that should have been found via previous color or proper motion searches can be identified for the first time, it having been missed because of unfortunate circumstances in earlier surveys. One example of this is WISEPC J150649.97+702736.0 (Kirkpatrick et al. 2011), an unusually bright and close ($J$ = 14.33 mag, d = $3.3^{+0.7}_{-0.3}$ pc; Marsh et al. 2012) T6 dwarf that should have been found with 2MASS but was not because its sky location at the 2MASS epoch left it confused with a bright background star. Another example is WISE 104915.57−531906.1 (Luhman 2013), an even brighter and closer ($J$ = 10.73 mag, d = 2.0±0.15 pc) object with two components, for which the brighter one at $i$ band is classified as an L8±1. At the 2MASS epoch, this object is easily seen but falls at a sky position nearly coincident with a background star seen in the earlier Digitized Sky

Survey; quality checking of candidates across surveys would likely have assumed this 2MASS source was a reddened object in the Galactic Plane.

WISE provides another set of data with which we can make the census of the Solar Neighborhood even more complete. By having an accurate and thorough understanding of this nearby census, astronomers can, via extrapolation, extend this knowledge to determine the spatial density and distribution of objects in the Milky Way and other galaxies.

In this paper, we present the discovery of forty-two new low-mass star and brown dwarf candidates and the recovery of twelve others using WISE. We explain our selection techniques in section 2 and discuss photometric and spectroscopic follow-up in section 3. In section 4 we present our spectral classifications and discuss objects of special interest. Brief conclusions are given in section 5.

## 2. *Brown Dwarf Candidate Selection*

To find missing members of the Solar Neighborhood, we have employed three different sets of selection criteria, as detailed further below. We present data on those objects for which we have obtained spectroscopic follow-up so far.

WISE and 2MASS photometry for our low-mass star and brown dwarf candidates is summarized in Table 1. Source names are shown in column 1. Photometry from the WISE All-Sky Source Catalog in the W1, W2, and W3 bands along with the WISE W1-W2 color are given in columns 2 through 5. Photometry from 2MASS ($J$, $H$, and $K_S$ magnitudes along with $J$-$H$ and $J$-$K_S$ colors) is given in columns 6 through 10. Finder charts are presented in Figure 1 for forty-two newly discovered objects. The twelve objects not included in Figure 1 are those that have been previously discovered, as noted in Table 1.

The candidates listed in Table 1 were drawn from three different searches. We name the three different selection methods as the "bright L and T dwarf search", the "late-T dwarf search", and the "proper motion search," and discuss each below. Codes for the search method used to uncover each candidate, along with the catalog providing the WISE photometry of each source, are listed in column 11 of Table 1.

*2.1. The Bright L and T Dwarf Search*

Using query services at the NASA/IPAC Infrared Science Archive[1] (IRSA), we used the following eight criteria on lists of source detections from the WISE coadded images to search for bright, previously missed L and T dwarfs. The first round of searches was run on internal source lists made as part of the first-pass processing of the WISE data, and subsequent runs were made on the later processing that formed the basis of the WISE Preliminary and All Sky public data releases[2]:

1. The W1-W2 color must be greater than 0.4 mag. This criterion is used to identify brown dwarfs of type mid-L and later. See Figure 1 of Kirkpatrick et al. (2011).

2. The signal-to-noise ratio at W2 must be greater than 20. This criterion is used to isolate the brightest, closest objects and to help eliminate spurious sources.

3. There must not be a source from the 2MASS All-Sky Point Source Catalog within 3 arcsec of the WISE source. This criterion was used to retain only those objects showing substantial motion – greater than ~0.3 arcsec/yr – during the decade-long time span between 2MASS and WISE. In a few cases, there is a weakly detected source within 3 arcsec on the 2MASS *images* but not in the 2MASS catalog itself; in these instances, the source was retained because the implied *J*-W2 color is large enough to suggest a late spectral type. See Figure 7 of Kirkpatrick et al. (2011).

---

[1] http://irsa.ipac.caltech.edu.

[2] Due both to improvements in the processing pipeline between the early and later runs and to the addition of more data into the coadd stacks for the later runs, some of the objects chosen in the first round of searches would no longer meet all of the criteria outlined below if the WISE All-Sky Source Catalog alone was used.

4. The W1-W2 color must satisfy the relation (W1−W2) > 0.96(W2−W3) − 0.96. As explained in section 2.2 of Kirkpatrick et al. (2011), this criterion is used to eliminate the bulk of extragalactic contaminants.

5. A source must have been detected at least seven times in the individual W2 frames at a signal-to-noise value greater than 3. This criterion is used to eliminate cosmic rays and other spurious objects.

6. The galactic latitude must be at least 20° away from the galactic plane. This criterion is used to avoid the most confused regions of the sky where our color and motion selections are likely to produce many false candidates.

7. A source must not be flagged as a known solar system object.

8. A source meeting all of the above criteria must not have a counterpart within 3 arcsec of the United States Naval Observatory B1 (USNO-B1; Monet et al. 2003) Catalog[3]. This criterion further ensures that the source is either very red (in its optical minus infrared color limit) or moving.

Once these constraints were applied, we created finder charts of each candidate. These charts showed for each source the WISE coadded images in all four WISE bands compared to images of identical locations in the sky taken in the three bands of 2MASS, the three bands of the Digitized Sky Survey (DSS), and (when there was coverage) the five bands of SDSS. These charts were used as the final quality check to ensure that our candidates had the photometric properties expected for cold stars and brown dwarfs.

[3] Also available at IRSA.

Eighteen objects resulting from this search have been discussed previously in Kirkpatrick et al. (2011). More recent spectroscopic follow-up has been acquired on an additional thirty-seven candidates from this search; six confirmed T dwarfs have been recently reported in Mace et al. (2013), and the other thirty-one with types earlier than T0 are presented here.

*2.2. The Late-T Dwarf Search*

This search methodology has been discussed in Kirkpatrick et al. (2012). Briefly, it differs from the "Bright L and T Dwarf Search" above in that it concentrates only on objects with W1-W2 > 2.0 mag (types ≥T6), employs a magnitude cut that varies as a function of W1-W2 color (because the goal is to create a volume-limited sample), uses an additional color cut of W2-W3 < 3.5 mag to eliminate background sources, and probes all of the galactic plane except that nearest the galactic center. See section 2.1 of Kirkpatrick et al. (2012) for more details. Six new objects from this search are presented here.

*2.3. The Proper Motion Search*

The methodology for this search is driven by the desire to identify nearby low-mass stars and brown dwarfs independent of color selections. Specifically, we have searched for objects exhibiting large proper motions between the imaging data comprising the WISE All-Sky Data Release and that comprising the WISE 3-Band Cryo Data Release. These latter data were taken between 06 August 2010 and 29 September 2010 when the inner cryogen tank still had hydrogen (enabling observations in the W3 band) but the outer cryogen tank had been exhausted (thus disabling observations at W4). These 3-band cryogenic observations represent a second-epoch of coverage six months after the 4-band imaging of the same area and cover roughly 30% of the total sky.

The criteria applied to the 3-Band Cryo Working Database were as follows:

1. The W1 profile-fit magnitude is brighter than 14.5 mag. This selects objects with very high signal-to-noise while still going deep enough to include nearby, late-T dwarfs.

2. The chi2 value from the W1 profile fit falls between 0.5 and 3.0 in order to eliminate extended sources and blended double stars.

3. In this same vein, the number of blend components used in the profile fitting is required to be 1, meaning there is no deblending.

4. There is no associated 2MASS All-Sky Point Source Catalog entry within 3 arcsec of the WISE source.

5. The radial offset between the 3-Band Cryo Working Database source and the associated WISE All-Sky Release Catalog source is greater than 0.3 arcsec. This assures that the source is moving greater than ~0.6 arcsec/yr.

6. The source has no confusion or contamination flags listed in any of the four WISE bands. This improves the probability that the source is a true, astrophysical object.

7. A source meeting all of the above criteria must not have a counterpart within 3 arcsec of the USNO-B1 Catalog. This criterion further ensures that the source is either very red (in its optical minus infrared color limit) or moving.

Once all of these constraints were applied and the source list of objects created, finder charts were made for each candidate, as described in the previous section. These charts were used as the final quality check to ensure that our candidates had the photometric properties expected for cold stars and brown dwarfs. Although all candidates were found to have proper motion, several were found to have motions considerably smaller than the values implied by the WISE data alone. Sixteen objects in Table 1 were identified using these criteria.

## *3. Follow-up Observations*

### *3.1. Photometric Follow-up Observations*

To characterize our candidates further and to prioritize targets for spectroscopic follow-up, we observed a number of our fainter candidates at ground-based observatories to provide $J$, $H$, and $K_s$ photometry. A few of our candidates were also observed in the two usable channels – at 3.6 (hereafter, ch1) and 4.5 (ch2) μm – using the warm Infrared Array Camera (IRAC) on the *Spitzer* Space Telescope.

#### *3.1.1. Ground-based Measurements*

A few of the objects in Table 1 were observed with the Peters Automated Imaging Telescope (PAIRITEL) on Mt. Hopkins, Arizona; the NOAO Extremely Wide-Field Infrared Imager (NEWFIRM) on the 4m Blanco Telescope at Cerro Tololo, Chile; the Wide-field Infrared Camera (WIRC) on the Hale 5m Telescope at Palomar Mountain, California; the 2MASS camera at the Kuiper 61-inch Telescope on Mt. Bigelow, Arizona; or the Ohio State Infrared Imager/Spectrometer (OSIRIS) on the 4.1m Southern Astrophysical Research (SOAR) Telescope at Cerro Pachón, Chile. The observing setup and data reduction methodologies for all of these instruments are identical to those described in Kirkpatrick et al. (2012), Kirkpatrick et al. (2011), and Cushing et al. (2011).

*3.1.2. Spitzer/IRAC*

Photometry from *Spitzer*/IRAC Cycle 7 and 8 programs 70062 and 80109 (Kirkpatrick, PI) for twenty-five of our candidates is listed in Table 2. The observing methodology been discussed elsewhere (Kirkpatrick et al. 2011, Griffith et al. 2012) and the reduction procedures were identical to those outlined in Kirkpatrick et al. (2011).

*3.2. Spectroscopic Follow-up Observations*

In order to verify that these sources are cold stars and brown dwarfs, we acquired follow-up spectroscopic observations using four different instruments – the Triple Spectrograph (TSpec) at the 5m Hale Telescope at Palomar Mountain, the 0.8-5.5 μm Medium-Resolution Infrared Spectrograph (SpeX) at the 3m NASA Infrared Telescope Facility (IRTF) on Mauna Kea, the Triple Spectrograph (TSpec) on the Astrophysical Research Consortium 3.5m telescope at Apache Point Observatory (APO) in New Mexico, and the Near-Infrared Spectrometer (NIRSPEC) at the 10m W.M. Keck Observatory on Mauna Kea. Details of the observing runs are given in Table 3. Source names are shown in column 1, near-infrared spectral types in column 2, and the instrument, observation date, and telluric corrector star in columns 3 through 5. Instrument setups and reduction steps are discussed below.

*3.2.1. IRTF/SpeX*

SpeX is a medium-resolution spectrograph and imager that uses a 1024×1024 InSb array for its spectroscopic observations (Rayner et al. 2003). Observations were conducted in prism mode with a 0.5 arcsec wide slit, which achieves a resolving power ($\lambda/\Delta\lambda$) of ~150 over the range 0.8-2.5 μm. A series of 120s or 180s exposures were typically obtained at two different nod positions along the 15 arcsec long slit. Stars of type A0 V at a similar airmass to the target were observed near in time for telluric correction and flux calibration. Observations were typically obtained with the slit oriented to the parallactic angle to minimize slit loses and spectral slope variations due to differential atmospheric refraction. A set of exposures of internal flat field and argon arc lamps were obtained for flat fielding and wavelength calibration. The data were reduced using Spextool, the IDL-based data reduction package for SpeX (Cushing et al. 2004). Raw images were corrected for non-linearity, then pair-subtracted and flat fielded. If the sources were faint, multiple pair-subtracted images were averaged in order to perform tracing. The spectra were then optimally extracted (e.g., Horne 1986) and wavelength calibrated using the argon lamp exposures. Then multiple spectra were averaged and the resulting spectrum corrected for telluric absorption and, employing the same technique as described in Vacca et al. (2003), flux calibrated using the A0 V stars. Thirty-seven objects were observed using SpeX (Table 3).

*3.2.2. Palomar/TSpec*

TSpec has a 1024×2048 HAWAII-2 array to cover the spectral range from 1.0 to 2.45 μm (Herter et al. 2008). Use of a 1×30-arcsec slit gives a resolving power of ~2700. Observations were acquired in an ABBA nod sequence with an exposure time per nod position not exceeding 300s to lessen issues resulting from ever-changing OH background levels. Observations of A0 V stars were taken for telluric correction and flux calibration and were both near in time and near in airmass to the targets. In addition, dome flats were taken to calculate the pixel-to-pixel response of the detector. As with the IRTF/SpeX data discussed above, observations were typically obtained with the slit oriented to the parallactic angle. For reductions, a modified version of Spextool was used, as described in Kirkpatrick et al. (2011). Ten objects were observed using TSpec (Table 3).

*3.2.3. Keck/NIRSPEC*

Observations with NIRSPEC (McLean et al. 2000) used the N3 setting (approximately J-band) with a slit width of 0.57 arcsec, which results in a spectral resolution at 1.27 μm of R ~ 1500. Data reduction made use of the publicly available REDSPEC package (McLean et al. 2003), with modifications to remove residuals from the sky-subtracted pairs prior to 1-D spectral extraction. Other reduction details can be found in Kirkpatrick et al. (2011). Five objects were observed using NIRSPEC, as noted in Table 3.

*3.2.4. APO/TSpec*

The Apache Point Observatory version of Triplespec (hereafter, APO/TSpec; Wilson et al. 2004, Herter et al. 2008) was used to take 0.9-2.4 μm spectra like those acquired by Palomar/TSpec above. Data acquisition was done with nodded pairs having a total integration time of 300s at each nod position. Data reduction made use of the Triplespectool[4], which is a version of Spextool modified specifically for use on APO/TSpec. Three objects were observed using APO/TSpec, as noted in Table 3.

*4. Analysis*

To classify our spectra, we performed by-eye fits to near-infrared spectral standards. For spectra falling midway between established integral classes, a half type was used. For M dwarfs and L dwarfs, we used the near-infrared standards established in

---

[4] Available for download at http://www.astro.virginia.edu/~mfs4n/tspec/user/tspectool/ or at http://www.apo.nmsu.edu/arc35m/Instruments/TRIPLESPEC/#7

Kirkpatrick et al. (2010). For T dwarfs, we used the near-infrared standards established in Burgasser et al. (2006) for T0-T8 and Cushing et al. (2011) for T9. In some cases where an object was clearly not a normal M, L, or T dwarf or we wanted to compare to higher-resolution data than those presented in the above papers, we also consulted the IRTF Spectral Library[5] (Rayner et al. 2009) or the NIRSPEC Brown Dwarf Spectroscopic Survey[6] (McLean et al. 2003). Comparison of our spectra to these standards is shown in Figures 2-16 and 18-20, and classifications are given in Table 3.

In the subsections below, we discuss the M, L, and T dwarfs in sequence followed by objects classified as non main-sequence stars. To further characterize our sources, we have tabulated WISE and 2MASS astrometry in Table 4 and computed proper motions, spectrophotometric distance estimates, and implied tangential velocities in Table 5. For M dwarfs, the spectrophotometric distance estimates used the *J*-band magnitudes and spectral types of our objects compared to the mean $M_J$ magnitudes presented in Table 3 of Kirkpatrick & McCarthy (1994). For L0-T8 dwarfs, we used the measured *J*-band magnitudes and the polynomial relation provided in Table 3 of Looper et al. (2008). For our sole T9 dwarf, we estimated the distance using an average of the *H*-band and W2-band spectrophotometric distances derived from the relations (that omit WISE 1828+2650 in their fits) in section 4.3 of Kirkpatrick et al. (2012).

For previously discovered objects, we present a comparison of our measurement of spectral type, proper motion, and/or distance to previously published values in Table 6. There is generally excellent agreement, with the exception of late-L dwarf classifications, which are discussed more fully below.

[5] Available at http://irtfweb.ifa.hawaii.edu/~spex/IRTF_Spectral_Library/
[6] Available at http://www.astro.ucla.edu/~mclean/BDSSarchive/

*4.1. M Dwarfs and Subdwarfs*

The spectra of dwarfs classified as M type are illustrated in Figures 2-6. In these and subsequent figures, the spectrum of the object is shown by the heavy black line, and overplots of the nearest spectral standard(s) is (are) shown in grey. Most of these are normal dwarfs, but a few warrant special mention.

SDSS 0750+4454 (Figure 5) best fits the M8 dwarf standard at *J*-band but is overall bluer than the standard at *H* and *K* bands and has deeper $H_2O$ bands. It is classified as "M8 pec". WISE 1411-1403 (Figure 5) shares these same traits, though to a lesser degree, and is thus classified as "M8 pec?". The causes for these peculiarities is not known, as they do not seem to match the late-M hallmarks of either low-gravity or low-metallicity (Figures 14, 21, and 22 of Kirkpatrick et al. 2010).

PM 1522-0244 (Figure 4) does not match any of the dwarf standards well because its *H*- and *K*-band flux is noticeably depressed but is found to be a near duplicate of the sdM6 dwarf LHS 1074. Two other objects, PM 1713-4535 and WISE 1529-4513 (both in Figure 2), show similar depressions at *H* and *K* relative to the standards but have types near M0. We do not have near-infrared spectra of early-M subdwarfs nor could we find any in available archives, so we tentatively classify both of these objects as "sdM0?". If these two objects were normal M0 dwarfs, the implied tangential velocities would be 1100±100 and 690±70 km/s for estimated distances of 500±30 pc and 560±30 pc for PM 1713-4535 and WISE 1529-4513, respectively. Both of these velocities are well in excess of the Galactic escape velocity of ~525 km/s (Carney & Latham 1987). Using equation 11 in Gizis & Reid (1999) along with optical colors and magnitudes in Gizis (1997) and tabulated *J* magnitudes in the 2MASS All-Sky Point Source Catalog for three sdM0 dwarfs (LHS 12, 42, and 211), we find that the average $M_J$ value for sdM0 dwarfs is ~7.91 mag. Re-classification of our objects as sdM0 dwarfs implies distances of ~200 and ~223 pc and tangential velocities of ~436 and ~275 km/s (Table 5) for PM 1713-4535 and WISE 1529-4513, respectively, which are much more reasonable values.

One other object, the M9 dwarf WISE 1557+5914, appears to be a companion to the high proper motion K5 dwarf (Stephenson 1986) G 225-36 (Gliese 605, MCC 755; Vyssotsky 1956), located 122 arcsec away. Table 7 compares our measured proper motion and estimated distance of WISE 1557+5914 with published values of motion and distance for G 225-36. The values are identical to within 1σ for all quantities, leading us to conclude that this is likely a true common proper motion binary system. In fact, Pinfield et al. (2006) were the first to discover the companion object and associate it as co-moving with G 225-36; our independent discovery is the first to establish the companion's spectral type. The projected separation between components is 3800 AU. This separation is not unprecedented for a mid-K/late-M pair, although few such systems are currently recognized (Caballero et al. 2012). Using data in Lopez-Morales (2007), we can assume a mass of ~0.7 $M_{\odot}$ for the K5 component and ~0.1 $M_{\odot}$ for the M9 component to compute a gravitational binding energy, $-U_g$*, of the system of ~$3\times10^{34}$ J, which is within the range of energies occupied by the most fragile binaries known (Cabellero 2009).

*4.2. L Dwarfs*

The spectra of dwarfs classified as L type are illustrated in Figures 5-16. Most are normal L dwarfs, but ten stand out as having peculiar spectroscopic features. The first, WISE 1951−3311, is only mildly peculiar relative to the L1 spectral standard (Figure 7) in that it has deeper water bands and a less flat H-band peak. We classify this one as "L1 pec?" to reflect its slightly odd nature.

A second object, WISE 1851+5935 best fits the L9 spectral standard (Figure 9) but shows peculiarities at *J*, *H* and *K* bands leading us to a classification of "L9 pec". Specifically, the *J*-band is noticeably suppressed shortward of 1.1 μm relative to the standard and has a much deeper FeH band at 9986 Å; the flat peak of *H*-band has a divot near 1.6 μm; and the shape of the *K*-band is odd relative to the standard. A closer look at the 1.15-1.35 μm spectrum (Figure 12) shows much stronger FeH bands at 1.20 and 1.24 μm than either an L9 or a T0 dwarf. The spectrum of this object is reminiscent of the oddities seen in the L9 near-infrared spectrum of SDSS J080531.84+481233.0, which Burgasser (2007) showed could be better understood as a composite binary with types of ~L4.5 and ~T5. Our analysis (Figure 15) shows that the spectral morphology of WISE 1851+5935 can be explained as the composite of an ~L7 dwarf and a ~T2 dwarf.

The next three objects – WISE 1658+5103, WISE 1552+5033, and WISE 0230-0225 – are like WISE 1851+5935 in that their peculiarities may also be explained by unresolved binarity. Specifically, we classify WISE 1658+5103 (Figures 8 and 11) as "L6 pec" and find that its spectral morphology can be better explained as the composite of an ~L4 dwarf and a ~T0 dwarf (Figure 16). We classify WISE 1552+5033 (Figure 10) as "L9: pec" and find that it can be interpreted as an unresolved ~L6 and ~T5 binary (Figure 17). Finally, WISE 0230-0225 (Figure 9), although having a noisy spectrum ("L8: pec") that lacks adequate data over most of the *J*-band region, nonetheless has the *H*-band divot near 1.6 μm as well as indications of methane at 2.2 μm – both hallmarks of an unresolved T dwarf companion. We find that the spectrum may be better explained as a binary composed of ~L7 and ~T2 components (Figure 18), although higher signal-to-noise data on this object are needed to better discern this.

The sixth and seventh peculiar objects -- WISE 0446-2429 and WISE 0807+4130 (both shown in Figure 8) – are blue compared to the best fitting standard. In the case of the former, the discrepancy is large, resulting in a classification as "L5 pec (blue)" for WISE 0446-2429. The case of WISE 0807+4130 is less extreme, so we merely classify it as "L8 pec". For both of these objects, the peculiarities do not match those seen for the possible composite binaries discussed above although all of these objects show suppressed *H*- and *K*-bands relative to the standards. Such intrinsically blue L dwarfs are comprised of two subclasses: the L subdwarfs (section 6.3.3 of Kirkpatrick et al. 2010), whose spectral morphology can be ascribed to low metallicity (e.g., Burgasser et al. 2003), and the "unusually blue L dwarfs" (section 6.4 of Kirkpatrick et al. 2010), whose morphology appears to be related not to metallicity but to thin and/or large-grained clouds (Burgasser et al. 2008, Radigan et al. 2008), although the underlying physical cause is still a mystery. We ascribe WISE 0446-2429 to the latter category because it does not show the obvious hallmarks of low-metallicity (e.g., overly strong FeH bands). The 2MASS-measured $J$-$K_s$ color of 1.30±0.20 mag for this object further confirms what the spectrum reveals – that this object is abnormal; only a small percentage of L5 dwarfs have $J$-$K_s$ colors this blue (Figure 14 of Kirkpatrick et al. 2008). Thus, WISE 0446-2429 provides another example of this rare class and can be used to further study the blue L dwarf phenomenon.

The eighth, ninth, and tenth objects – WISE 1358+1458 (Figures 6 and 7), WISE 1733+3144 (Figure 7), and WISE 2335+4511 (Figures 13 and 14) – are red compared to the best fitting standard. Intrinsically red L dwarfs, like their blue L dwarf counterparts, are comprised of two subclasses: young L dwarfs (section 6.1 of Kirkpatrick et al. 2010), whose spectral morphology can be attributed to the low surface gravities that are a consequence of the fact that these objects are still contracting to their final radii (e.g., Kirkpatrick et al. 2006, Cruz et al. 2009), and "unusually red L dwarfs"(section 6.2 of Kirkpatrick et al. 2010), whose atmospheres appear to be markedly more dusty than normal (e.g., Looper et al. 2008), although the underlying physical cause is still under debate (Gizis et al. 2012). WISE 1358+1458 does not show the hallmarks of low-gravity in its K I lines (Figure 6), and given the fact that it is significantly redder than the best matching standard (Figure 7) we classify it as "L4 pec (red)". Likewise, WISE 1733+3144 does not differ markedly at *J*-band from its best fitting standard, but is clearly redder than the standard and *H*- and *K*-bands. We thus classify it as "L2 pec (red)".

WISE 2335+4511, on the other hand, represents a more extreme case. Its *J*-band spectrum (Figure 14), though noisy, does not show the signs of weaker K I lines expected in a low-gravity object. Its *H*- and *K*-band spectrum (Figure 13) is wildly redder than the best fitting standard at *J*-band, although its $J$-$K_s$ color (<2.24 mag; Table 1) is not as extreme as either WISE 1738+6142 (2.55±0.16 mag) or WISE 0047+6803 (2.55±0.08 mag), also shown in Figure 13. We classify the new WISE object as "L9 pec (v red)" but note that a unified classification scheme for these objects has not yet been devised. Gizis et al. (2013) classify WISE 0047+6803 as "L7.5 pec" following the lead from Looper et al. (2008) in classifying another red L dwarf, 2MASS J21481628+4003593. Mace et al. (2013) classify WISE 1738+6142 merely as "extremely red" with no spectral class attached, to emphasize the fact that that object cannot strictly be classified as either an L dwarf or a T dwarf given its unusual spectral morphology.

It should be noted that our Palomar/TSpec spectrum of WISE 0047+6803 has much better resolution (R~2700) than the SpeX prism spectrum (R~150) presented in the Gizis et al. (2012) discovery paper. A blow-up of the *J*-band portion of the TSpec spectrum is shown in Figure 14. The two pairs of K I doublets are clearly much weaker than in the normal L9 dwarf overplotted. The measured equivalent widths of these lines compared to the equivalent widths of the same lines for a selection of normal L dwarfs from McLean et al. (2003) is shown in Table 8. The weakness of these lines is a hallmark of lower gravity (e.g., Kirkpatrick et al. 2006), so we take this to mean that WISE 0047+6803 is a low-gravity (young) object[7], which at late-L type would make it much lower in mass than a normal late-L dwarf. Figure 10 in Kirkpatrick et al. (2006) shows the theoretical $T_{eff}$ vs log(g) plane for objects with $T_{eff}$ values between 1000 and 3000K. Assuming that WISE 0047+6803 has a $T_{eff}$ value of ~1400K like other late-L dwarfs (Figure 8 of Kirkpatrick 2005) and that our ability to discern a lower-gravity corresponds to an age likely below 100 Myr, then the implied mass is below ~15 $M_{Jup}$. Gizis et al. (2012) note that model fitting to the spectrum of WISE 0047+6803 does a poor job of constraining the temperature, with values anywhere between 1100 and 1600K possible.

[7] We have explored the possibility that weaker K I lines may instead be caused by higher metallicity (e.g., Looper et al. 2008). Using BT-Settl models from the Lyon grid (Allard et al. 2003; http://phoenix.ens-lyon.fr), we find that for fixed gravity (log(g)=4.5) and effective temperature, the K I lines are somewhat weaker at [M/H] = +0.5 relative to those at [M/H] = 0.0 for the 2000K 1900K, and 1800K models, but at lower temperatures (1700K < Teff < 1100K) the K I lines are either stronger in the higher metallicity model or weak at both values of [M/H]. As we believe that these lower temperatures are the ones more applicable to WISE 0047+6803, higher metallicity is not likely to be the physical cause.

In this case, the implied upper mass is anywhere between ~25 to ~12 $M_{Jup}$. A young object at such high northern declination does not rule out membership in one of the young moving groups primarily located in the southern hemisphere, however, because, as one example, the AB Dor moving group is close enough and dispersed enough that the Sun is currently moving through it (see Table 5 of Zuckerman & Song 2004). Gizis et al. (2012) suggest that a distance as close as 9-10 pc for WISE 0047+6803 would give it a $v_{tan}$ value consistent with membership in the β Pic Association, implying an age of 12-40 Myr and a mass below ~15 $M_{Jup}$. Our identification of this object as a low-gravity dwarf implies that such a low mass may indeed be warranted.

WISE has been particularly adept at uncovering red L dwarfs. This is largely because the more extreme examples are undetectable at *J*-band in 2MASS, thus eliminating them from any 2MASS based $J$-$K_s$ color selection. With more examples now in hand, we can explore the red L dwarf population in more detail. Table 9 lists $J$, $K_s$, W1, W2, and W3 magnitudes for all known brown dwarfs having $J$-$K_s$ colors greater than 2.0 mag. The objects are separated into three groups: those thought to be normal L dwarfs[8], red L dwarfs whose peculiarities are believed to be due to low gravity, and "unusually red L dwarfs" whose peculiarities are of unknown cause. Figure 17 shows the J-W2 versus W1-W2 plot with the three types of objects highlighted. Even though the reddest of these fall in either the "low-g" or "unusually red L" categories, examples of all three categories are seen to occupy phase space at bluer colors. The inclusion of W1 data appears not to delineate these types in the color-color plane or help with their physical interpretation.

[8] This subset includes a number of objects with optical spectra only, and some of these may be revealed as peculiar when near-infrared spectra are obtained.

Finally, Kirkpatrick et al. (2010) stated that the first "unusually red L dwarfs" identified appear to have substantially larger overall space motions than young, low-gravity M and L dwarfs, meaning that they are drawn from an older population. We can revisit this using our larger sample, although we have only tangential velocity estimates and no radial velocity measurements. Table 10 lists the $v_{tan}$ estimates for objects with $J$-$K_s$ > 2.0 mag in both the low-gravity L dwarf class and the "unusually red L dwarfs" class. Low-gravity L dwarfs have an average $v_{tan}$ value of 24 km/s with a tangential velocity dispersion of 25 km/s. The unusually red L dwarfs have an average $v_{tan}$ value of 32 km/s with a tangential velocity dispersion of 35 km/s. The slightly tighter velocity dispersion for the low-g L dwarfs along with their slightly smaller average $v_{tan}$ values may indicate that the two populations are distinct, with the low-g population (as concluded in Faherty et al. 2009) being younger. Reid (1997) finds that stars in the Galactic disk have an average $v_{tan}$ value of 37 km/s, and Faherty et al. (2009) find that the 20 pc sample of late-M through late-T dwarfs has a median $v_{tan}$ value of 29 km/s with a dispersion of 25 km/s. The "unusually red L dwarfs" are therefore similar in kinematic age to average stars in the disk and do not share the same young kinematics as low-gravity L dwarfs. Measuring full space motions through the acquisition of trigonometric parallaxes and radial velocities are still badly needed to clarify this picture further, however, because the $v_{tan}$ values in Table 10 are based solely on distance estimates based on the spectral type versus absolute magnitude relations for normal L dwarfs.

### *4.3. T Dwarfs*

The spectra of dwarfs classified as T type are illustrated in Figures 20 and 21. All six of these are normal dwarfs ranging in type from T7 to T9. Based on our spectrophotometric distance estimates in Table 5, four of these are expected to be newly identified members of the 20 pc census, including the T9 dwarf WISE 1243+8445, which is the latest of the six.

### *4.4. Other Objects*

The spectra of our other discoveries are illustrated in Figure 22. WISE 0200+8742 has a very red, featureless continuum and is likely either an embedded object or a heavily reddened object seen through a column of severe obscuration. We note that it lies only 30 arcsec away from an object identified as a dense core inside a molecular cloud (Ward-Thompson et al. 2010).

WISE 1749-3804 shows a mostly featureless continuum that best matches an early-K or even earlier type object. The photometry presented in Table 1 shows that it has a magnitude of ~13.4 in all bands except *J*, where it is measured to be somewhat fainter. Pending an optical spectrum to investigate this further, we tentatively classify this object as "white dwarf?" although a subdwarf classification at a type earlier than M is also possible.

The final three objects – WISE 0528+0901, WISE 1307-4630, and WISE 1507-3440 – show water absorption between the *J* and *H* bands and between the *H* and *K* bands, but none of these have morphologies consistent with M, L, or T dwarfs. We interpret these to be M giant stars. Their proper motion measurements from Table 5 are zero within 1.0-1.5σ, adding further evidence that these objects are probably not members of the immediate Solar Neighborhood.

## *5. Conclusions*

We have presented spectra for a collection of nearby M, L, and T dwarf candidates selected from WISE data. These objects – as varied in type from M subdwarfs to objects on the cusp of the Y dwarf class – represent previously missed members of the immediate solar vicinity, some estimated to be as close as 10-20 pc from the Sun. On-going searches that combine WISE's unique near- and mid-infrared coverage of the entire sky at epoch ~2010 with the shorter wavelength data of earlier, all-sky surveys such as DSS and 2MASS, will continue to allow researchers to refine our knowledge of the local low-mass stellar and substellar census.

## *6. Acknowledgments*

We thank David Ciardi and Gerard van Belle for their help in acquiring the 2013 Feb 21 Keck/NIRSPEC observations, and we thank our anonymous referee for a timely and helpful report. This publication makes use of data products from the Wide-field Infrared Survey Explorer, which is a joint project of the University of California, Los Angeles, and the Jet Propulsion Laboratory/California Institute of Technology, funded by the National Aeronautics and Space Administration (NASA). This publication also makes use of data products from 2MASS, SDSS, and DSS. 2MASS is a joint project of the University of Massachusetts and the Infrared Processing and Analysis Center/California Institute of Technology, funded by NASA and the National Science Foundation. SDSS is funded by the Alfred P. Sloan Foundation, the Participating Institutions, the National Science Foundation, the U.S. Department of Energy, the National Aeronautics and Space Administration, the Japanese Monbukagakusho, the Max Planck Society, and the Higher

Education Funding Council for England. The DSS were produced at the Space Telescope Science Institute under U.S. Government grant NAG W-2166. The images of these surveys are based on photographic data obtained using the Oschin Schmidt Telescope on Palomar Mountain and the UK Schmidt Telescope. This work is based in part on observations made with the *Spitzer* Space Telescope, which is operated by the Jet Propulsion Laboratory, California Institute of Technology, under a contract with NASA. Support for this work was provided by NASA through an award issued to programs 70062 and 80109 by JPL/Caltech. This research has made use of the NASA/ IPAC Infrared Science Archive, which is operated by the Jet Propulsion Laboratory, California Institute of Technology, under contract with NASA. This research has also benefitted from the M, L, and T dwarf compendium housed at DwarfArchives.org; from the SIMBAD database, operated at CDS, Strasbourg, France; and from NASA's Astrophysics Data System. We acknowledge support from the Steward/Mt. Bigelow staff for use of the 2MASS imager at the Bigelow 61-inch telescope. We acknowledge use of PAIRITEL, which is operated by the Smithsonian Astrophysical Observatory (SAO) and was made possible by a grant from the Harvard University Milton Fund, the camera loaned from the University of Virginia, and the continued support of the SAO and UC Berkeley. The PAIRITEL project is supported by NASA Grant NNG06GH50G. This material is based on work supported by the National Science Foundation under Award No. AST-0847170, a PAARE Grant for the California-Arizona Minority Partnership for Astronomy Research and Education (CAMPARE). Any opinions, findings, and conclusions or recommendations expressed in this material are those of the authors and do not necessarily reflect the views of the National Science Foundation.

*Table 1: Photometry of Candidates with Confirming Spectra*

| Designation | W1 (mag) | W2 (mag) | W3 (mag) | W1-W2 (mag) | J (mag) | H (mag) | Ks (mag) | J-H (mag) | J-Ks (mag) | Code |
|---|---|---|---|---|---|---|---|---|---|---|
| WISE J003110.04 +574936.3 | 12.41± 0.02 | 11.84± 0.02 | 11.30± 0.10 | 0.57± 0.03 | 14.95± 0.04 | 13.78± 0.04 | 13.22± 0.03 | 1.18± 0.06 | 1.74± 0.05 | b,1a |
| WISEP J004701.06 +680352.1[i] | 11.88± 0.02 | 11.27± 0.02 | 10.33± 0.07 | 0.61± 0.03 | 15.60± 0.07 | 13.97± 0.04 | 13.05± 0.03 | 1.64± 0.08 | 2.55± 0.07 | b,1a |
| WISE J020011.40 +874207.2[m] | 13.27± 0.02 | 12.85± 0.02 | >13.20 | 0.42± 0.03 | >17.16 | 14.77± 0.09 | 13.75± 0.05 | >2.39 | >3.42 | b,1a |
| WISE J023038.90 -022554.0 | 14.23± 0.03 | 13.63± 0.04 | 11.85± 0.19 | 0.60± 0.05 | 16.68± 0.11 | 15.72± 0.11 | 14.93± 0.11 | 0.96± 0.16 | 1.75± 0.16 | b,1a |
| WISE J033713.43 +114824.5 | 13.87± 0.03 | 13.53± 0.04 | 12.03± 0.32 | 0.34± 0.05 | 14.99± 0.05 | 14.55± 0.07 | 14.27± 0.08 | 0.44± 0.09 | 0.72± 0.09 | b,1a |
| WISE J044633.45 -242956.8 | 14.27± 0.03 | 13.77± 0.04 | 12.42± 0.33 | 0.51± 0.05 | 16.43± 0.12 | 15.53± 0.13 | 15.14± 0.16 | 0.90± 0.18 | 1.30± 0.20 | b,1a |
| WISE J052857.68 +090104.4 | 14.21± 0.03 | 13.64± 0.04 | 11.29± 0.16 | 0.58± 0.05 | 16.26± 0.11[p] | 15.44± 0.12 | 14.97± 0.11 | 0.82± 0.17* | 1.29± 0.16* | b,1b |
| WISEP J060738.65 +242953.4[h] | 11.54± 0.02 | 10.95± 0.02 | 10.14± 0.08 | 0.59± 0.03 | 14.22± 0.03 | 13.04± 0.03 | 12.47± 0.02 | 1.18± 0.04 | 1.75± 0.04 | --, 1a |
| " | " | " | " | " | 14.13± 0.14 | 13.18± 0.18 | --- | 0.95± 0.23 | --- | --, 2 |
| SDSS J075054.74 +445418.7[a] | 13.80± 0.03 | 13.45± 0.03 | 12.78± 0.51 | 0.35± 0.04 | 15.13± 0.06 | 14.46± 0.06 | 14.25± 0.07 | 0.68± 0.08 | 0.89± 0.10 | b,1a |
| " | " | " | " | " | 15.04± 0.04 | 14.47± 0.07 | 14.08± 0.06 | 0.57± 0.08 | 0.96± 0.07 | b, 2 |
| WISE J080700.23 +413026.8 | 13.47± 0.03 | 13.11± 0.03 | 12.09± 0.31 | 0.36± 0.04 | 15.84± 0.09 | 14.78± 0.07 | 14.35± 0.08 | 1.05± 0.11 | 1.48± 0.11 | b, 1a |
| WISE J083450.79 +642526.8 | 14.38± 0.03 | 13.93± 0.04 | 11.92± 0.20 | 0.45± 0.05 | 15.63± 0.07 | 15.12± 0.10 | 14.58± 0.09 | 0.51± 0.12 | 1.06± 0.11 | b, 1a |
| " | " | " | " | " | 15.64± 0.04 | 15.09± 0.03 | 14.60± 0.05 | 0.55± 0.05 | 1.04± 0.06 | b, 2 |
| 2MASSI J0859254 -194926[b] | 12.88± 0.02 | 12.38± 0.03 | 10.99± 0.10 | 0.50± 0.03 | 15.53± 0.05 | 14.44± 0.04 | 13.75± 0.06 | 1.09± 0.07 | 1.78± 0.08 | b, 1a |
| WISE J100926.40 +354137.5 | 13.74± 0.03 | 13.39± 0.03 | >12.27 | 0.35± 0.04 | 15.11± 0.04 | 14.47± 0.05 | 14.13± 0.07 | 0.64± 0.06 | 0.99± 0.08 | b, 1a |
| ULAS J1029 +0935[n] | 16.84± 0.13 | 14.29± 0.08 | 11.58± 0.33 | 2.55± 0.15 | 17.32± 0.01 | 17.63± 0.02 | --- | -0.31 ±0.02 | --- | t, 4 |
| WISE J105257.95 -194250.2 | 16.87± 0.15 | 14.20± 0.05 | 12.34± 0.37 | 2.67± 0.16 | 17.07± 0.20 | >16.79 | >15.34 | <0.28 | <1.73 | t, 1b |
| " | " | " | " | " | 17.15± 0.06 | 17.06± 0.12 | 16.61± 0.25 | 0.09± 0.13 | 0.54± 0.28 | t, 2 |
| WISE J113949.24 -332425.1 | 17.98± 0.31 | 14.83± 0.07 | >12.87 | 3.15± 0.31 | 17.97± 0.07 | 17.94± 0.08 | --- | 0.03± 0.11 | -- | t, 6 |
| SDSS J115553.86 +055957.5[c] | 13.30± 0.03 | 12.85± 0.03 | 12.42± 0.49 | 0.45± 0.04 | 15.66± 0.08 | 14.70± 0.07 | 14.12± 0.07 | 0.96± 0.10 | 1.54± 0.11 | b, 1a |
| WISE J124309.61 +844547.8 | >18.79 | 15.48± 0.09 | 12.37± 0.22 | >3.31 | 18.86± 0.03 | 19.21± 0.07 | --- | -0.35 ±0.08 | --- | t, 4 |
| WISE J125448.52 -072828.4 | 17.21± 0.18 | 14.85± 0.08 | >12.14 | 2.36± 0.20 | 17.30± 0.01 | 17.63± 0.03 | --- | -0.33 ±0.03 | --- | t, 4 |
| WISE J130740.45 -463035.1 | 15.00± 0.03 | 12.93± 0.03 | 11.58± 0.14 | 2.07± 0.04 | 17.17± 0.22 | 16.03± 0.18 | 15.40± 0.18 | 1.13± 0.28 | 1.77± 0.28 | b, 1b |
| " | " | " | " | " | 16.97± 0.03 | 16.24± 0.02 | --- | 0.73± 0.04 | --- | b, 3 |
| WISE J135307.51 -085712.0 | 13.73± 0.03 | 13.35± 0.03 | >12.24 | 0.39± 0.04 | >14.88 | >14.63 | 14.20± 0.08 | --- | >0.68 | b, 1a |
| " | " | " | " | " | 15.04± 0.07 | 14.45± 0.09 | 14.07± 0.10 | 0.59± 0.11 | 0.97± 0.13 | b, 2 |
| 2MASS J13580384 +1458204[d] | 13.98± 0.03 | 13.57± 0.03 | 13.04± 0.45 | 0.41± 0.04 | 16.37± 0.11 | 15.25± 0.11 | 14.66± 0.08 | 1.13± 0.16 | 1.71± 0.14 | b, 1a |
| WISE J141144.13 -140300.5 | 13.61± 0.03 | 13.24± 0.03 | 12.15± 0.26 | 0.37± 0.04 | 14.92± 0.04 | 14.41± 0.05 | 13.97± 0.07 | 0.51± 0.07 | 0.95± 0.08 | b, 1a |
| WISE J144127.48 -515807.6 | 13.16± 0.03 | 12.91± 0.03 | >12.05 | 0.25± 0.04 | 14.35± 0.03 | 13.89± 0.02 | 13.49± 0.04 | 0.46± 0.04 | 0.87± 0.04 | p, 1a |
| WISE J144806.48 -253420.3 | >18.28 | 15.03± 0.09 | >12.90 | >3.25 | 19.21± 0.12 | 18.91± 0.12 | --- | 0.30± 0.17 | --- | t, 6 |
| WISE J150406.66 -455223.9 | 14.15± 0.03 | 13.81± 0.05 | >12.34 | 0.34± 0.06 | 16.56± 0.11 | 15.43± 0.08 | 14.83± 0.10 | 1.13± 0.14 | 1.74± 0.15 | b, 1a |
| WISE J150711.06 -344026.0 | 15.63± 0.06 | 14.03± 0.05 | >11.94 | 1.60± 0.08 | >17.59 | 16.59± 0.27 | >16.65 | >1.00 | --- | b, 1b |
| WISE J151314.61 +401935.6 | 14.39± 0.03 | 13.86± 0.03 | 12.68± 0.25 | 0.54± 0.04 | >17.17 | 16.09± 0.21 | 15.09± 0.16 | >1.08 | >2.07 | b, 1b |
| " | " | " | " | " | 17.08± 0.09 | 16.15± 0.07 | 15.22± 0.10 | 0.93± 0.11 | 1.86± 0.13 | b, 2 |
| PM J15217 -2713[e] | 10.97± 0.02 | 10.78± 0.02 | 10.56± 0.09 | 0.20± 0.03 | 11.80± 0.02 | 11.34± 0.02 | 11.08± 0.03 | 0.46± 0.03 | 0.72± 0.03 | p, 1a |
| PM J15229 -0244[j] | 13.32± 0.03 | 13.06± 0.03 | 11.99± 0.24 | 0.26± 0.04 | 14.28± 0.03 | 13.80± 0.03 | 13.46± 0.05 | 0.47± 0.04 | 0.82± 0.06 | p, 1a |
| WISE | 13.90± | 13.77± | >12.15 | 0.13± | 14.65± | 14.06± | 14.04± | 0.59± | 0.61± | p, 1a |
| J152915.47 -451348.1 | 0.03 | 0.05 | | 0.06 | 0.03 | 0.05 | 0.06 | 0.06 | 0.08 | |
| WISE J154047.02 -050658.4 | 10.19± 0.02 | 10.01± 0.02 | 9.92± 0.05 | 0.18± 0.03 | 11.17± 0.02 | 10.60± 0.02 | 10.34± 0.02 | 0.58± 0.03 | 0.84± 0.03 | p, 1a |
| WISE J155254.84 +503307.6 | 15.00± 0.03 | 14.41± 0.04 | 13.50± 0.47 | 0.59± 0.05 | 17.13± 0.24 | >15.90 | 15.46± 0.16 | <1.23 | 1.66± 0.29 | b, 1b |
| " | " | " | " | " | 17.50± 0.10 | 16.08± 0.13 | 15.53± 0.13 | 1.42± 0.16 | 1.97± 0.18 | b, 2 |
| WISE J155755.29 +591425.3[l] | 12.80± 0.02 | 12.52± 0.02 | 12.40± 0.15 | 0.28± 0.03 | 14.32± 0.03 | 13.61± 0.04 | 13.12± 0.03 | 0.71± 0.05 | 1.20± 0.04 | b[q], 1a |
| WISE J160357.51 -044340.2 | 11.75± 0.02 | 11.51± 0.02 | 11.74± 0.26 | 0.24± 0.03 | 12.74± 0.03 | 12.23± 0.02 | 11.94± 0.03 | 0.51± 0.04 | 0.80± 0.04 | p, 1a |
| WISE J160603.47 -014525.0 | 11.30± 0.02 | 11.09± 0.02 | 10.81± 0.10 | 0.21± 0.03 | 12.21± 0.02 | 11.67± 0.02 | 11.48± 0.03 | 0.54± 0.03 | 0.73± 0.04 | p, 1a |
| WISE J160656.14 -020014.5 | 11.37± 0.03 | 11.16± 0.02 | 10.93± 0.12 | 0.21± 0.04 | 12.33± 0.03 | 11.81± 0.02 | 11.54± 0.02 | 0.52± 0.04 | 0.79± 0.04 | p, 1a |
| SDSS J161459.98 +400435.1[f] | 14.58± 0.03 | 14.24± 0.04 | >13.61 | 0.34± 0.04 | 16.57± 0.12 | 15.84± 0.15 | 15.01± 0.12 | 0.73± 0.19 | 1.56± 0.17 | b, 1a |
| " | " | " | " | " | 16.51± 0.07 | 15.43± 0.06 | 15.18± 0.10 | 1.08± 0.09 | 1.33± 0.12 | b, 2 |
| WISE J162359.70 -050811.4 | 13.21± 0.03 | 12.94± 0.03 | 11.64± 0.22 | 0.27± 0.04 | 14.94± 0.04 | 14.08± 0.03 | 13.56± 0.04 | 0.87± 0.05 | 1.39± 0.06 | p, 1a |
| WISE J164031.21 -010313.9 | 10.08± 0.02 | 9.92± 0.02 | 9.92± 0.06 | 0.16± 0.03 | 11.00± 0.02 | 10.47± 0.02 | 10.20± 0.02 | 0.53± 0.03 | 0.80± 0.04 | p, 1a |
| WISE J165842.56 +510335.0 | 13.18± 0.02 | 12.80± 0.02 | 12.26± 0.19 | 0.38± 0.03 | 15.06± 0.04 | 14.18± 0.03 | 13.66± 0.04 | 0.89± 0.05 | 1.41± 0.05 | b, 1a |
| " | " | " | " | " | 15.09± 0.12 | 14.09± 0.08 | 13.59± 0.11 | 1.00± 0.14 | 1.50± 0.16 | b, 2 |
| WISE J170353.06 -033748.7 | 10.84± 0.03 | 10.66± 0.02 | 10.40± 0.08 | 0.18± 0.04 | 11.88± 0.03 | 11.34± 0.03 | 11.02± 0.02 | 0.54± 0.04 | 0.87± 0.04 | p, 1a |
| PM J17137 -4535[k] | 13.52± 0.04 | 13.64± 0.06 | >11.04 | -0.12± 0.07 | 14.42± 0.03 | 13.92± 0.04 | 13.74± 0.04 | 0.49± 0.05 | 0.68± 0.05 | p, 1a |
| WISE J173332.50 +314458.3 | 13.58± 0.03 | 13.27± 0.03 | >12.99 | 0.31± 0.04 | 15.87± 0.07 | 14.89± 0.06 | 14.30± 0.06 | 0.98± 0.09 | 1.57± 0.09 | p, 1a |
| WISE J174928.57 -380401.6 | 13.42± 0.04 | 13.48± 0.07 | >12.47 | -0.06± 0.08 | 14.16± 0.03 | 13.59± 0.02 | 13.46± 0.04 | 0.56± 0.04 | 0.69± 0.05 | p, 1a |
| WISEPA J183058.57 +454257.9[g] | 14.81± 0.03 | 14.17± 0.04 | >13.17 | 0.65± 0.05 | >18.75 | 16.08± 0.18 | 15.37± 0.18 | >2.67 | >3.38 | b, 1b |
| " | " | " | " | " | >17.95 | 16.12± 0.13 | 14.98± 0.11 | >1.83 | >2.97 | b, 2 |
| WISE | 10.25± | 10.15± | 10.24± | 0.10± | 11.23± | 10.70± | 10.47± | 0.52± | 0.76± | p, 1a |
| J183921.35 -374431.0 | 0.05 | 0.05 | 0.09 | 0.07 | 0.02 | 0.02 | 0.02 | 0.03 | 0.03 | |
| WISE J185101.83 +593508.6 | 12.65± 0.02 | 12.18± 0.02 | 11.23± 0.07 | 0.47± 0.03 | 14.94± 0.04 | 13.97± 0.04 | 13.46± 0.05 | 0.97± 0.05 | 1.48± 0.06 | b, 1a |
| " | " | " | " | " | 14.94± 0.06 | 13.96± 0.05 | 13.40± 0.10 | 0.98± 0.08 | 1.54± 0.12 | b, 2 |
| WISE J191915.54 +304558.4 | 13.39± 0.03 | 12.94± 0.03 | 11.72± 0.19 | 0.45± 0.04 | 15.57± 0.05 | 14.60± 0.05 | 13.95± 0.05 | 0.97± 0.07 | 1.62± 0.07 | b, 1a |
| WISE J195113.62 -331116.7 | 14.06± 0.03 | 13.71± 0.04 | >12.73 | 0.35± 0.05 | 15.71± 0.06 | 15.02± 0.07 | 14.51± 0.07 | 0.69± 0.09 | 1.20± 0.09 | b, 1a |
| " | " | " | " | " | 15.79± 0.06 | 14.97± 0.07 | 14.46± 0.09 | 0.82± 0.09 | 1.33± 0.11 | b, 2 |
| WISE J200403.17 -263751.7 | 13.97± 0.03 | 13.60± 0.04 | >12.70 | 0.37± 0.05 | 15.19± 0.04 | 14.64± 0.05 | 14.09± 0.05 | 0.55± 0.06 | 1.10± 0.06 | b, 1a |
| " | " | " | " | " | 15.17± 0.06 | 14.57± 0.07 | 13.99± 0.09 | 0.60± 0.09 | 1.18± 0.11 | b, 2 |
| WISE J222219.93 +302601.4 | 14.00± 0.03 | 13.58± 0.04 | >12.55 | 0.42± 0.05 | 16.55± 0.11 | 15.60± 0.10 | 15.18± 0.14 | 0.95± 0.15 | 1.38± 0.18 | b, 1a |
| " | " | " | " | " | 15.89± 0.06 | 15.49± 0.06 | 14.91± 0.07 | 0.40± 0.08 | 0.98± 0.09 | b, 5 |
| WISE J233527.07 +451140.9 | 13.48± 0.03 | 12.93± 0.03 | 12.72± 0.54 | 0.55± 0.04 | 16.70± 0.19 | >15.26 | >14.46 | <1.44 | <2.24 | b, 1b |
| WISE J234755.42 +683319.4 | 13.26± 0.05 | 12.95± 0.04 | 12.40± 0.50 | 0.31± 0.06 | 14.26± 0.03 | 13.77± 0.03 | 13.59± 0.04 | 0.49± 0.04 | 0.68± 0.05 | p, 1a |

[a] Discovered by West et al. (2008). Also known as WISE J075054.65+445416.2.

[b] Discovered by Cruz et al. (2003). Also known as WISE J085925.22-194927.9.

[c] Discovered by Knapp et al. (2004). Also known as WISE J115553.58+055957.0.

[d] Discovered by Sheppard & Cushing (2009). Also known as WISE J135803.55+145822.7.

[e] Discovered by Lepine (2008). Also known as WISE J152145.89-271309.0.

[f] Discovered by Zhang et al. (2009). Also known as WISE J161459.80+400436.4.

[g] Discovered by Kirkpatrick et al. (2011). Also known as WISE J183058.56+454257.4.

[h] Discovered by Castro & Gizis (2012), not independently discovered by us, but listed/re-observed for comparison purposes. Also known as WISE J060738.65+242953.5.

[i] Discovered by Gizis et al. (2012). Also known as WISE J004701.07+680352.1.

[j] Discovered by Lepine (2008). Also known as WISE J152258.92-024456.5.

[k] Discovered by Lepine (2008). Also known as WISE J171343.45-453601.4 .

[l] Companion to G 225-36. See section 4.1.

[m] Lies 30 arcsec away from [WKA2010] Core 5, a dense core inside a molecular cloud.

[n] Discovered by Burningham et al. (2013). Also known as WISE J102940.51+093514.1

[p] The *J*-band magnitude of this source comes primarily from a latent artifact caused by a bright star, so this measurement should be treated with caution.

[q] This source was selected from the WISE Pass-2 4-Band Level 1 Source Table using criteria paralleling those of the bright L and T dwarf search.

Notes: The final column is comprised of two codes separated by commas. The first indicates the WISE search (see section 2) from which the object was drawn: (b) the bright L and T dwarf search, (t) the late-T dwarf search, or (p) the proper motion search. The second indicates the catalog or instrument from which the $JHK_S$ photometry was taken: (1a) 2MASS All-Sky Point Source Catalog, (1b) 2MASS Survey Point Source Reject Table, (2) PAIRITEL, (3) CTIO4m/NEWFIRM, (4) Palomar/WIRC, (5) Bigelow/2MASS, (6) SOAR/OSIRIS.

*Table 2: Spitzer/IRAC Photometry of Objects in Table 1*

| Abbreviated Designation | ch 1 (mag) | ch2 (mag) | ch1-ch2 (mag) | Program Number |
|---|---|---|---|---|
| WISE 0200+8742 | 13.10±0.02 | 12.93±0.02 | 0.17±0.03 | 70062 |
| WISE 0230-0225 | 13.82±0.02 | 13.65±0.02 | 0.17±0.03 | 70062 |
| WISE 0337+1148 | 13.71±0.02 | 13.63±0.02 | 0.08±0.03 | 70062 |
| SDSS 0750+4454 | 13.54±0.02 | 13.46±0.02 | 0.08±0.03 | 70062 |
| WISE 0807+4130 | 13.14±0.02 | 13.13±0.02 | 0.01±0.03 | 70062 |
| WISE 0834+6425 | 14.14±0.02 | 14.06±0.02 | 0.08±0.03 | 70062 |
| ULAS 1009+3541 | 13.54±0.02 | 13.43±0.02 | 0.11±0.03 | 70062 |
| WISE 1029+0935 | 16.08±0.03 | 14.46±0.02 | 1.62±0.04 | 80109 |
| WISE 1052-1942 | 15.66±0.03 | 14.22±0.02 | 1.44±0.04 | 70062 |
| WISE 1139-3324 | 16.89±0.05 | 15.00±0.02 | 1.90±0.05 | 80109 |
| WISE 1243+8445 | 17.36±0.06 | 15.40±0.02 | 1.95±0.06 | 80109 |
| WISE 1254-0728 | 16.33±0.04 | 14.82±0.02 | 1.51±0.04 | 80109 |
| WISE 1307-4630 | 14.31±0.02 | 13.07±0.02 | 1.24±0.03 | 70062 |
| WISE 1353-0857 | 13.52±0.02 | 13.41±0.02 | 0.11±0.03 | 70062 |
| WISE 1411-1403 | 13.37±0.02 | 13.26±0.02 | 0.11±0.03 | 70062 |
| WISE 1448-2534 | 17.20±0.06 | 15.09±0.02 | 2.11±0.06 | 80109 |
| WISE 1507-3440 | 15.94±0.03 | 14.70±0.02 | 1.24±0.03 | 70062 |
| WISE 1513+4019 | 14.05±0.02 | 13.89±0.02 | 0.16±0.03 | 70062 |
| SDSS 1614+4004 | 14.26±0.02 | 14.26±0.02 | 0.00±0.03 | 70062 |
| WISE 1658+5103 | 12.87±0.02 | 12.83±0.02 | 0.04±0.03 | 70062 |
| WISE 1830+4542 | 14.29±0.02 | 14.13±0.02 | 0.16±0.03 | 70062 |
| WISE 1851+5935 | 12.28±0.02 | 12.20±0.02 | 0.08±0.03 | 70062 |
| WISE 1951-3311 | 13.78±0.02 | 13.79±0.02 | -0.01±0.03 | 70062 |
| WISE 2004-2637 | 13.71±0.02 | 13.60±0.02 | 0.11±0.03 | 70062 |
| WISE 2222+3026 | 13.83±0.02 | 13.74±0.02 | 0.09±0.03 | 70062 |

*Table 3: Spectroscopic Observation Log*

| Abbreviated Designation | Near-infrared Spectral Type[a] | Telescope/ Instrument | Observation Date (UT) | Telluric Corrector Star |
|---|---|---|---|---|
| WISE 0031+5749 | L8 | Palomar/TSpec | 2012 Jan 06 | HD 223386 |
| WISE 0047+6803 | L9 pec (v. red) | Palomar/TSpec | 2012 Jan 07 | HD 1287 |
| WISE 0200+8742 | embedded star? | APO/TSpec | 2012 Feb 14 | HD 8991 |
| WISE 0230-0225 | L8: pec | APO/TSpec | 2012 Feb 14 | HD 16140 |
| WISE 0337+1148 | M7 | IRTF/SpeX | 2012 Jan 31 | HD 25175 |
| WISE 0446-2429 | L5 pec (blue) | IRTF/SpeX | 2012 Jan 31 | HD 29433 |
| WISE 0528+0901 | late-M giant | IRTF/SpeX | 2012 Jan 31 | HD 35036 |
| WISE 0607+2429 | L9 | IRTF/SpeX | 2012 Feb 01 | HD 43607 |
| SDSS 0750+4454 | M8 pec | IRTF/SpeX | 2012 Feb 12 | HD 58296 |
| WISE 0807+4130 | L8 pec | IRTF/SpeX | 2012 Jan 31 | HD 71906 |
| WISE 0834+6425 | M8 | IRTF/SpeX | 2012 Jan 31 | HD 237611 |
| 2MASS 0859-1949 | L8 | IRTF/SpeX | 2012 Jan 31 | HD 82724 |
| WISE 1009+3541 | M8 | IRTF/SpeX | 2012 Feb 12 | HD 89239 |
| ULAS 1029+0935 | T8 | IRTF/SpeX | 2012 Dec 24 | HD 85504 |
| WISE 1052-1942 | T7.5 | IRTF/SpeX | 2012 Dec 24 | HD 90606 |
| WISE 1139-3324 | T7 | Keck/NIRSPEC | 2013 Feb 21 | HD 101169 |
| SDSS 1155+0559 | L8.5 | Palomar/TSpec | 2012 Jan 07 | HD 111744 |
| WISE 1243+8445 | T9 | Keck/NIRSPEC | 2013 Feb 21 | HD 107193 |
| WISE 1254-0728 | T7 | IRTF/SpeX | 2012 Dec 24 | HD 109309 |
| WISE 1307-4630 | mid-M giant? | IRTF/SpeX | 2012 May 13 | HD 115527 |
| WISE 1353-0857 | L0<br>L0 | Palomar/TSpec<br>IRTF/SpeX | 2011 Jul 13<br>2012 Jul 19 | HD 134013<br>HD 122749 |
| 2MASS 1358+1458 | L4 pec (red) | Palomar/TSpec | 2012 Jan 07 | HD 131951 |
| WISE 1411-1403 | M8 pec? | IRTF/SpeX | 2011 Jul 28 | HD 126818 |
| WISE 1441-5158 | M7 | IRTF/SpeX | 2012 Jul 19 | HD 120077 |
| WISE 1448-2534 | T8 | Keck/NIRSPEC | 2013 Feb 21 | HD 129544 |
| WISE 1504-4552 | L8 | IRTF/SpeX | 2012 Feb 13 | HD 132302 |
| WISE 1507-3440 | mid-M giant? | IRTF/SpeX | 2011 Jul 29 | HD 134685 |
| WISE 1513+4019 | L8 | IRTF/SpeX | 2012 Feb 12 | HD 128039 |
| PM 1521-2713 | M3 | IRTF/SpeX | 2011 Jul 29 | HD 131885 |
| PM 1522-0244 | sdM6 | IRTF/SpeX | 2012 Jul 23 | HD 144980 |
| WISE 1529-4513 | sdM0? | IRTF/SpeX | 2012 Jul 19 | HD 137957 |
| WISE 1540-0506 | M3.5 | IRTF/SpeX | 2012 Jul 17 | HD 14698 |
| WISE 1552+5033 | L9: pec | IRTF/SpeX | 2012 Feb 12 | HD 143187 |
| WISE 1557+5914 | M9 | IRTF/SpeX | 2011 Jul 23 | HD 143187 |
| WISE 1603-0443 | M5 | IRTF/SpeX | 2012 Jul 23 | HD 144980 |
| WISE 1606-0145 | M3 | IRTF/SpeX | 2012 Jul 23 | HD 144980 |
| WISE 1606-0200 | M3.5 | IRTF/SpeX | 2012 Jul 23 | HD 144980 |
| SDSS 1614+4004 | L2 | IRTF/SpeX | 2012 Feb 12 | HD 143187 |
| WISE 1623-0508 | L1 | IRTF/SpeX | 2012 Jul 23 | HD 148968 |
| WISE 1640-0103 | M3 | IRTF/SpeX | 2012 Jul 17 | HD 148968 |
| WISE 1658+5103 | L6 pec | Palomar/TSpec | 2011 Jul 14 | HD 179933 |
| WISE 1703-0337 | M4.5 | IRTF/SpeX | 2012 Jul 17 | HD 159008 |
| PM 1713-4535 | sdM0? | IRTF/SpeX | 2012 Jul 19 | HD 161706 |
| WISE 1733+3144 | L2 pec (red) | IRTF/SpeX | 2012 Jul 23 | HD 167163 |
| WISE 1749-3804 | white dwarf? | IRTF/SpeX | 2012 Jul 23 | HD 159312 |
| WISE 1830+4542 | L9 | Palomar/TSpec | 2011 Jul 13 | HD 165029 |
| WISE 1839-3744 | M0 | IRTF/SpeX | 2012 Jul 23 | HD 159312 |
| WISE 1851+5935 | L9 pec | Palomar/TSpec | 2011 Jul 13 | HD 143187 |
| WISE 1919+3045 | L6 | Keck/NIRSPEC | 2012 Sep 25 | HD 205314 |
| WISE 1951-3311 | L1 pec? | IRTF/SpeX | 2011 Jul 23 | HD 193130 |
| WISE 2004-2637 | L0 | Keck/NIRSPEC | 2012 Sep 25 | HD 193130 |
| WISE 2222+3026 | L9 | Palomar/TSpec | 2011 Jul 13 | HD 210501 |
| WISE 2335+4511 | L9 pec (v. red) | Palomar/TSpec | 2012 Jan 06 | HD 223386 |
| WISE 2347+6833 | M7 | APO/TSpec | 2012 Jul 17 | HD 2904 |

[a] All objects are dwarfs unless otherwise noted.

*Table 4: Astrometry of Candidates*

| Abbrev. Designation | WISE RA | WISE Dec | WISE MJD | 2MASS RA | 2MASS Dec | 2MASS MJD |
|---|---|---|---|---|---|---|
| WISE 0031+5749 | 7.7918363d ±0.08″ | 57.8267566d ±0.08″ | 55313.1 | 7.788698d ±0.06″ | 57.826797d ±0.06″ | 51173.2 |
| WISE 0047+6803 | 11.7544608d ±0.07″ | 68.0644947d ±0.07″ | 55231.4 | 11.751611d ±0.06″ | 68.065102d ±0.07″ | 51497.2 |
| WISE 0200+8742 | 30.0475123d ±0.09″ | 87.7020071d ±0.09″ | 55265.8 | 30.047993d ±0.09″ | 87.701996d ±0.09″ | 51810.3 |
| WISE 0230-0225 | 37.6621154d ±0.12″ | -2.4316704d ±0.11″ | 55306.4 | 37.661059d ±0.13″ | -2.431835d ±0.13″ | 51087.3 |
| WISE 0337+1148 | 54.3059668d ±0.12″ | 11.8068188d ±0.13″ | 55235.7 | 54.305836d ±0.08″ | 11.807791d ±0.09″ | 51520.3 |
| WISE 0446-2429 | 71.6393766d ±0.06″ | -24.4991260d ±0.06″ | 55245.7 | 71.640094d ±0.19″ | -24.500395d ±0.22″ | 51169.1 |
| WISE 0528+0901 | 82.2403666d ±0.13″ | 9.0178898d ±0.14″ | 55263.0 | 82.240407d ±0.09″ | 9.017970d ±0.10″ | 51569.2 |
| WISE 0607+2429 | 91.9110660d ±0.08″ | 24.4982019d ±0.08″ | 55272.0 | 91.912850d ±0.06″ | 24.499292d ±0.07″ | 50821.2 |
| SDSS 0750+4454 | 117.7277475d ±0.11″ | 44.9045105d ±0.12″ | 55292.1 | 117.728297d ±0.07″ | 44.905720d ±0.07″ | 50930.1 |
| WISE 0807+4130 | 121.7509686d ±0.10″ | 41.5074472d ±0.11″ | 55295.8 | 121.751035d ±0.07″ | 41.508476d ±0.07″ | 51293.1 |
| WISE 0834+6425 | 128.7116332d ±0.13″ | 64.4241249d ±0.14″ | 55293.0 | 128.712544d ±0.07″ | 64.424881d ±0.07″ | 51191.4 |
| 2MASS 0859-1949 | 134.8550863d ±0.09″ | -19.8244414d ±0.09″ | 55325.9 | 134.856160d ±0.06″ | -19.824135d ±0.07″ | 51219.2 |
| WISE 1009+3541 | 152.3600375d ±0.11″ | 35.6937538d ±0.11″ | 55323.0 | 152.359742d ±0.08″ | 35.694874d ±0.13″ | 50896.3 |
| ULAS 1029+0935 | 157.4188151d ±0.22″ | 9.5872643d ±0.25″ | 55338.9 | not seen | not seen | --- |
| WISE 1052-1942 | 163.2414590d ±0.33″ | -19.7139546d ±0.35″ | 55355.5 | 163.240496d ±0.25″ | -19.713057d ±0.22″ | 51242.1 |
| WISE 1139-3324 | 174.9551693d ±0.46″ | -33.4069796d ±0.50″ | 55364.0 | not seen | not seen | --- |
| SDSS 1155+0559 | 178.9732639d ±0.10″ | 5.9991826d ±0.10″ | 55359.9 | 178.974543d ±0.16″ | 5.999362d ±0.10″ | 51611.3 |
| WISE 1243+8445 | 190.7900466d ±0.62″ | 84.7632977d ±0.56″ | 55284.8 | not seen | not seen | --- |
| WISE 1254-0728 | 193.7021720d ±0.47″ | -7.4745611d ±0.52″ | 55333.4 | not seen | not seen | --- |
| WISE 1307-4630 | 196.9185443d ±0.11″ | -46.5097703d ±0.11″ | 55322.5 | 196.918616d ±0.23″ | -46.509792d ±0.26″ | 51306.1 |
| WISE 1353-0857 | 208.2813131d ±0.11″ | -8.9533420d ±0.11″ | 55313.1 | 208.282433d ±0.11″ | -8.953315d ±0.13″ | 51238.3 |
| 2MASS 1358+1458 | 209.5148273d ±0.11″ | 14.9729891d ±0.11″ | 55304.5 | 209.516001d ±0.15″ | 14.972354d ±0.11″ | 51530.6 |
| WISE 1411-1403 | 212.9338949d ±0.10″ | -14.0501475d ±0.11″ | 55319.1 | 212.934120d ±0.07″ | -14.048441d ±0.07″ | 50900.4 |
| WISE 1441-5158 | 220.3645410d ±0.10″ | -51.9687854d ±0.10″ | 55241.8 | 220.368966d ±0.06″ | -51.968002d ±0.06″ | 51642.1 |
| WISE 1448-2534 | 222.0270415d ±0.56″ | -25.5723322d ±0.62″ | 55365.9 | not seen | not seen | --- |
| WISE 1504-4552 | 226.0277634d ±0.15″ | -45.8733150d ±0.16″ | 55243.4 | 226.031017d ±0.09″ | -45.872227d ±0.08″ | 51334.2 |
| WISE 1507-3440 | 226.7961225d ±0.23″ | -34.6739109d ±0.26″ | 55240.7 | 226.795999d ±0.29″ | -34.673786d ±0.29″ | 50995.0 |
| WISE 1513+4019 | 228.3109124d ±0.11″ | 40.3265730d ±0.11″ | 55320.5 | 228.310568d ±0.28″ | 40.326389d ±0.26″ | 50926.4 |
| PM 1521-2713 | 230.4412098d ±0.07″ | -27.2191864d ±0.07″ | 55241.6 | 230.442335d ±0.06″ | -27.217533d ±0.06″ | 50992.1 |
| PM 1522-0244 | 230.7455367d ±0.10″ | -2.7490508d ±0.10″ | 55292.1 | 230.747169d ±0.07″ | -2.748082d ±0.06″ | 51233.4 |
| WISE 1529-4513 | 232.3144592d ±0.15″ | -45.2300549d ±0.16″ | 55247.6 | 232.315332d ±0.06″ | -45.229584d ±0.06″ | 51317.2 |
| WISE 1540-0506 | 235.1959207d ±0.07″ | -5.1162402d ±0.07″ | 55240.7 | 235.196841d ±0.06″ | -5.116612d ±0.06″ | 51256.3 |
| WISE 1552+5033 | 238.2285193d ±0.12″ | 50.5521274d ±0.12″ | 55333.8 | 238.228843d ±0.30″ | 50.552792d ±0.27″ | 51600.4 |
| WISE 1557+5914 | 239.4804110d ±0.08″ | 59.2403728d ±0.08″ | 55320.0 | 239.482053d ±0.14″ | 59.239784d ±0.06″ | 51601.4 |
| WISE 1603-0443 | 240.9896262d ±0.08″ | -4.7278419d ±0.08″ | 55246.0 | 240.989121d ±0.06″ | -4.726888d ±0.06″ | 51242.4 |
| WISE 1606-0145 | 241.5144615d ±0.08″ | -1.7569526d ±0.07″ | 55246.0 | 241.514127d ±0.07″ | -1.756210d ±0.06″ | 51242.4 |
| WISE 1606-0200 | 241.7339484d ±0.09″ | -2.0040494d ±0.08″ | 55246.2 | 241.734810d ±0.06″ | -2.004060d ±0.06″ | 51243.3 |
| SDSS 1614+4004 | 243.7492007d ±0.11″ | 40.0767799d ±0.12″ | 55316.5 | 243.750310d ±0.17″ | 40.076252d ±0.17″ | 50932.3 |
| WISE 1623-0508 | 245.9987560d ±0.10″ | -5.1365016d ±0.10″ | 55250.8 | 245.999230d ±0.06″ | -5.135186d ±0.06″ | 51258.3 |
| WISE 1640-0103 | 250.1300526d ±0.07″ | -1.0538825d ±0.07″ | 55254.1 | 250.129874d ±0.06″ | -1.052876d ±0.06″ | 51261.4 |
| WISE 1658+5103 | 254.6773644d ±0.09″ | 51.0597243d ±0.08″ | 55242.6 | 254.678760d ±0.06″ | 51.060654d ±0.06″ | 50978.3 |
| WISE 1703-0337 | 255.9710869d ±0.08″ | -3.6302011d ±0.08″ | 55260.2 | 255.971160d ±0.06″ | -3.629341d ±0.06″ | 51268.3 |
| PM 1713-4535 | 258.4310592d ±0.19″ | -45.6003972d ±0.21″ | 55267.9 | 258.431610d ±0.06″ | -45.599079d ±0.06″ | 51374.1 |
| WISE 1733+3144 | 263.3854377d ±0.05″ | 31.7495376d ±0.05″ | 55266.9 | 263.386723d ±0.06″ | 31.749222d ±0.06″ | 50914.5 |
| WISE 1749-3804 | 267.3690464d ±0.16″ | -38.0671275d ±0.18″ | 55274.1 | 267.369082d ±0.06″ | -38.065998d ±0.06″ | 51764.0 |
| WISE 1830+4542 | 277.7440380d ±0.07″ | 45.7159605d ±0.07″ | 55291.0 | 277.743778d ±0.26″ | 45.715626d ±0.26″ | 50969.3 |
| WISE 1839-3744 | 279.8389836d ±0.11″ | -37.7419704d ±0.20″ | 55284.2 | 279.838598d ±0.06″ | -37.741276d ±0.06″ | 51046.1 |
| WISE 1851+5935 | 282.7576554d ±0.08″ | 59.5857394d ±0.08″ | 55322.2 | 282.757433d ±0.07″ | 59.584461d ±0.07″ | 51337.3 |
| WISE 1919+3045 | 289.8147583d ±0.11″ | 30.7662323d ±0.09″ | 55305.1 | 289.813314d ±0.06″ | 30.764889d ±0.06″ | 50930.5 |
| WISE 1951-3311 | 297.8067590d ±0.13″ | -33.1879795d ±0.13″ | 55300.4 | 297.806376d ±0.06″ | -33.186848d ±0.06″ | 51374.3 |
| WISE 2004-2637 | 301.0132171d ±0.12″ | -26.6310349d ±0.13″ | 55304.3 | 301.012951d ±0.06″ | -26.629114d ±0.06″ | 51036.1 |
| WISE 2222+3026 | 335.5830545d ±0.12″ | 30.4337426d ±0.12″ | 55359.8 | 335.582884d ±0.20″ | 30.432877d ±0.19″ | 50982.4 |
| WISE 2335+4511 | 353.8627971d ±0.10″ | 45.1947010d ±0.10″ | 55387.2 | 353.863133d ±0.22″ | 45.194889d ±0.20″ | 51115.1 |
| WISE 2347+6833 | 356.9809348d ±0.14″ | 68.5553916d ±0.22″ | 55225.8 | 356.975530d ±0.06″ | 68.555443d ±0.07″ | 51716.4 |

*Table 5: Proper Motions and Estimates of Distances and Tangential Velocities*

| Abbreviated Designation | $\mu_{RA}$ (″/yr) | $\mu_{Dec}$ (″/yr) | $\mu_{total}$ (″/yr) | Distance Estimate (pc) | Tangential Velocity (km/s) |
|---|---|---|---|---|---|
| WISE 0031+5749 | 0.53±0.01 | -0.01±0.01 | 0.53±0.01 | 11±1 | 28±2 |
| WISE 0047+6803 | 0.37±0.01 | -0.21±0.01 | 0.43±0.01 | 15±1 | 30±2 |
| WISE 0200+8742 | -0.01±0.01 | 0.00±0.01 | 0.01±0.02 | --- | --- |
| WISE 0230-0225 | 0.33±0.02 | 0.05±0.01 | 0.33±0.02 | 25±2 | 38±3 |
| WISE 0337+1148 | 0.05±0.01 | -0.34±0.02 | 0.35±0.02 | 71±4 | 120±10 |
| WISE 0446-2429 | -0.21±0.02 | 0.41±0.02 | 0.46±0.03 | 35±2 | 76±7 |
| WISE 0528+0901 | -0.01±0.02 | -0.03±0.02 | 0.03±0.02 | --- | --- |
| WISE 0607+2429 | -0.48±0.01 | -0.32±0.01 | 0.58±0.02 | 8±1 | 21±1 |
| SDSS 0750+4454 | -0.12±0.01 | -0.36±0.01 | 0.38±0.02 | 60±4 | 110±10 |
| WISE 0807+4130 | -0.02±0.01 | -0.34±0.01 | 0.34±0.02 | 17±1 | 27±2 |
| WISE 0834+6425 | -0.13±0.01 | -0.24±0.01 | 0.27±0.02 | 79±5 | 100±10 |
| 2MASS 0859-1949 | -0.32±0.01 | -0.10±0.01 | 0.34±0.01 | 14±1 | 23±2 |
| WISE 1009+3541 | 0.07±0.01 | -0.33±0.01 | 0.34±0.02 | 62±4 | 100±8 |
| ULAS 1029+0935 | --- | --- | --- | 14±1 | --- |
| WISE 1052-1942 | 0.29±0.04 | -0.29±0.04 | 0.41±0.05 | 15±1 | 29±4 |
| WISE 1139-3324 | --- | --- | --- | 25±2 | --- |
| SDSS 1155+0559 | -0.45±0.02 | -0.06±0.01 | 0.45±0.02 | 15±1 | 32±2 |
| WISE 1243+8445 | --- | --- | --- | 16±3 | --- |
| WISE 1254-0728 | --- | --- | --- | 19±1 | --- |
| WISE 1307-4630 | -0.02±0.02 | 0.01±0.03 | 0.02±0.03 | --- | --- |
| WISE 1353-0857 | -0.36±0.01 | -0.01±0.02 | 0.36±0.02 | 44±3 | 75±6 |
| 2MASS 1358+1458 | -0.40±0.02 | 0.22±0.02 | 0.45±0.02 | 43±3 | 91±7 |
| WISE 1411-1403 | -0.06±0.01 | -0.51±0.01 | 0.51±0.01 | 57±3 | 137±9 |
| WISE 1441-5158 | -1.00±0.01 | -0.29±0.01 | 1.04±0.02 | 53±3 | 260±20 |
| WISE 1448-2534 | --- | --- | --- | 33±2 | --- |
| WISE 1504-4552 | -0.76±0.02 | -0.37±0.02 | 0.85±0.02 | 23±1 | 93±6 |
| WISE 1507-3440 | 0.03±0.03 | -0.04±0.03 | 0.05±0.05 | --- | --- |
| WISE 1513+4019 | 0.08±0.03 | 0.06±0.02 | 0.10±0.03 | 29±2 | 14±4 |
| PM 1521-2713 | -0.31±0.01 | -0.51±0.01 | 0.60±0.01 | 87±5 | 250±20 |
| PM 1522-0244 | -0.53±0.01 | -0.31±0.01 | 0.61±0.02 | ~81[a] | ~234[a] |
| WISE 1529-4513 | -0.21±0.02 | -0.16±0.02 | 0.26±0.02 | ~223[b] | ~275[b] |
| WISE 1540-0506 | -0.30±0.01 | 0.12±0.01 | 0.33±0.01 | 45±3 | 70±5 |
| WISE 1552+5033 | -0.07±0.03 | -0.23±0.03 | 0.25±0.04 | 35±2 | 41±7 |
| WISE 1557+5914 | -0.30±0.02 | 0.21±0.01 | 0.36±0.02 | 34±2 | 58±5 |
| WISE 1603-0443 | 0.17±0.01 | -0.31±0.01 | 0.35±0.01 | 61±4 | 101±7 |
| WISE 1606-0145 | 0.11±0.01 | -0.24±0.01 | 0.27±0.01 | 105±6 | 130±10 |
| WISE 1606-0200 | -0.28±0.01 | 0.00±0.01 | 0.28±0.01 | 76±5 | 101±7 |
| SDSS 1614+4004 | -0.25±0.02 | 0.16±0.02 | 0.30±0.02 | 70±4 | 99±9 |
| WISE 1623-0508 | -0.16±0.01 | -0.43±0.01 | 0.46±0.02 | 39±2 | 85±6 |
| WISE 1640-0103 | 0.06±0.01 | -0.33±0.01 | 0.34±0.01 | 60±4 | 97±6 |
| WISE 1658+5103 | -0.27±0.01 | -0.29±0.01 | 0.39±0.01 | 15±1 | 28±2 |
| WISE 1703-0337 | -0.02±0.01 | -0.28±0.01 | 0.28±0.01 | 47±3 | 62±4 |
| PM 1713-4535 | -0.13±0.02 | -0.45±0.02 | 0.46±0.03 | ~200[b] | ~436[b] |
| WISE 1733+3144 | -0.33±0.01 | 0.10±0.01 | 0.34±0.01 | 52±3 | 84±6 |
| WISE 1749-3804 | -0.01±0.02 | -0.42±0.02 | 0.42±0.03 | --- | --- |
| WISE 1830+4542 | 0.06±0.02 | 0.10±0.02 | 0.12±0.03 | 43±3 | 24±6 |
| WISE 1839-3744 | 0.09±0.01 | -0.22±0.02 | 0.24±0.02 | 115±7 | 130±10 |
| WISE 1851+5935 | 0.04±0.01 | 0.42±0.01 | 0.42±0.01 | 11±1 | 21±1 |
| WISE 1919+3045 | 0.37±0.01 | 0.40±0.01 | 0.55±0.01 | 19±1 | 49±3 |
| WISE 1951-3311 | 0.11±0.01 | -0.38±0.01 | 0.39±0.02 | 55±3 | 102±8 |
| WISE 2004-2637 | 0.07±0.01 | -0.59±0.01 | 0.60±0.02 | 47±3 | 134±9 |
| WISE 2222+3026 | 0.04±0.02 | 0.26±0.02 | 0.26±0.03 | 17±1 | 20±3 |
| WISE 2335+4511 | -0.07±0.02 | -0.06±0.02 | 0.09±0.03 | 24±2 | 10±3 |
| WISE 2347+6833 | 0.74±0.02 | -0.02±0.02 | 0.74±0.03 | 51±3 | 180±10 |

[a] Estimates use $M_J$ ~ 9.74 mag, which is the value measured for the sdM6.5 dwarf LHS 1166, the closest match in spectral type with an available parallax (van Altena et al. 1995).

[b] Estimates, which assume a tentative classification of sdM0, are discussed in section 4.1.

*Table 6: Proper Motion Measurements and Distance Estimates for Re-discoveries*

| Abbrev. Designation | Our $\mu_{RA}$ (″/yr) | Our $\mu_{Dec}$ (″/yr) | Our Dist. (pc) | Our Spec. Type | Pub. $\mu_{RA}$ (″/yr) | Pub. $\mu_{Dec}$ (″/yr) | Pub. Dist. (pc) | Pub. Spec. Type | Ref. |
|---|---|---|---|---|---|---|---|---|---|
| WISE 0047+6803 | 0.37 ±0.01 | -0.21 ±0.01 | 15±1 | L9 pec (v. red) | 0.38 ±0.012 | -0.212 ±0.012 | ~10-20 | L7.5 pec | 7 |
| WISE 0607+2429 | -0.48 ±0.01 | -0.32 ±0.01 | 8±1 | L9 | -0.47 ±0.01 | -0.33 ±0.02 | 7.8±1.4 | L8[a] | 8 |
| SDSS 0750+4454 | -0.12 ±0.01 | -0.37 ±0.01 | 65±4 | M8 pec | 0±0 | 0±0 | 79 | L0[a] | 1 |
| 2MASS 0859-1949 | -0.32 ±0.01 | -0.10 ±0.01 | 16±1 | L8 | -0.325 ±0.004 | -0.100 ±0.004 | 19±2 | L7[a] | 2, 3 |
| ULAS 1029+0935 | --- | --- | 14±1 | T8 | --- | --- | 6-24 | T8 | 10 |
| SDSS 1155+0559 | -0.45 ±0.02 | -0.06 ±0.01 | 18±1 | L8.5 | -0.421 ±0.006 | -0.055 ±0.005 | 17±3 | L7.5 | 2 |
| 2MASS 1358+1458 | -0.40 ±0.02 | 0.22 ±0.01 | 50±3 | L4 pec (red) | -0.45 | 0.239 | 90 | L0 | 4 |
| PM J15217-2713 | -0.31 ±0.01 | -0.51 ±0.01 | | M3 | -0.323 ±0.019 | -0.499 ±0.019 | -- | -- | 5 |
| PM J15229 -0244 | -0.53 ±0.01 | -0.31 ±0.01 | 65±4 | sdM6 | -0.524 ±0.008 | -0.314 ±0.008 | -- | -- | 5 |
| SDSS 1614+4004 | -0.25 ±0.02 | 0.16 ±0.02 | 72±5 | L2 | 0.270 | 0.132 | ~25-100 | L1 | 6 |
| PM J17137 -4535 | -0.13 ±0.02 | -0.45 ±0.02 | 500±30 | white dwarf? | -0.143 ±0.024 | -0.444 ±0.024 | -- | -- | 5 |
| WISE 1830+4542 | 0.06 ±0.02 | 0.10 ±0.02 | 43±3 | L9 | 0.056 ±0.022 | 0.107 ±0.022 | 32.8 | L9 | 9 |

References: (1) West et al. 2008, (2) Faherty et al. 2012, (3) Faherty et al. 2009, (4) Sheppard & Cushing 2009, (5) Lepine 2008, (6) Zhang et al. 2009, (7) Gizis et al. 2012, (8) Castro & Gizis 2012, (9) Kirkpatrick et al. 2011, (10) Burningham et al. 2013.

[a] Optical-based spectral type

*Table 7: Comparison of Measurements for the G 225-36 (Gliese 605) Binary*

| Object | Spectral Type | $\mu_{RA}$ (″/yr) | $\mu_{Dec}$ (″/yr) | Dist. (pc) | Reference |
|---|---|---|---|---|---|
| G 225-36 | M0 | −0.296±0.001 | 0.203±0.002 | 31±1 | van Leeuwen (2007) |
| WISE 1557+5914 | M9 | −0.30±0.02 | 0.21±0.01 | 34±2 | this paper |

*Table 8: Comparison of Equivalent Width Measurements for the K I Doublets*[a]

| Object | Spec. Type[b] | 1.168 µm EW (Å) | 1.177 µm EW(Å) | 1.243 µm EW (Å) | 1.254 µm EW (Å) |
|---|---|---|---|---|---|
| 2MASSI J0103320+193536 | L6 opt | 5.8±0.3 | 6.2±0.4 | 4.3±0.3 | 5.3±0.4 |
| 2MASSs J0850359+105716 | L6 opt | 6.5±0.4 | 9.2±0.5 | 4.2±0.4 | 5.4±0.4 |
| DENIS-P J0205.4-1159 | L7 opt L5.5 NIR | 6.5±0.6 | 9.4±0.4 | 4.3±0.3 | 6.0±0.4 |
| 2MASSW J1728114+394859 | L7 opt | 6.7±0.7 | 9.5±0.4 | 5.6±0.3 | 7.1±0.5 |
| Gl 584C | L8 opt L8 NIR | 7.0±0.8 | 8.5±0.8 | 3.2±0.2 | 5.6±0.5 |
| 2MASSW J1632291+190441 | L8 opt L8 NIR | 4.7±0.3 | 6.3±0.4 | 3.0±0.1 | 3.5±0.2 |
| 2MASSW J0310599+164816 | L8 opt L9 NIR | 5.9±0.6 | 8.0±0.5 | 3.9±0.1 | 5.3±0.1 |
| WISE 0047+6803 | L9 pec (v. red) NIR | 2.5±0.4 | 3.3±0.4 | 2.4±0.2 | 3.9±0.4 |

[a] Equivalent width measurements for all sources except WISE 0047+6803 are from McLean et al. (2003).

[b] Spectral types are labeled as either optical (opt) or near-infrared (NIR).

## *Table 9: L Dwarfs with $J-K_s > 2.00$ mag*

| Discovery Name | Disc. Ref. | J (mag) | $K_s$ (mag) | $J-K_s$ (mag) | W1 (mag) | W2 (mag) | W3 (mag) | Opt Typ | Opt Ref | NIR Typ | NIR Ref |
|---|---|---|---|---|---|---|---|---|---|---|---|
| (a) Normal (?) L dwarfs | | | | | | | | | | | |
| 2MASSI J0103320+193536 | 3 | 16.29 ±0.08 | 14.15 ±0.06 | 2.14 ±0.10 | 13.18 ±0.02 | 12.70 ±0.03 | 12.23 ±0.33 | L6 | 3 | -- | -- |
| SDSSp J010752.33+004156.1 | 4 | 15.82 ±0.06 | 13.71 ±0.04 | 2.12 ±0.07 | 12.69 ±0.02 | 12.17 ±0.03 | 11.45 ±0.20 | L8 | 22 | L5.5 | 8 |
| 2MASSW J0129122+351758 | 2 | 16.78 ±0.16 | 14.70 ±0.08 | 2.08 ±0.18 | 14.07 ±0.03 | 13.71 ±0.03 | 12.86 ±0.41 | L4 | 2 | -- | -- |
| 2MASSW J0205034+125142 | 3 | 15.68 ±0.06 | 13.67 ±0.04 | 2.01 ±0.07 | 12.93 ±0.02 | 12.55 ±0.03 | 12.11 ±0.29 | L5 | 3 | -- | -- |
| 2MASS J03185403-3421292 | 13 | 15.57 ±0.06 | 13.51 ±0.04 | 2.06 ±0.07 | 12.62 ±0.02 | 12.13 ±0.02 | 11.03 ±0.07 | L7 | 13 | -- | -- |
| 2MASS J03264225-2102057 | 6 | 16.13 ±0.09 | 13.92 ±0.07 | 2.21 ±0.12 | 12.95 ±0.02 | 12.44 ±0.02 | 12.17 ±0.20 | L4 | 12 | L | 6 |
| 2MASSW J0337036-175807 | 3 | 15.62 ±0.06 | 13.58 ±0.04 | 2.04 ±0.07 | 12.83 ±0.03 | 12.46 ±0.03 | 12.08 ±0.26 | L4.5 | 3 | -- | -- |
| 2MASS J03421621-6817321 | 12 | 16.85 ±0.14 | 14.541 ±0.09 | 2.31 ±0.17 | 13.95 ±0.03 | 13.46 ±0.03 | >13.07 | L2: | 12 | -- | -- |
| 2MASS J03582255-4116060 | 14 | 15.85 ±0.09 | 13.84 ±0.05 | 2.01 ±0.10 | 12.90 ±0.02 | 12.45 ±0.02 | 11.72 ±0.10 | L5 | 14 | -- | -- |
| 2MASSI J0512063-294954 | 7 | 15.46 ±0.06 | 13.29 ±0.04 | 2.18 ±0.07 | 12.38 ±0.02 | 11.92 ±0.02 | 11.33 ±0.11 | L4.5 | 13 | -- | -- |
| SDSS J074007.30+200921.9 | 8 | 17.29 ±0.25 | 15.08 ±0.12 | 2.21 ±0.28 | 14.45 ±0.04 | 13.98 ±0.05 | 12.50 ±0.52 | -- | -- | L6 ±1.5 | 10 |
| SDSS J080959.01+443422.2 | 8 | 16.44 ±0.11 | 14.42 ±0.06 | 2.02 ±0.13 | 13.34 ±0.03 | 12.81 ±0.03 | 11.79 ±0.21 | -- | -- | L6 | 10 |
| 2MASSW J0820299+450031 | 3 | 16.28 ±0.11 | 14.22 ±0.07 | 2.06 ±0.13 | 13.52 ±0.03 | 13.16 ±0.03 | >12.10 | L5 | 3 | -- | -- |
| 2MASSI J0825196+211552 | 3 | 15.10 ±0.03 | 13.03 ±0.03 | 2.07 ±0.04 | 12.08 ±0.02 | 11.57 ±0.02 | 10.39 ±0.07 | L7.5 | 3 | L6 | 8 |
| 2MASSW J0829570+265510 | 3 | 17.11 ±0.19 | 14.96 ±0.10 | 2.15 ±0.21 | 13.90 ±0.03 | 13.40 ±0.04 | >12.63 | L6.5 | 3 | -- | -- |
| 2MASSI J0835425-081923 | 7 | 13.17 ±0.02 | 11.14 ±0.02 | 2.03 ±0.03 | 10.39 ±0.02 | 10.04 ±0.02 | 9.47 ±0.03 | L5 | 7 | -- | -- |
| SDSSp J085758.45+570851.4 | 4 | 15.04 ±0.04 | 12.96 ±0.03 | 2.08 ±0.05 | 12.02 ±0.02 | 11.42 ±0.02 | 10.38 ±0.06 | L8 | 13 | L8 ±1 | 4 |
| 2MASSW J0951054+355801[b] | 3 | 17.23 ±0.21 | 15.14 ±0.13 | 2.09 ±0.25 | -- | -- | -- | L6 | 3 | -- | -- |
| 2MASSW J1102337-235945 | 3 | 16.72 ±0.16 | 14.59 ±0.10 | 2.13 ±0.18 | 14.12 ±0.03 | 13.75 ±0.04 | >12.57 | L4.5 | 3 | -- | -- |
| 2MASSW J1343167+394508 | 3 | 16.16 ±0.07 | 14.15 ±0.05 | 2.01 ±0.09 | 13.23 ±0.02 | 12.84 ±0.03 | 12.28 ±0.27 | L5 | 3 | -- | -- |
| 2MASSW J1553214+210907 | 2 | 16.70 ±0.16 | 14.68 ±0.11 | 2.03 ±0.20 | 13.79 ±0.03 | 13.31 ±0.03 | 12.20 ±0.29 | L5.5 | 2 | -- | -- |
| 2MASSI J1711457+223204 | 3 | 17.09 ±0.18 | 14.73 ±0.10 | 2.36 ±0.20 | 14.35 ±0.03 | 13.81 ±0.04 | >12.60 | L6.5 | 3 | -- | -- |
| 2MASSW J1728114+394859 | 3 | 15.99 ±0.08 | 13.91 ±0.05 | 2.08 ±0.09 | 13.11 ±0.02 | 12.64 ±0.02 | 11.87 ±0.13 | L7 | 3 | -- | -- |
| WISEPA J183058.57+454257.9 | 17 | >17.95 | 14.98 ±0.11 | >2.97 | 14.81 ±0.03 | 14.17 ±0.04 | >13.17 | -- | -- | L9 | 17 |
| 2MASS J21163374-0729200 | 18 | 17.20 ±0.21 | 14.98 ±0.13 | 2.22 ±0.25 | 14.62 ±0.04 | 14.33 ±0.06 | >12.23 | -- | -- | L6 | 18 |
| 2MASS J 21512543-2441000 | 11 | 15.75 ±0.08 | 13.65 ±0.05 | 2.10 ±0.10 | 13.06 ±0.03 | 12.74 ±0.03 | >12.34 | L3 | 12 | -- | -- |
| 2MASSW J2224438-015852 | 3 | 14.07 ±0.03 | 12.02 ±0.02 | 2.05 ±0.04 | 11.36 ±0.02 | 11.12 ±0.02 | 10.65 ±0.09 | L4.5 | 3 | L3.5 | 8 |

(b) Peculiar L dwarfs for which peculiarities are due to low gravity

| | | | | | | | | | | | |
|---|---|---|---|---|---|---|---|---|---|---|---|
| WISEP J004701.06+680352.1[f] | 19 | 15.60 ±0.07 | 13.05 ±0.03 | 2.55 ±0.08 | 11.88 ±0.02 | 11.27 ±0.02 | 10.33 ±0.07 | -- | -- | L7.5 pec | 19 |
| 2MASS J03552337+1133437 | 14 | 14.05 ±0.02 | 11.53 ±0.02 | 2.52 ±0.03 | 10.53 ±0.02 | 9.94 ±0.02 | 9.29 ±0.04 | L5γ | 23 | -- | -- |
| 2MASS J04210718-6306022 | 12 | 15.57 ±0.05 | 13.45 ±0.04 | 2.12 ±0.07 | 12.56 ±0.02 | 12.14 ±0.02 | 11.60 ±0.10 | L5β | 23 | -- | -- |
| 2MASS J05012406-0010452 | 14 | 14.98 ±0.04 | 12.96 ±0.04 | 2.02 ±0.05 | 12.05 ±0.02 | 11.52 ±0.02 | 10.95 ±0.11 | L4γ | 23 | -- | -- |
| AB Pic b | 9 | 16.18 ±0.10 | 14.14 ±0.08 | 2.04 ±0.13 | -- | -- | -- | -- | -- | L1 ±1.5 | 9 |
| G 196-3 B | 1 | 14.83 ±0.05 | 12.78 ±0.03 | 2.05 ±0.06 | -- | -- | -- | L3β | 23 | -- | -- |
| 2MASSW J1207334-393254b | 26 | 20.0[a] ±0.2 | 16.93[a] ±0.11 | 3.07[a] ±0.23 | -- | -- | -- | -- | -- | L5-L9.5 | 26 |
| 2MASS J15515237+0941148 | 14 | 16.32 ±0.11 | 14.31 ±0.06 | 2.01 ±0.13 | 13.60 ±0.03 | 13.12 ±0.03 | 12.68 ±0.48 | L4γ | 25 | -- | -- |
| USco J160843.44-224516.0[c] | 15 | 18.59 ±0.10 | 16.26 ±0.05 | 2.33 ±0.11 | 15.43 ±0.06 | 14.67 ±0.10 | >12.10 | -- | -- | L1 | 15 |
| 2MASS J16154255+4953211 | 12 | 16.79 ±0.14 | 14.31 ±0.07 | 2.48 ±0.15 | 13.20 ±0.02 | 12.62 ±0.02 | 12.13 ±0.13 | L4γ | 25 | L6 | 18 |
| 2MASSI J1726000+153819 | 3 | 15.67 ±0.07 | 13.66 ±0.05 | 2.01 ±0.08 | 13.07 ±0.03 | 12.69 ±0.03 | 11.56 ±0.16 | L3β | 23 | -- | -- |
| 2MASSW J2244316+204343[d] | 5 | 16.48 ±0.14 | 14.02 ±0.07 | 2.45 ±0.16 | 12.78 ±0.02 | 12.11 ±0.02 | 11.14 ±0.12 | L6.5 | 13 | L7.5 ±2 | 8 |
| SDSSp J224953.45+004404.2 | 4 | 16.59 ±0.13 | 14.36 ±0.07 | 2.23 ±0.14 | 13.58 ±0.03 | 13.14 ±0.05 | >11.28 | L4γ | 25 | L5 ±1.5 | 8 |

(c) Peculiar L dwarfs for which peculiarities are not obviously due to low gravity

| | | | | | | | | | | | |
|---|---|---|---|---|---|---|---|---|---|---|---|
| WISEPA J020625.26+264023.6 | 17 | 16.53 ±0.11 | 14.52 ±0.08 | 2.01 ±0.14 | 13.40 ±0.03 | 12.82 ±0.03 | 11.33 ±0.22 | -- | -- | L9p (red) | 17 |
| WISEPA J164715.59+563208.2 | 17 | 16.91 ±0.18 | 14.61 ±0.09 | 2.30 ±0.20 | 13.60 ±0.02 | 13.09 ±0.02 | 12.06 ±0.09 | -- | -- | L9p (red) | 17 |
| WISE J173859.27+614242.1[e] | 20 | 17.82 ±0.13 | 15.27 ±0.10 | 2.55 ±0.16 | 14.06 ±0.03 | 13.34 ±0.03 | 12.20 ±0.15 | -- | -- | extr. red | 20 |
| 2MASS J21481628+4003593 | 16 | 14.15 ±0.03 | 11.77 ±0.02 | 2.38 ±0.04 | 10.74 ±0.02 | 10.24 ±0.02 | 9.66 ±0.04 | L6 | 16 | L6p | 16 |
| WISE J233527.07+451140.9 | 21 | 16.70 ±0.19 | >14.46 | <2.24 | 13.48 ±0.03 | 12.93 ±0.03 | 12.72 ±0.54 | -- | -- | L9p (v red) | 21 |

References: (1) Rebolo et al. 1998, (2) Kirkpatrick et al. 1999, (3) Kirkpatrick et al. 2000, (4) Geballe et al. 2002, (5) Dahn et al. 2002, (6) Gizis et al. 2003, (7) Cruz et al. 2003, (8) Knapp et al. 2004, (9) Chauvin et al. 2005, (10) Chiu et al. 2006, (11) Liebert & Gizis 2006, (12) Cruz et al. 2007, (13) Kirkpatrick et al. 2008, (14) Reid et al. 2008, (15) Lodieu et al. 2008, (16) Looper et al. 2008, (17) Kirkpatrick et al. 2011, (18) Geissler et al. 2011, (19) Gizis et al. 2012, (20) Mace et al. 2013, (21) this paper, (22) Hawley et al. 2002, (23) Cruz, Kirkpatrick, & Burgasser 2009, (24) Kirkpatrick et al. 2010, (25) Faherty et al. 2013, (26) Chauvin et al. 2004.

Photometry is from the 2MASS All-Sky Point Source Catalog or 2MASS Reject Table unless otherwise noted.

[a] J-band magnitude is from Mohanty et al. (2007) and K-band magnitude is from Chauvin et al. (2004).
[b] Possible common proper motion companion to LP 261-75 (Burgasser, Kirkpatrick & Lowrance 2005).
[c] $JHK_s$ photometry is from UKIDSS.
[d] Low-gravity features noted by Kirkpatrick et al. (2008). See also McLean et al. (2003).
[e] Object is a spectrum variable (Mace et al. 2013).
[f] Low-gravity features first noted in this paper.

*Table 10: Kinematics of Peculiar L Dwarfs with $J-K_s$>2.00 mag*

| Discovery Name | Proper Motion ("/yr) | Proper Motion Ref | Distance (pc) | Distance Ref | $V_{tan}$ (km/s) |
|---|---|---|---|---|---|
| (a) Peculiar L dwarfs for which peculiarities are due to low gravity | | | | | |
| WISEP J004701.06+680352.1 | 0.43±0.01 | 5 | 15±1 | 5 | 31±2 |
| 2MASS J03552337+1133437 | 0.662±0.007 | 1 | 8.2±0.8 | 1 | 26±3 |
| 2MASS J04210718-6306022 | 0.24±0.02 | 7 | 31±3 | 2 | 35±5 |
| 2MASS J05012406-0010452 | 0.21±0.02 | 7 | 24±5 | 2 | 24±8 |
| AB Pic b | 0.0474±0.0008 | 3 | 46±2 | 3 | 10.3±0.5 |
| G 196-3 B | 0.23±0.03 | 4 | 32±2 | 2 | 35±5 |
| 2MASSW J1207334-393254b | 0.067±0.002 | 11 | 54±2 | 11 | 17±1 |
| 2MASS J15515237+0941148 | 0.09±0.03 | 7 | 64±1 | 2 | 27±9 |
| USco J160843.44-224516.0 | -- | -- | ~145 | 6 | -- |
| 2MASS J16154255+4953211 | 0.05±0.06 | 8 | 71±15 | 8 | 17±21 |
| 2MASSI J1726000+153819 | 0.06±0.02 | 4 | 47±4 | 2 | 13±5 |
| 2MASSW J2244316+204343 | 0.33±0.02 | 4 | 19±2 | 2 | 30±4 |
| SDSSp J224953.45+004404.2 | 0.08±0.03 | 2 | 56±5 | 2 | 21±8 |
| (b) Peculiar L dwarfs for which peculiarities are not obviously due to low gravity | | | | | |
| WISEPA J020625.26+264023.6 | 0.43±0.01 | 9 | 18±2 | 9 | 37±4 |
| WISEPA J164715.59+563208.2 | 0.29±0.01 | 9 | 20±2 | 9 | 27±3 |
| WISE J173859.27+614242.1 | 0.28±0.04 | 10 | 42±4 | 5 | 56±10 |
| 2MASS J21481628+4003593 | 0.89±0.03 | 2 | 7±1 | 2 | 30±4 |
| WISE J233527.07+451140.9 | 0.09±0.03 | 5 | 24±2 | 5 | 10±3 |

References: (1) Faherty et al. 2013, (2) Faherty et al. 2009, (3) Perryman et al. 1997, (4) Jameson et al. 2008, (5) this paper, (6) Lodieu et al. 2007, (7) Faherty et al. 2008, (8) Schmidt et al. 2010, (9) Kirkpatrick et al. 2011, (10) Mace et al. 2013, (11) Gizis et al. 2007.

*Figure 1 (panels a through h): Finder charts for the forty-two discoveries discussed in this paper. For each object, the abbreviated WISE object name is given in the leftmost column followed by the DSS I-band image, the 2MASS J- and $K_S$-band images, the WISE W1 through W3 images, and finally, in the rightmost column, the WISE three-color image in which W1 is color coded as blue, W2 as green, and W3 as red. All images are two arcmin square, have north up and east to the left, and are centered at the position (red circle) of the WISE candidate. In most cases, the source was detected by 2MASS but at a slightly different position than that observed by WISE, confirming these objects to be nearby based on their proper motions.*

*Figure 2: Follow-up spectra of five early-type objects from Table 1 (heavy, black lines) compared to spectroscopic standards (light, grey lines). The flux of each spectrum is normalized to one at 1.28 μm and integral offsets added to the flux when necessary to separate spectra vertically. The near-infrared M dwarf standards are taken from Kirkpatrick et al. (2010). The K7 dwarf standard, the optically classified 61 Cyg B taken from Rayner et al. (2009), has been smoothed with an 11-pixel boxcar to better match the resolution of the IRTF/SpeX prism data shown elsewhere in the figure. Per Rayner et al. (2009), regions of telluric absorption are shown by the dark grey (atmospheric transmission < 20%) and light grey (20% < atmospheric transmission < 80%) zones in wavelength.*

*Figure 3: Follow-up spectra of five early- to mid-M dwarfs from Table 1 (heavy, black lines) compared to spectroscopic standards (light, grey lines). The near-infrared M dwarf standards are taken from Kirkpatrick et al. (2010). All data are from IRTF/SpeX in prism mode. Other details are the same as in Figure 2.*

*Figure 4: Follow-up spectra of one mid-M subdwarf and four late-M dwarfs from Table 1 (heavy, black lines) compared to spectroscopic standards and comparison objects (light, grey lines). The spectrum of WISE 2347+6833 has been smoothed with an 11-pixel boxcar; all other data are from IRTF/SpeX in prism mode and are shown at native resolution. The sdM6 comparison object, LHS 1074, along with the near-infrared M dwarf standards, are taken from Kirkpatrick et al. (2010). Other details are the same as in Figure 2.*

*Figure 5: Follow-up spectra of four late-M dwarfs and one early-L dwarf from Table 1 (heavy, black lines) compared to spectroscopic standards (light, grey lines). The Palomar/TSpec spectrum of WISE 1353-0857 has been smoothed with an 11-pixel boxcar; all other data are from IRTF/SpeX in prism mode and are shown at native resolution. The near-infrared M and L dwarf standards are taken from Kirkpatrick et al. (2010). Other details are the same as in Figure 2.*

*Figure 6: Details of the 1.15-1.35 μm region of four mid-M to mid-L dwarfs (heavy, black lines) having higher resolution data from Palomar/TSpec, APO/TSpec, or Keck/NIRSPEC. The Palomar/TSpec and APO/TSpec data have been smoothed with a 5-pixel boxcar to better match the Keck/NIRSPEC data. Comparison objects (light, grey lines) are the Kirkpatrick et al. (2010) near-infrared spectral standards observed by McLean et al. (2003) using Keck/NIRSPEC. Other details are the same as in Figure 2.*

*Figure 7: Follow-up spectra of five early- to mid-L dwarfs from Table 1 (heavy, black lines) compared to spectroscopic standards (light, grey lines). The Palomar/TSpec spectrum of WISE 1358+1458 has been smoothed with an 11-pixel boxcar; all other data are from IRTF/SpeX in prism mode and are shown at native resolution. The near-infrared L dwarf standards are taken from Kirkpatrick et al. (2010). Other details are the same as in Figure 2.*

*Figure 8: Follow-up spectra of five mid- to late-L dwarfs from Table 1 (heavy, black lines) compared to spectroscopic standards (light, grey lines). The Palomar/TSpec spectrum of WISE 1658+5103 has been smoothed with an 11-pixel boxcar; all other data are from IRTF/SpeX in prism mode and are shown at native resolution. The near-infrared L dwarf standards are taken from Kirkpatrick et al. (2010). Other details are the same as in Figure 2.*

*Figure 9: Follow-up spectra of five late-L dwarfs from Table 1 (heavy, black lines) compared to spectroscopic standards (light, grey lines). The APO/TSpec spectrum of WISE 0230-0225 and the Palomar/TSpec spectra of WISE 0031+5749, SDSS 1155+0559, and WISE 1851+5935 has been smoothed with an 11-pixel boxcar; all other spectra are from IRTF/SpeX in prism mode and are shown at native resolution. The near-infrared L dwarf standards are taken from Kirkpatrick et al. (2010). Other details are the same as in Figure 2.*

*Figure 10: Follow-up spectra of four L9 dwarfs from Table 1 (heavy, black lines) compared to spectroscopic standards (light, grey lines). The Palomar/TSpec spectrum of WISE 2222+3026 has been smoothed with an 11-pixel boxcar; all other spectra are from IRTF/SpeX in prism mode and are shown at native resolution. The near-infrared L dwarf standards are taken from Kirkpatrick et al. (2010). Other details are the same as in Figure 2.*

*Figure 11: Details of the 1.15-1.35 μm region of four mid- to late-L dwarfs (heavy, black lines) having higher resolution data from Palomar/TSpec or Keck/NIRSPEC. The Palomar/TSpec data have been smoothed with a 5-pixel boxcar to better match the Keck/NIRSPEC data. Comparison objects (light, grey lines) are the Kirkpatrick et al. (2010) L7 and L8 near-infrared spectral standards observed by McLean et al. (2003) using Keck/NIRSPEC along with comparison spectra of 2MASS J15074769-1627386 (L5) and 2MASS J03105986+1648155 (L9) from McLean et al. (2003) using Keck/NIRSPEC . Other details are the same as in Figure 2.*

*Figure 12: Details of the 1.15-1.35 μm region of three late-L dwarfs (heavy, black lines) having higher resolution data from Palomar/TSpec. The spectra of comparison objects (light, grey lines) 2MASS J03105986+1648155 (L9) and SDSSp J042348.57-041403.5 (T0) are from McLean et al. (2003) using Keck/NIRSPEC . The Palomar/TSpec data have been smoothed with a 5-pixel boxcar to better match the Keck/NIRSPEC data. Other details are the same as in Figure 2.*

*Figure 13: Follow-up spectra of two peculiar L9 dwarfs from Table 1 (heavy, black lines) compared to the peculiar L9 WISE 173859.27+614242.1 from Mace et al. (2013) and the L9 spectroscopic standard (light, grey line) from Kirkpatrick et al. (2010). The standard is plotted only once and normalized to match the spectrum of WISE 2335+4511 at 1.28 μm. The Palomar/TSpec spectra of the objects in black have been smoothed with an 11-pixel boxcar to better match the IRTF/SpeX prism-mode spectrum of the standard. Other details are the same as in Figure 2.*

*Figure 14: Details of the 1.15-1.35 μm region of the three peculiar L9 dwarfs (heavy, black lines) from Figure 13. These data, all from Palomar/TSpec , have been smoothed with a 5-pixel boxcar to better match the Keck/NIRSPEC data of the comparison object, 2MASS J03105986+1648155 (L9 — light grey line), from McLean et al. (2003) . Other details are the same as in Figure 2.*

*Figure 15: Interpretation of the WISE 1851+5935 spectrum as an unresolved binary. All spectra are normalized to one at 1.28 μm and half-integral offsets have been added to separate spectra vertically where needed. The top two spectra show the spectral templates used for the primary (L7) and secondary (T2) components of our synthesized composite system. Below this is the observed spectrum of our object, WISE 1851+5935, overplotted with the best fitting spectral standard. At the bottom is the observed spectrum of our object overplotted with the synthesized composite spectrum using L7 and T2 components. The relative scaling between components in the synthetic binary is set by the absolute value vs. spectral type relation given in Dupuy & Liu (2012), since the two components are assumed to be a physically associated pair equidistant from the Sun. The synthetic binary spectrum provides a much improved fit to the observed data.*

*Figure 16: Interpretation of the WISE 1658+5103 spectrum as an unresolved binary. All spectra are normalized to one at 1.28 μm and half-integral offsets have been added to separate spectra vertically where needed. The top two spectra show the spectral templates used for the primary (L4) and secondary (T0) components of our synthesized composite system. Below this is the observed spectrum of our object, WISE 1658+5103, overplotted with the best fitting spectral standard. At the bottom is the observed spectrum of our object overplotted with the synthesized composite spectrum using L4 and T0 components. The relative scaling between components in the synthetic binary is set by the absolute value vs. spectral type relation given in Dupuy & Liu (2012), since the two components are assumed to be a physically associated pair*

*equidistant from the Sun. The synthetic binary spectrum provides a much improved fit to the observed data.*

*Figure 17: Interpretation of the WISE 1552+5033 spectrum as an unresolved binary. All spectra are normalized to one at 1.28 μm and half-integral offsets have been added to separate spectra vertically where needed. The top two spectra show the spectral templates used for the primary (L6) and secondary (T5) components of our synthesized composite system. Below this is the observed spectrum of our object, WISE 1552+5033, overplotted with the best fitting spectral standard. At the bottom is the observed spectrum of our object overplotted with the synthesized composite spectrum using L6 and T5 components. The relative scaling between components in the synthetic binary is set by the absolute value vs. spectral type relation given in Dupuy & Liu (2012), since the two components are assumed to be a physically associated pair equidistant from the Sun. The synthetic binary spectrum provides a much improved fit to the observed data.*

*Figure 18: Interpretation of the WISE 0230-0225 spectrum as an unresolved binary. All spectra are normalized to one at 1.28 μm and integral offsets have been added to separate spectra vertically where needed. The top two spectra show the spectral templates used for the primary (L7) and secondary (T2) components of our synthesized composite system. Below this is the observed spectrum of our object, WISE 0230-0225, overplotted with the best fitting spectral standard. At the bottom is the observed spectrum of our object overplotted with the synthesized composite spectrum*

*using L7 and T2 components. The relative scaling between components in the synthetic binary is set by the absolute value vs. spectral type relation given in Dupuy & Liu (2012), since the two components are assumed to be a physically associated pair equidistant from the Sun. The synthetic binary spectrum provides a somewhat improved fit to the observed data.*

*Figure 19: A plot of the W1-W2 color vs. the J-W2 color for brown dwarfs with J-*$K_s$ *colors greater than 2.0 mag. Normal dwarfs from Table 8 are shown by small dots (black), low-gravity objects are shown by open triangles (green), and peculiar dwarfs whose redness is not obviously caused by low gravity are shown by open squares (red). Locations of specific examples in the latter class are marked with their abbreviated names.*

*Figure 20: Follow-up spectra of the three SpeX-observed T dwarfs from Table 1 (heavy, black lines) compared to the T7 and T8 spectroscopic standards (light, grey lines) from Burgasser et al. (2006). Other details are the same as in Figure 2.*

*Figure 21: Follow-up spectra of the three NIRSPEC-observed T dwarfs from Table 1 (heavy, black lines) compared to the T7, T8, and T9 spectroscopic standards (light, grey lines) from Burgasser et al. (2006) and Cushing et al. (2011). Other details are the same as in Figure 2.*

*Figure 22: Follow-up spectra of the five remaining objects from Table 1 (heavy, black lines). Comparison objects (light, grey lines) are the K3 dwarf Gliese 892, the mid-M giant R Lyr, and the late-M giants R Cnc and BK Vir, all from Rayner et al. (2009) . The APO/TSpec spectrum of WISE 0200+8742 and the spectra from Rayner et al. (2009) spectra have been smoothed with an 11-pixel boxcar to better match the IRTF/SpeX prism-mode spectra of the others. Other details are the same as in Figure 2.*

*Figure 1a:*

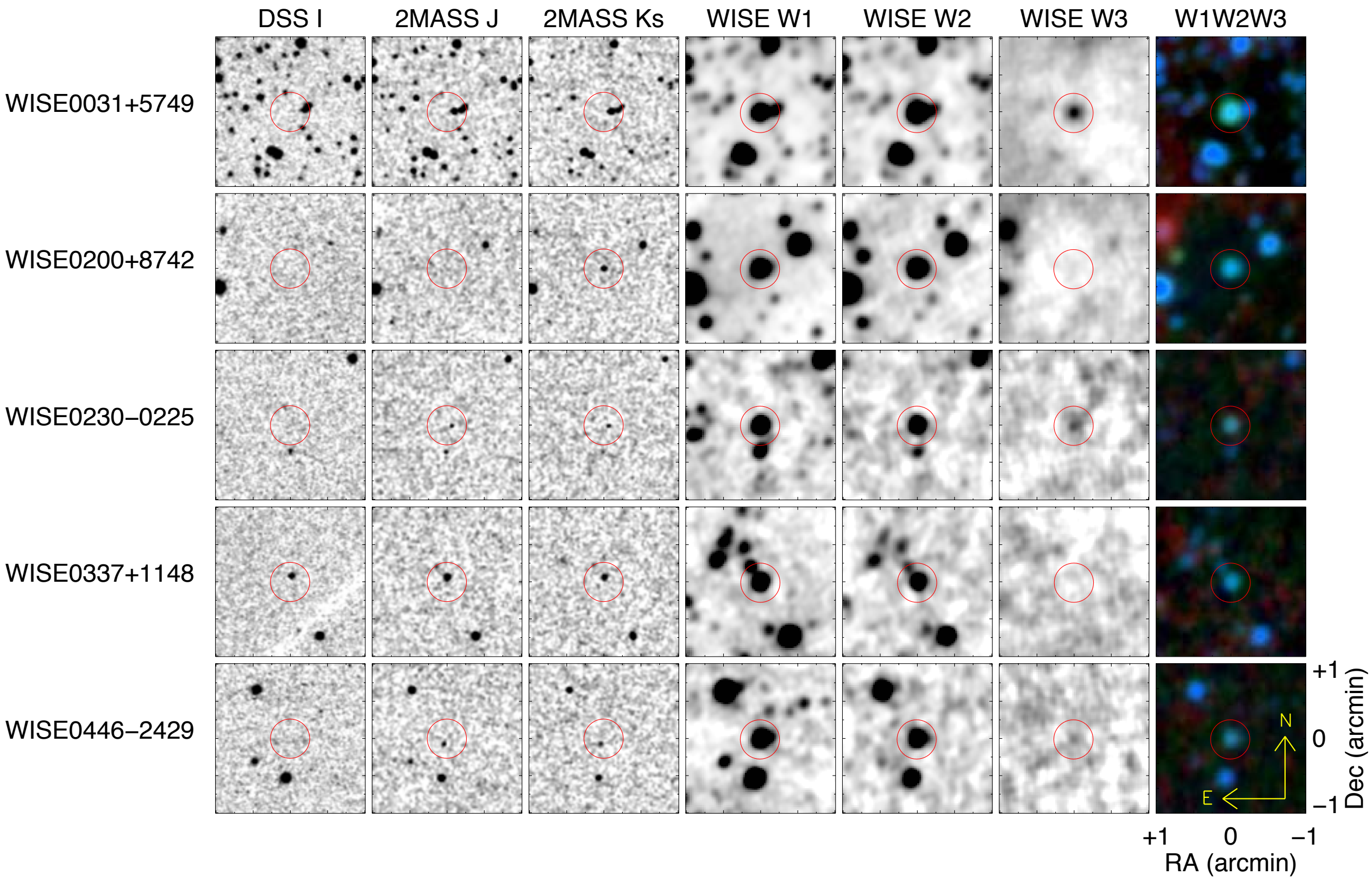

*Figure 1b:*

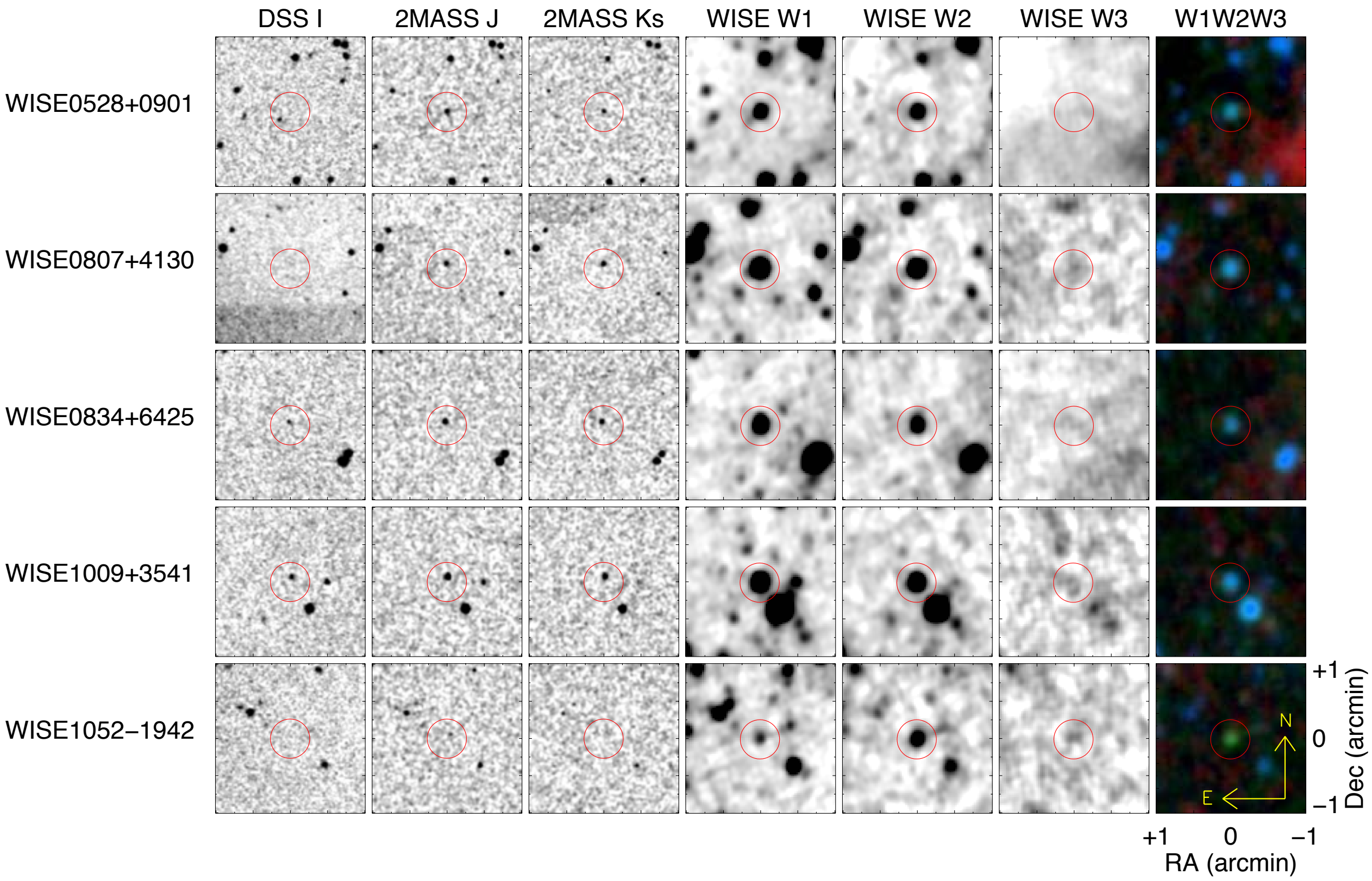

*Figure 1c:*

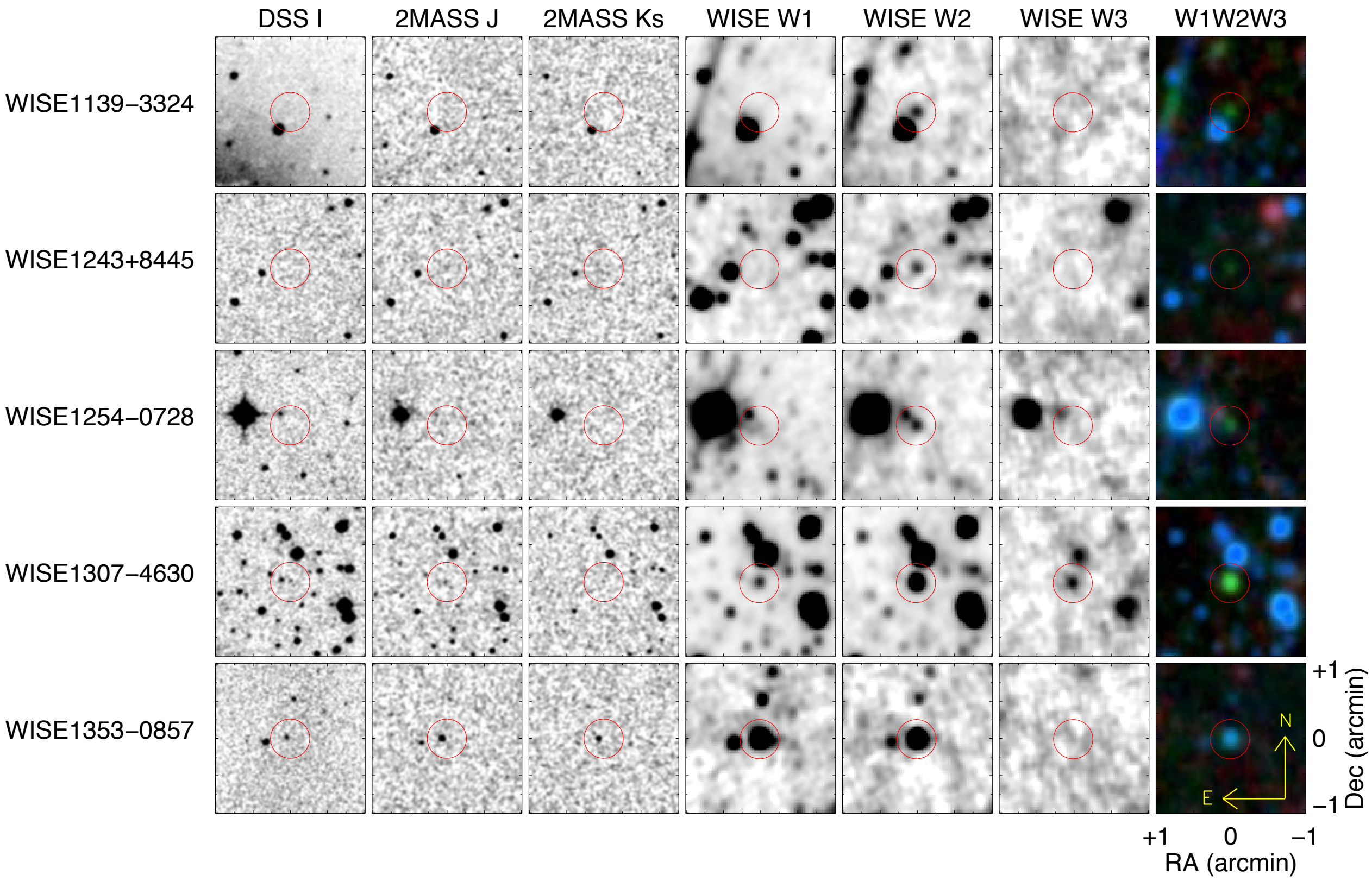

*Figure 1d:*

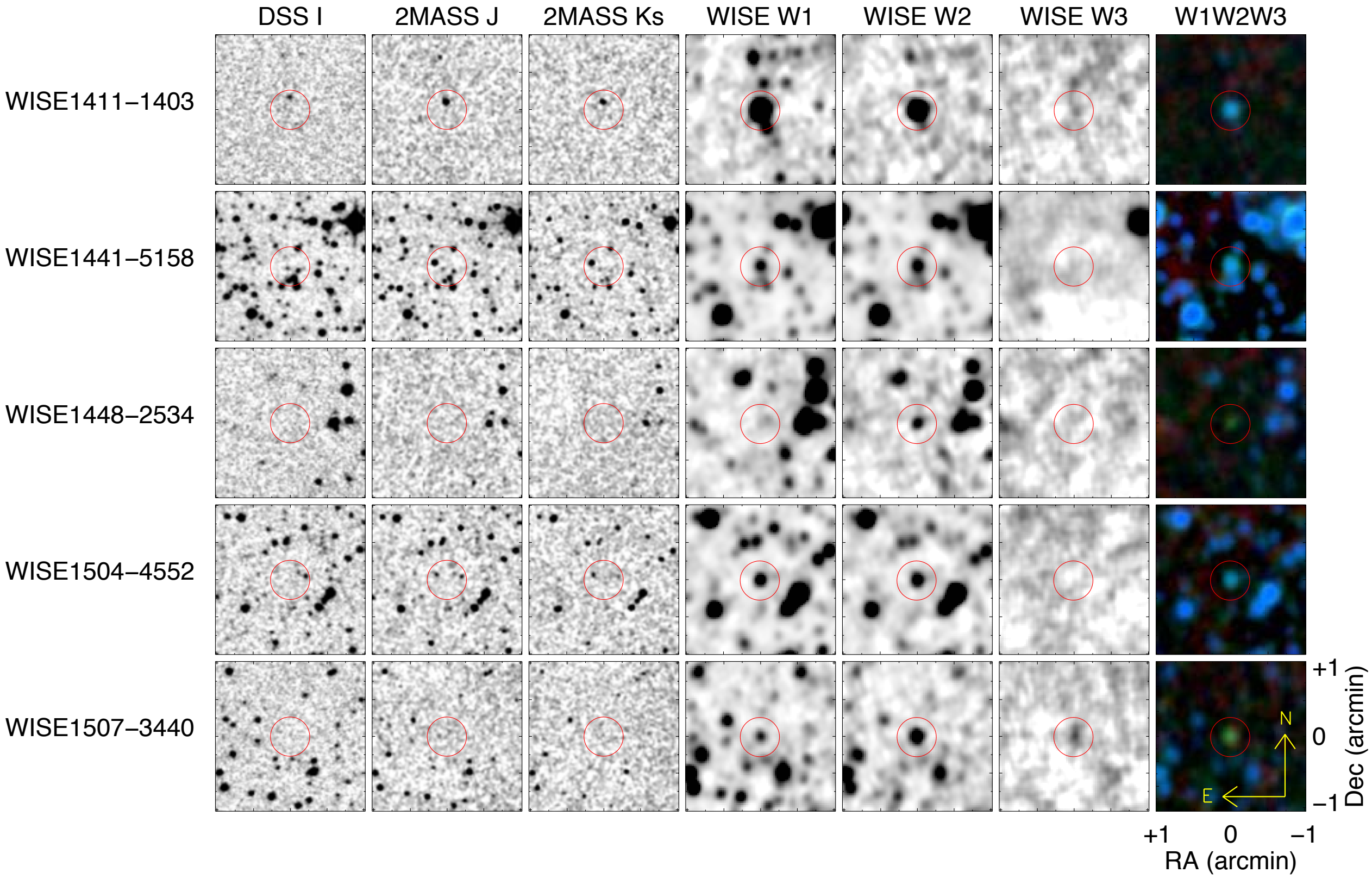

*Figure 1e:*

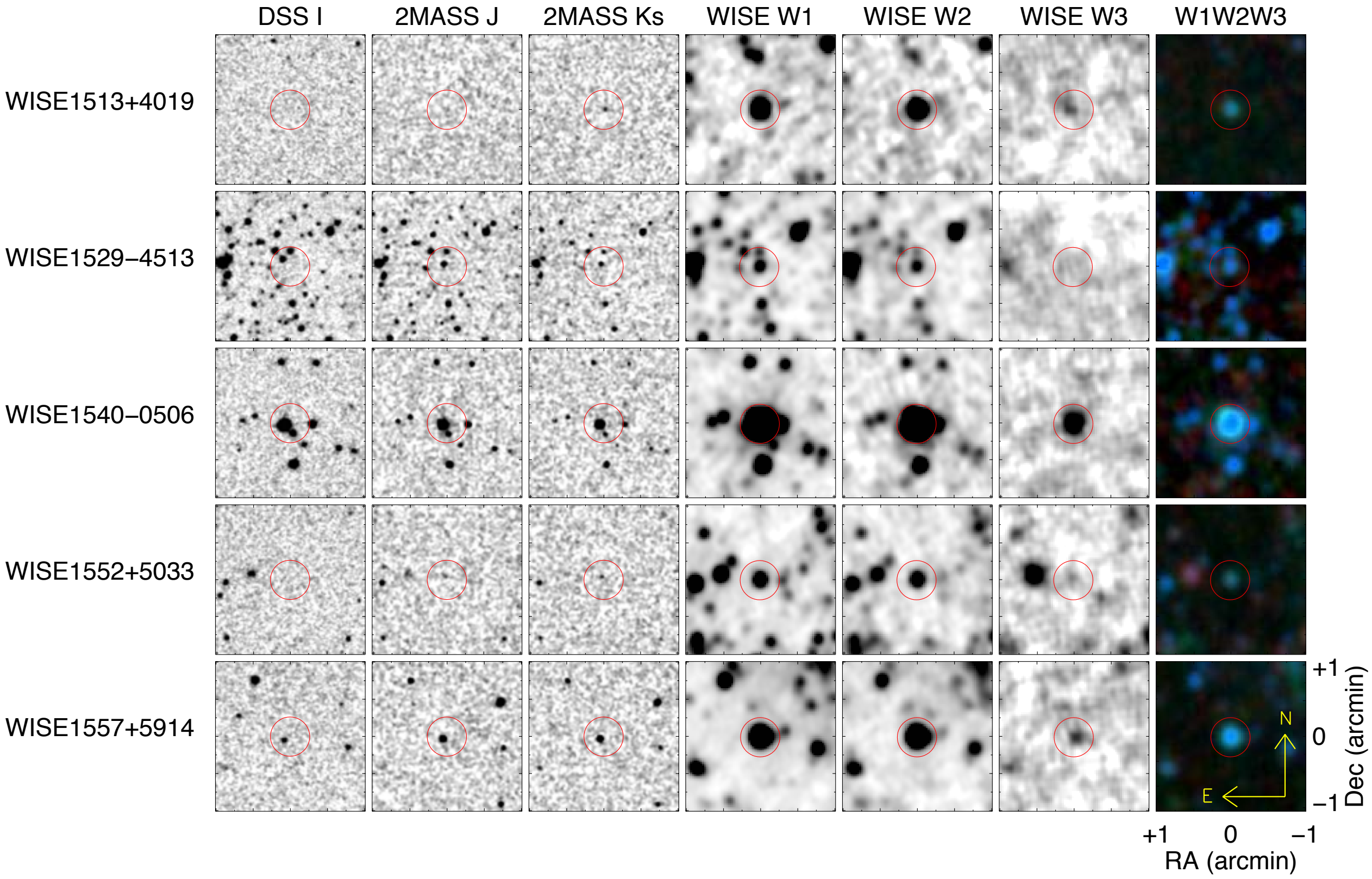

*Figure 1f:*

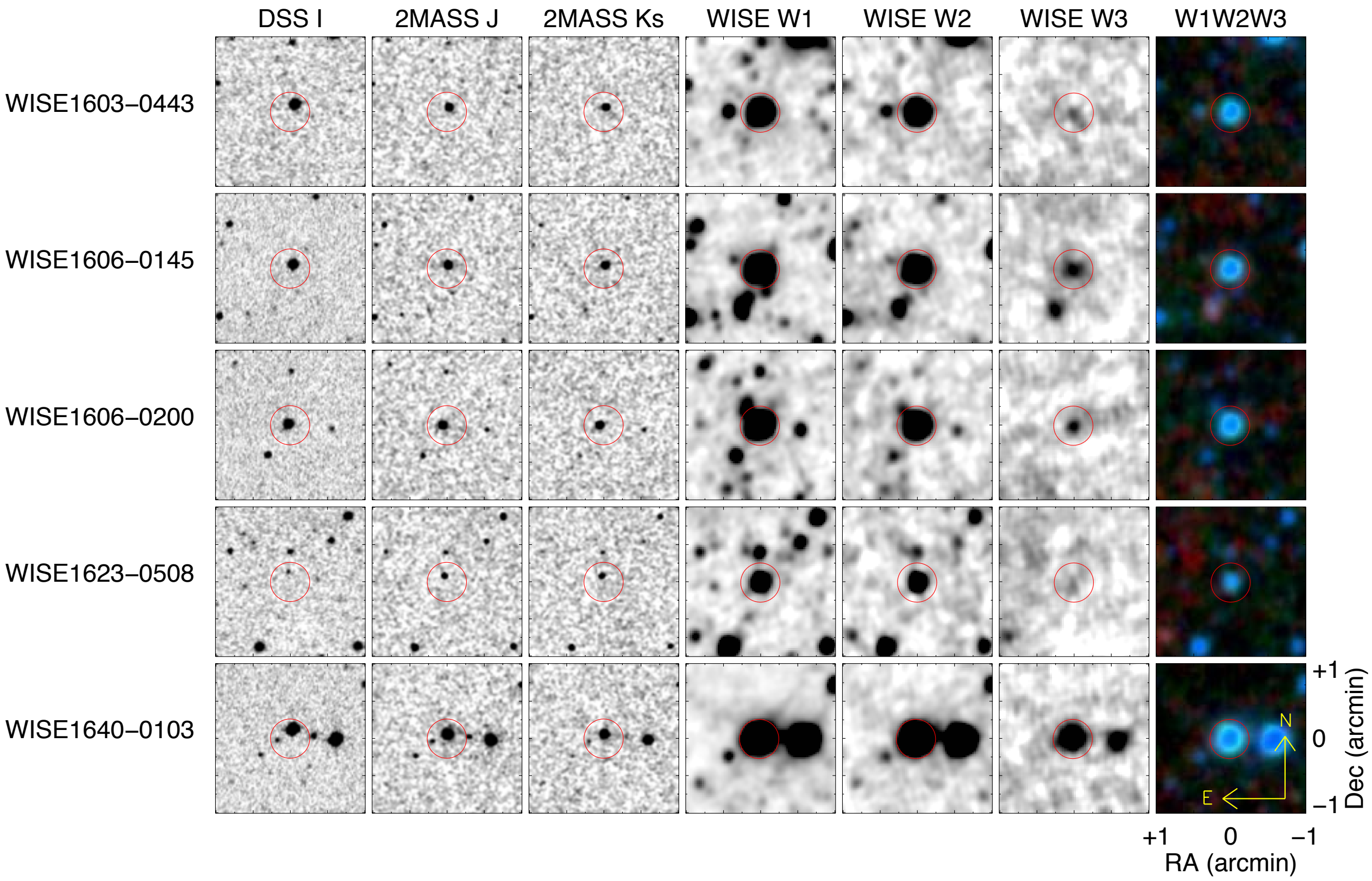

*Figure 1g:*

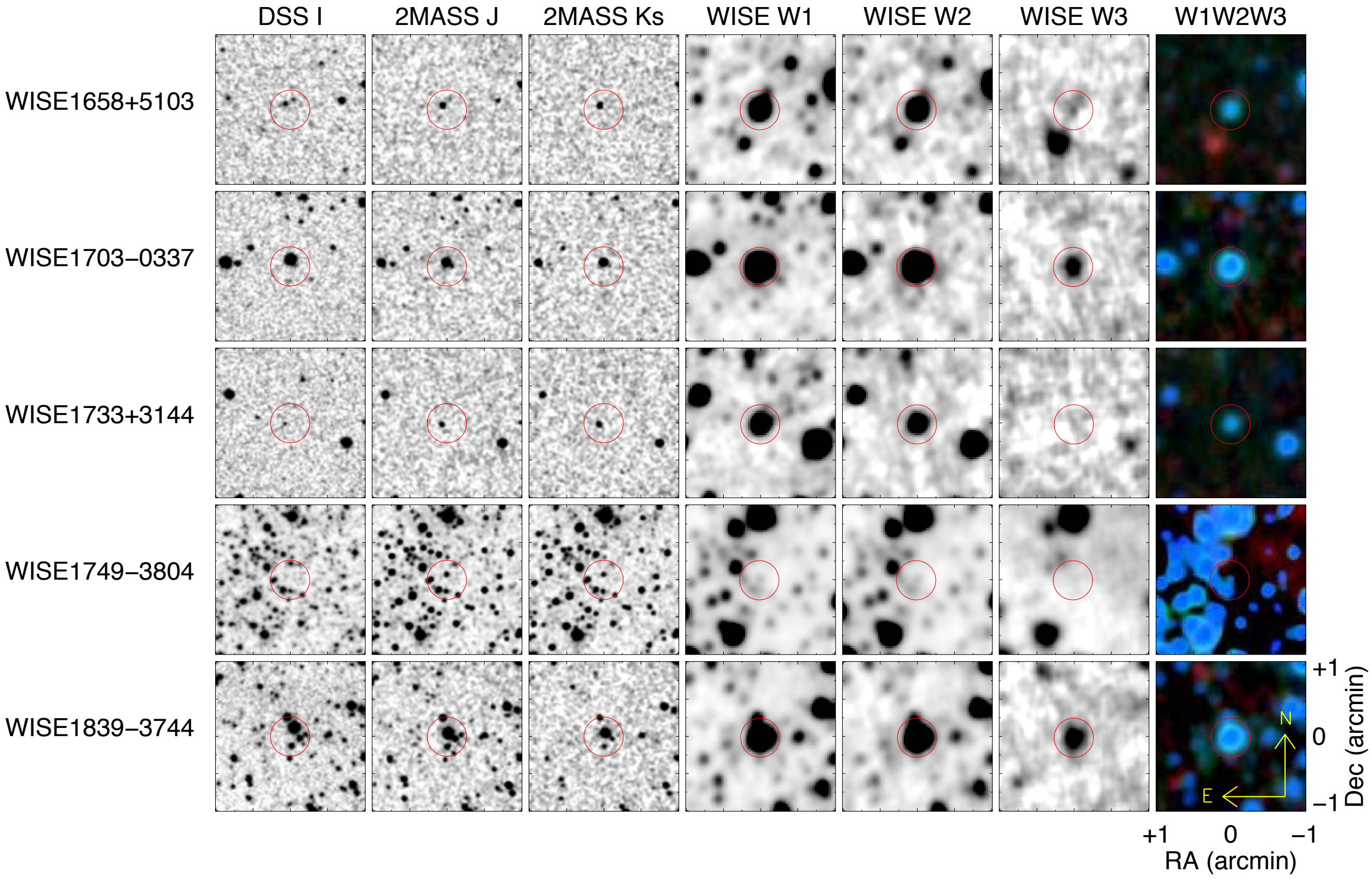

*Figure 1h:*

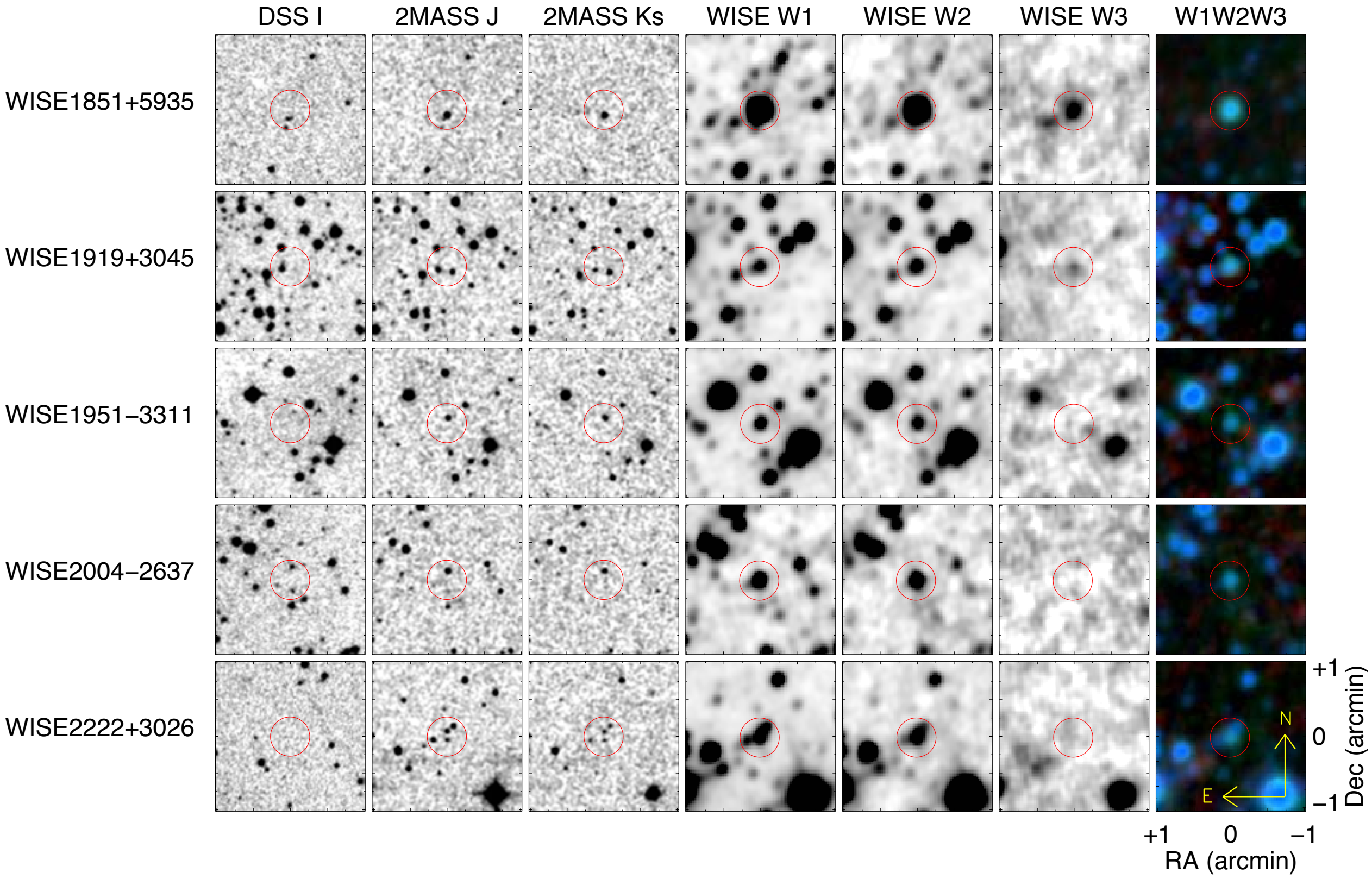

*Figure 1i:*

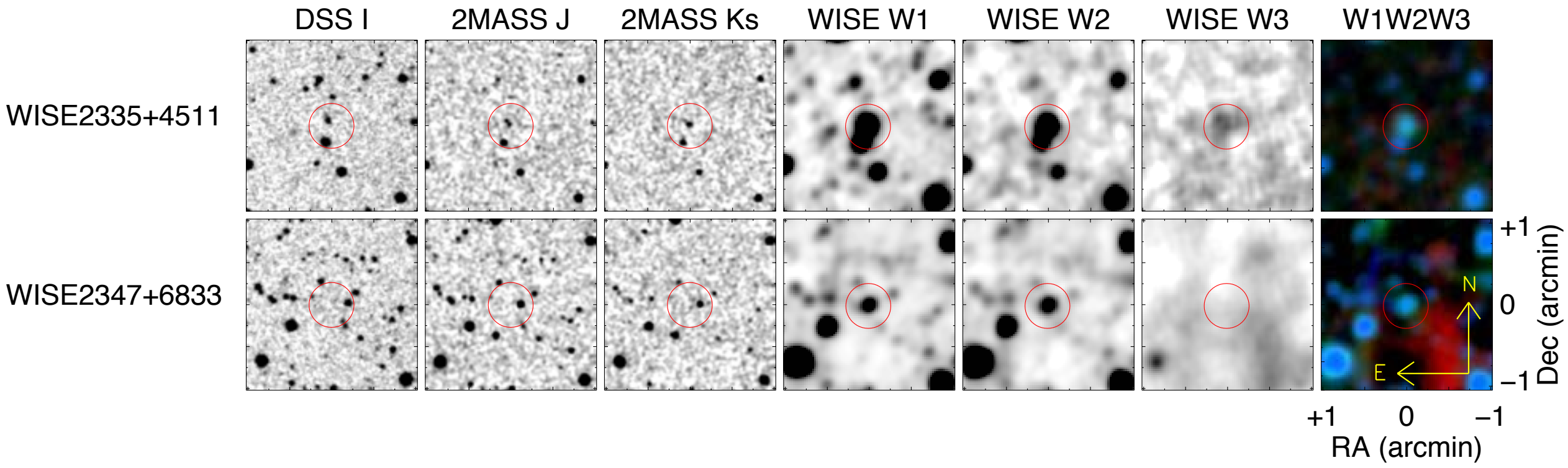

*Figure 2:*

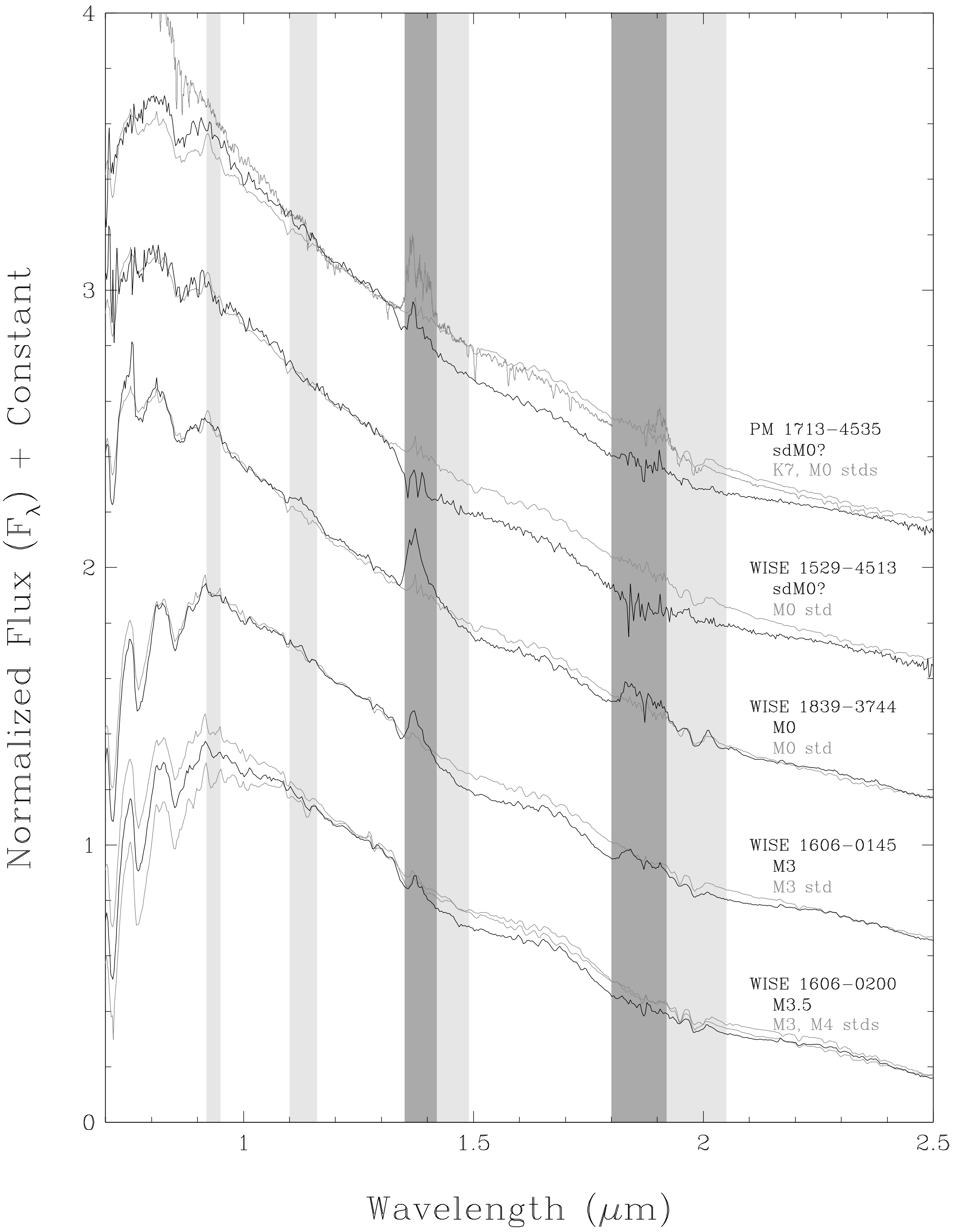

*Figure 3:*

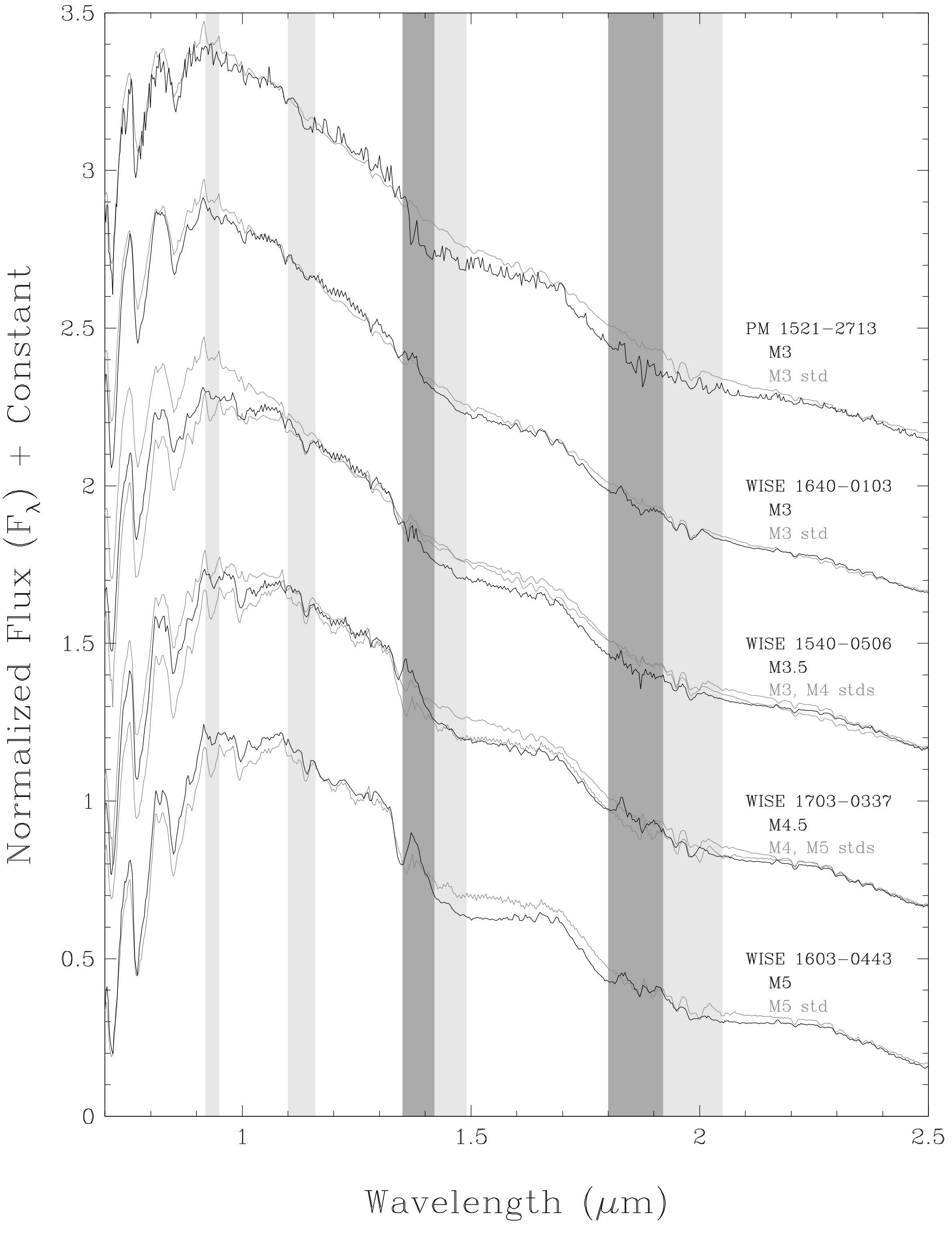

*Figure 4:*

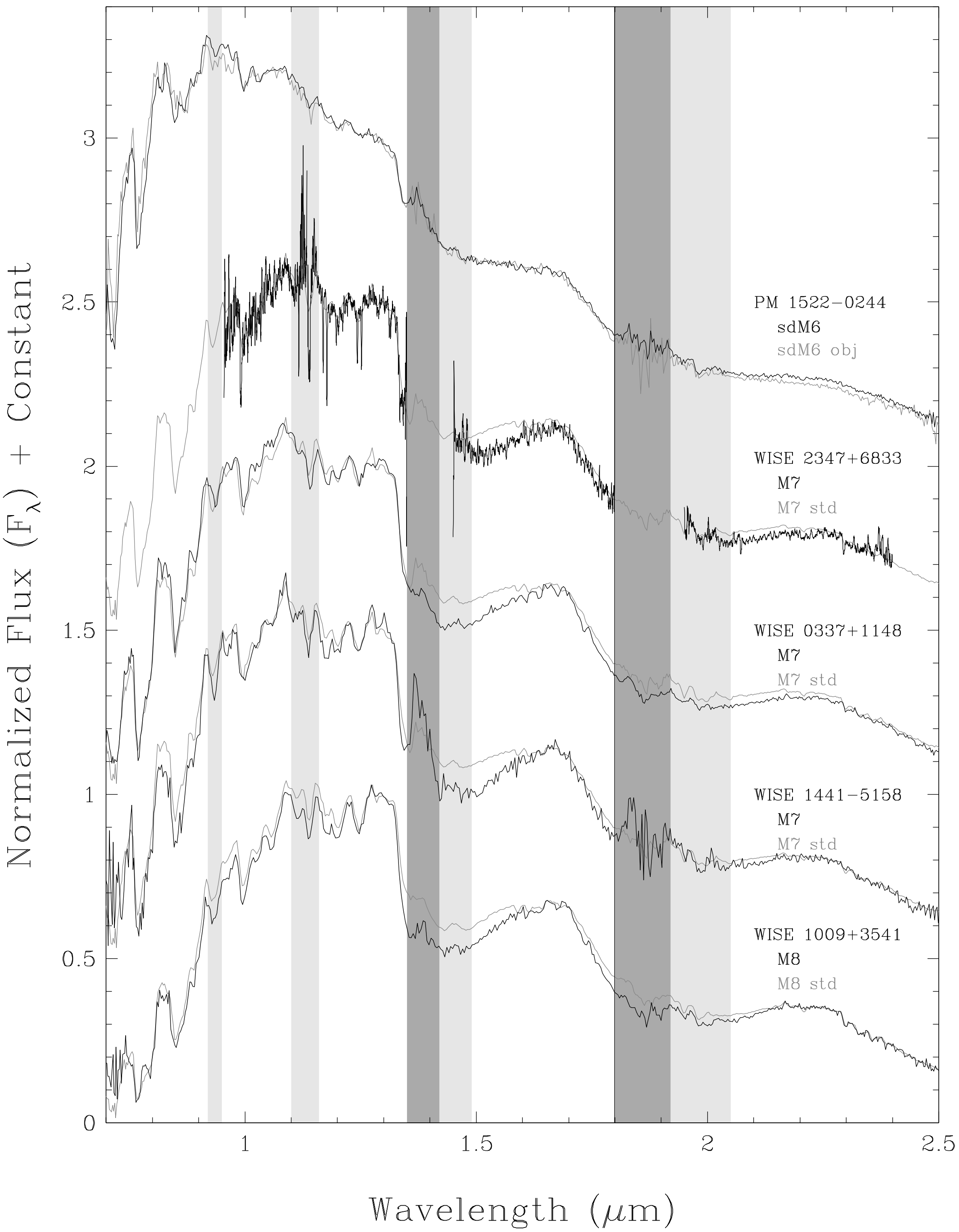

*Figure 5:*

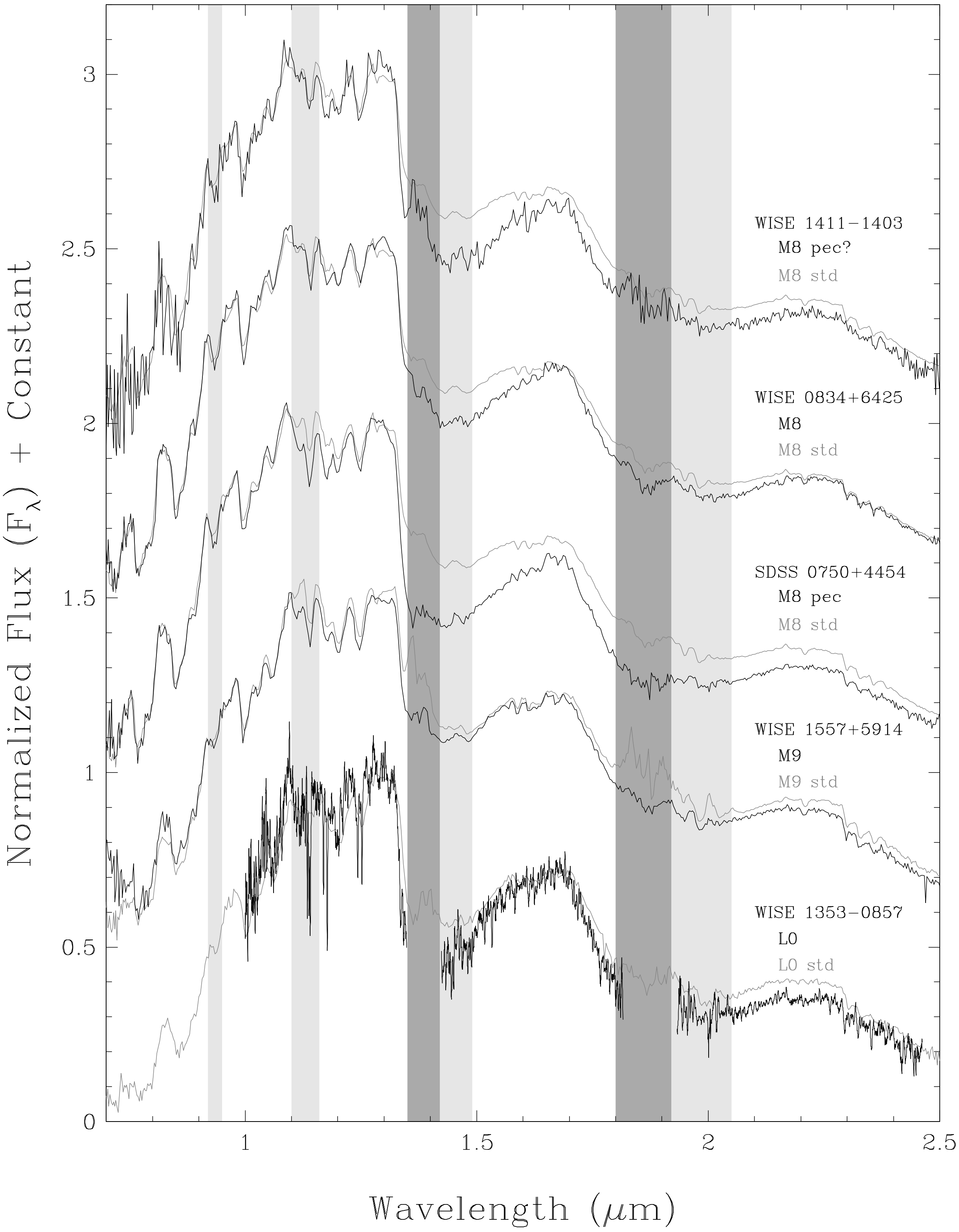

*Figure 6:*

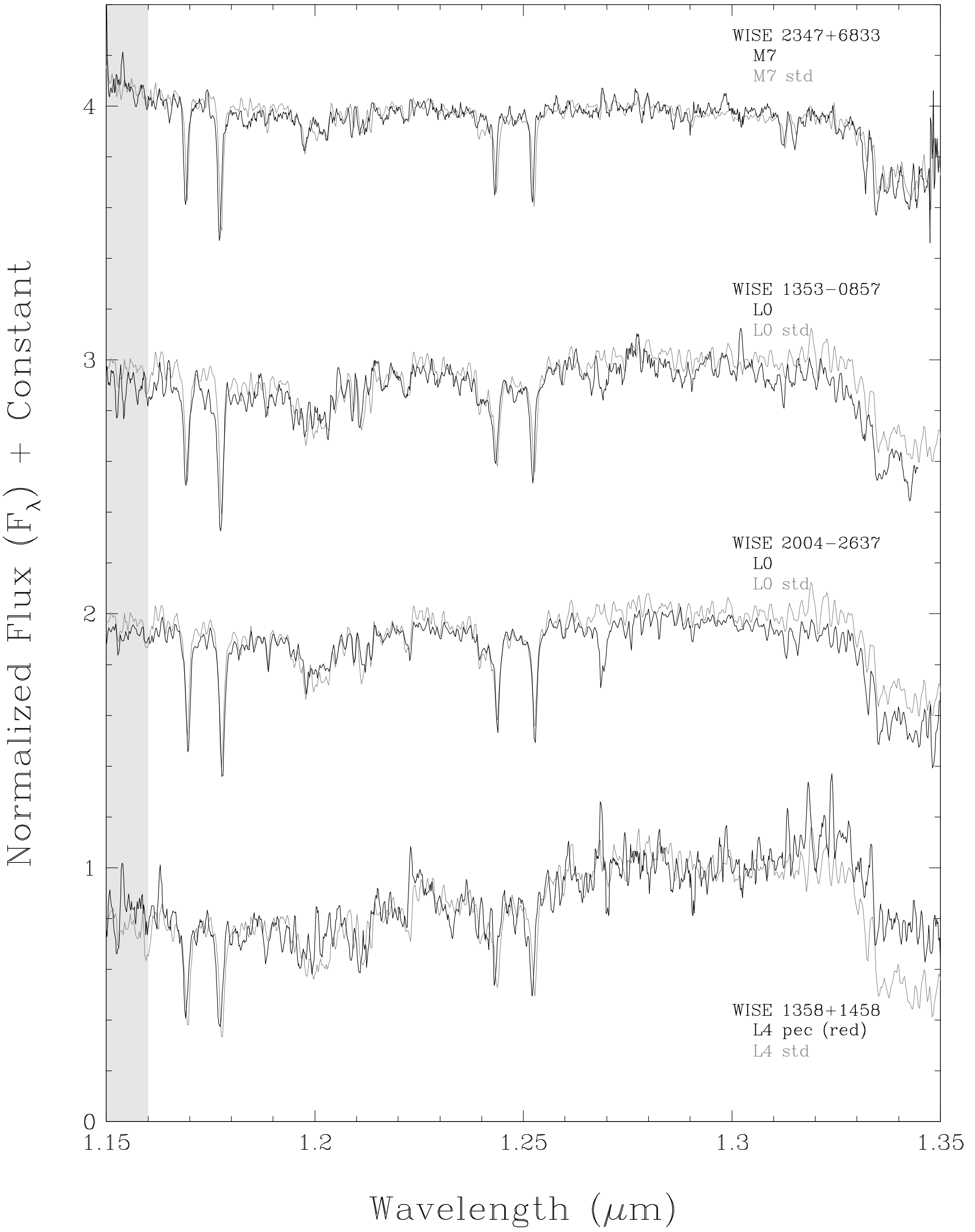

*Figure 7:*

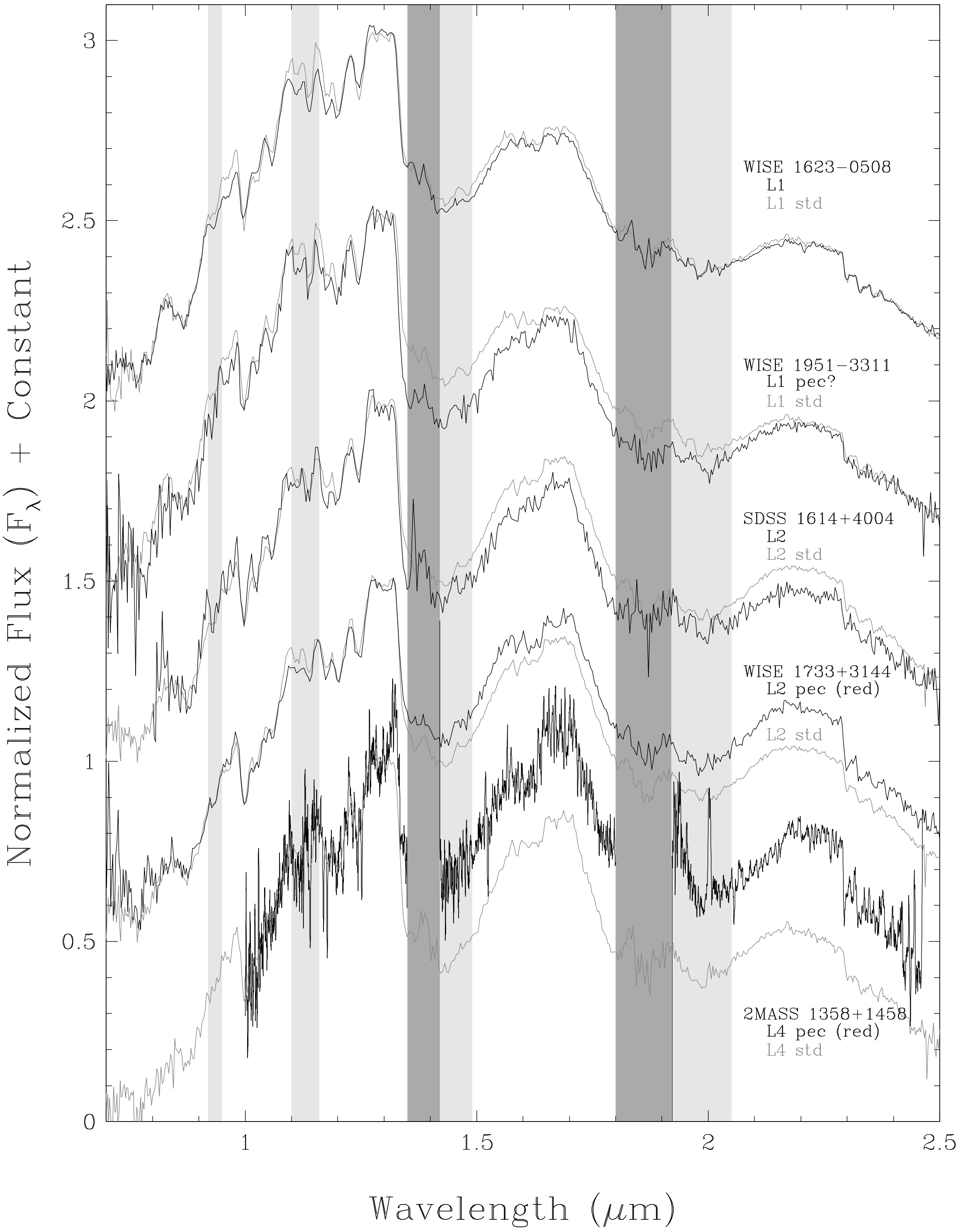

*Figure 8:*

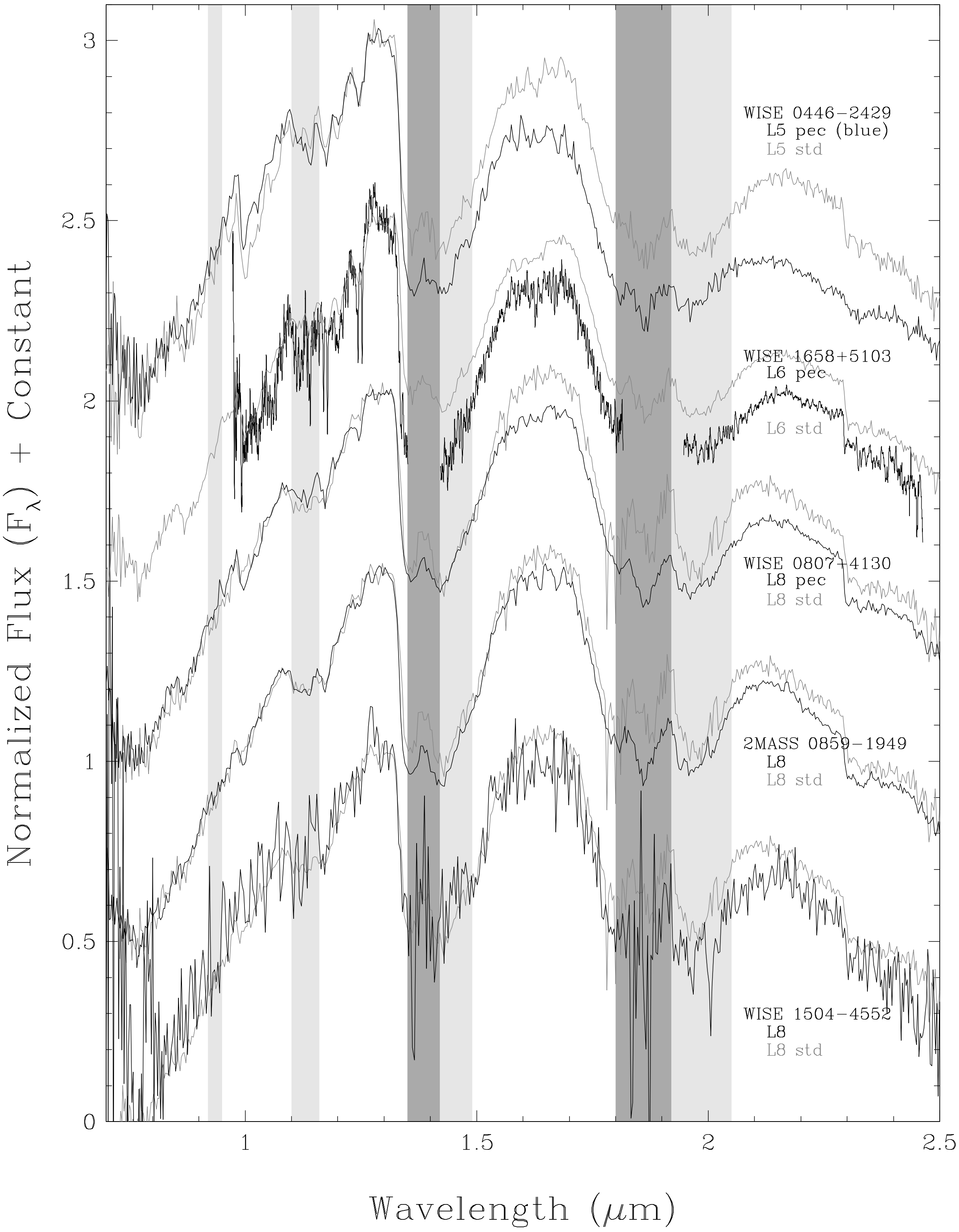

*Figure 9:*

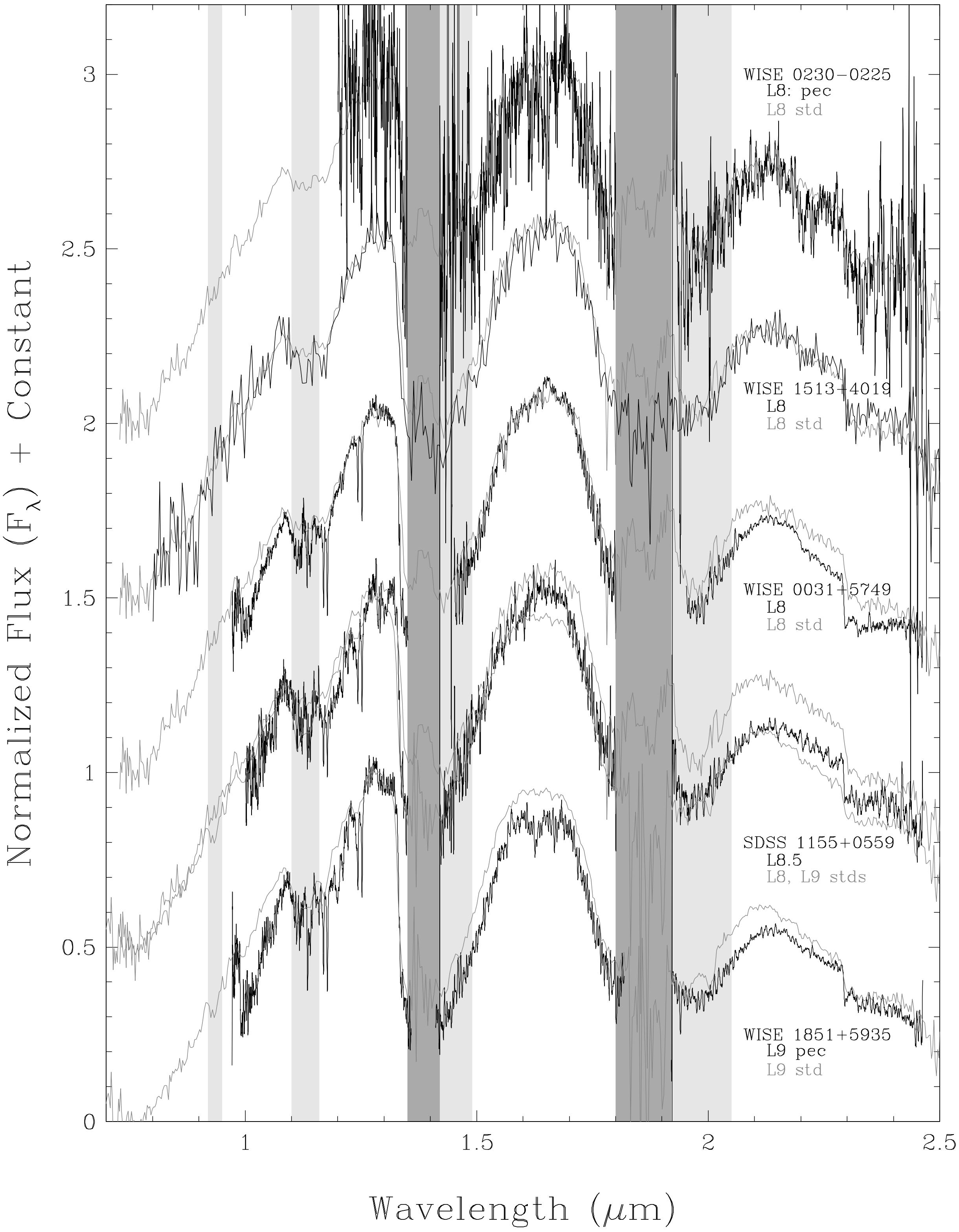

*Figure 10:*

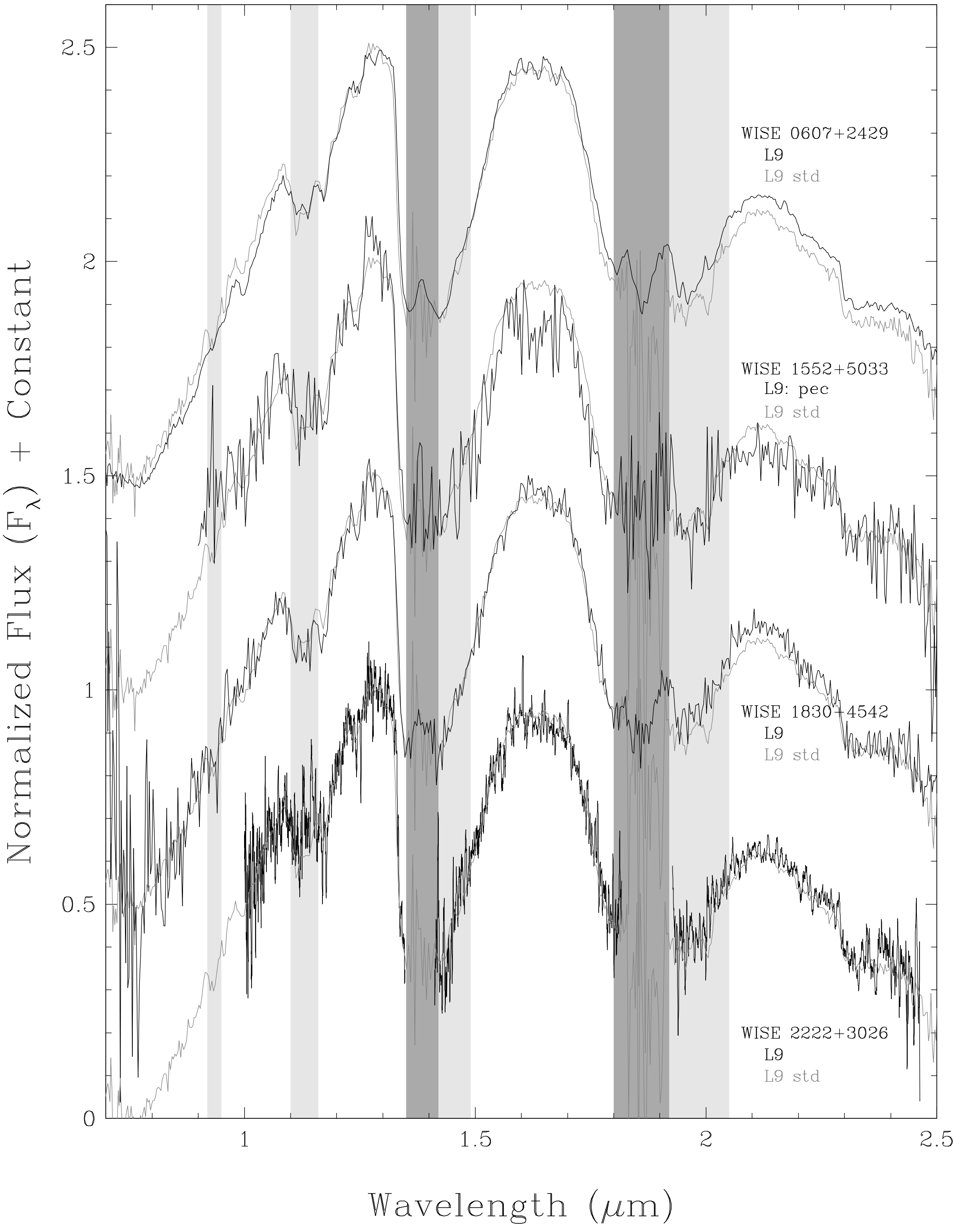

*Figure 11:*

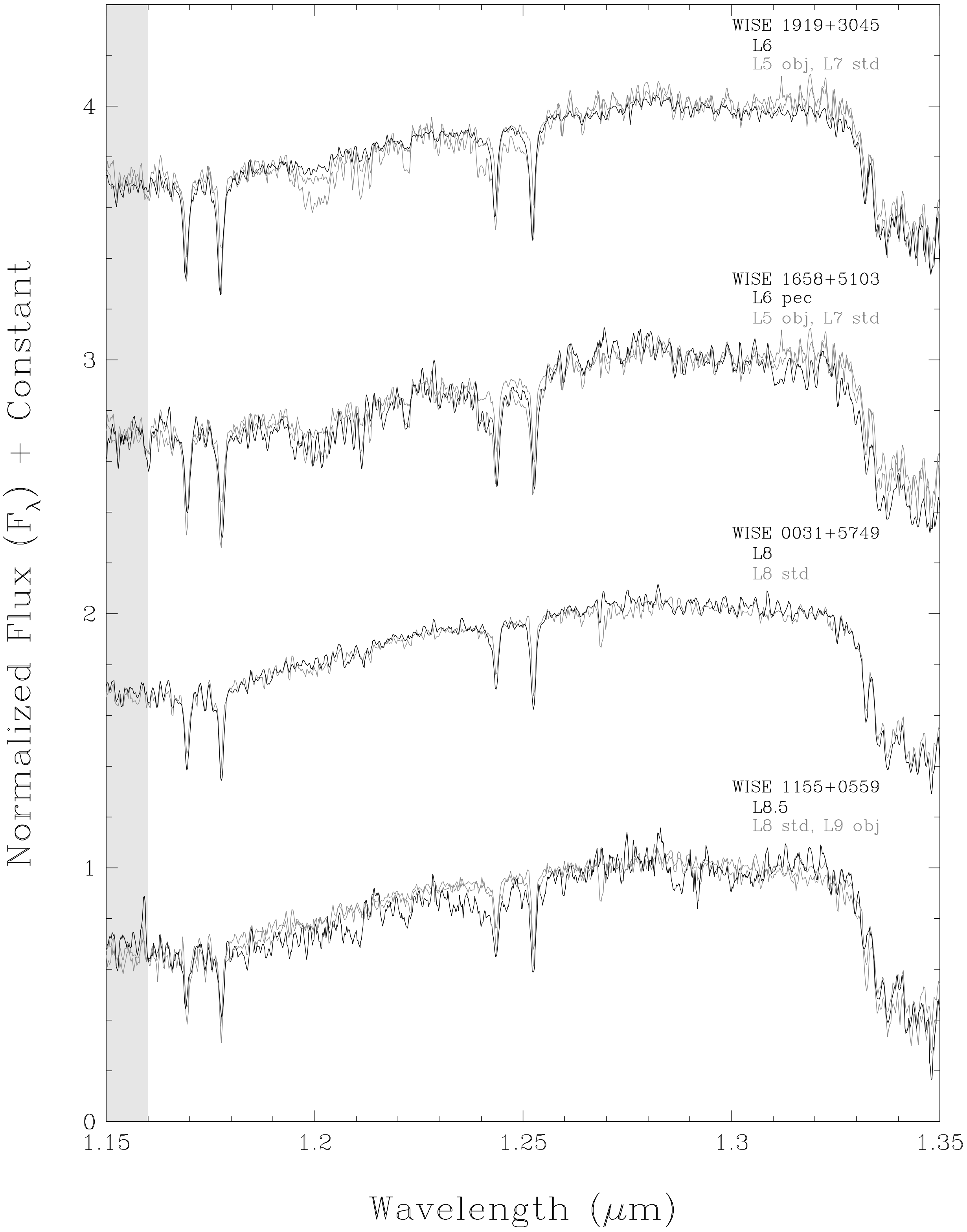

*Figure 12:*

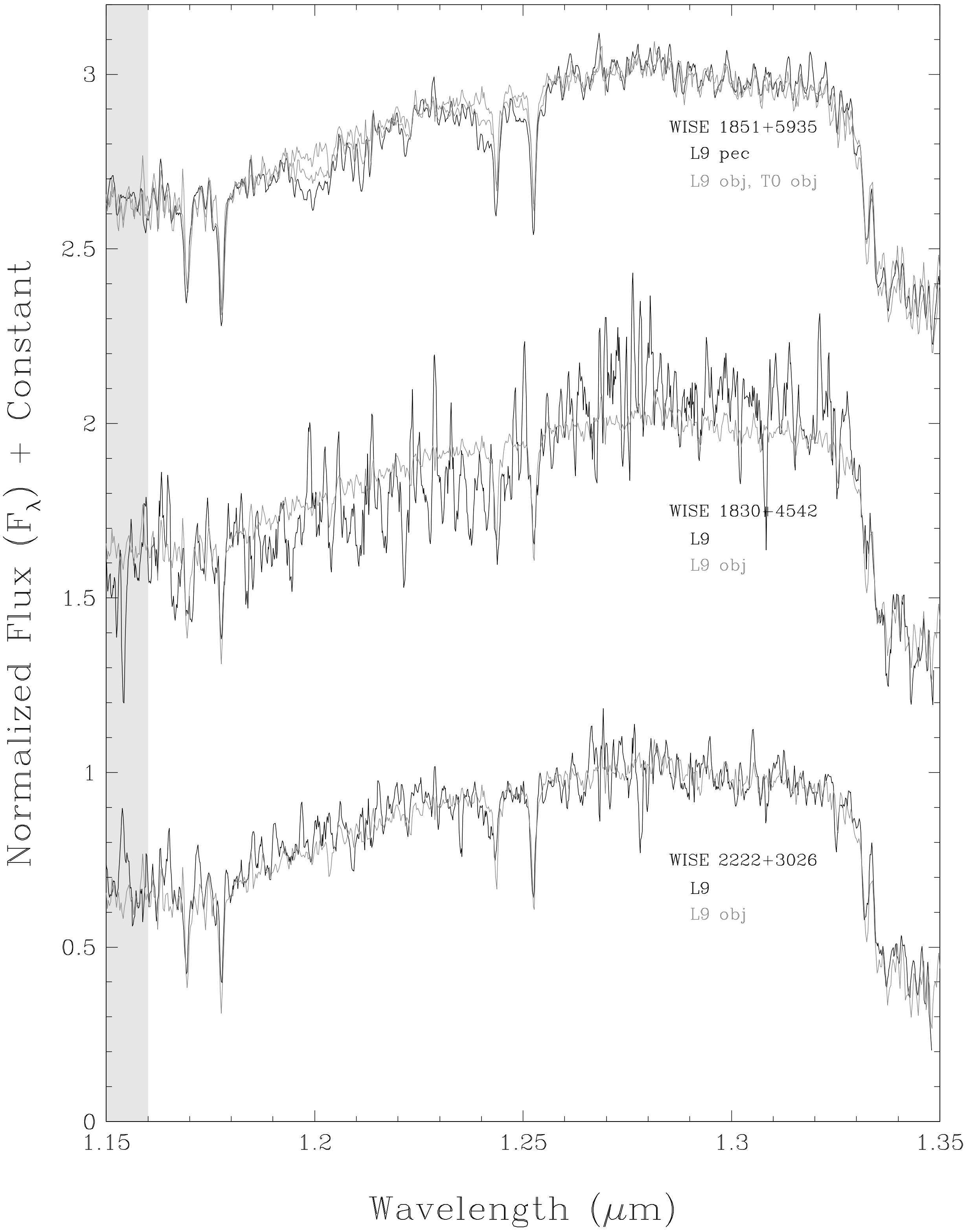

*Figure 13:*

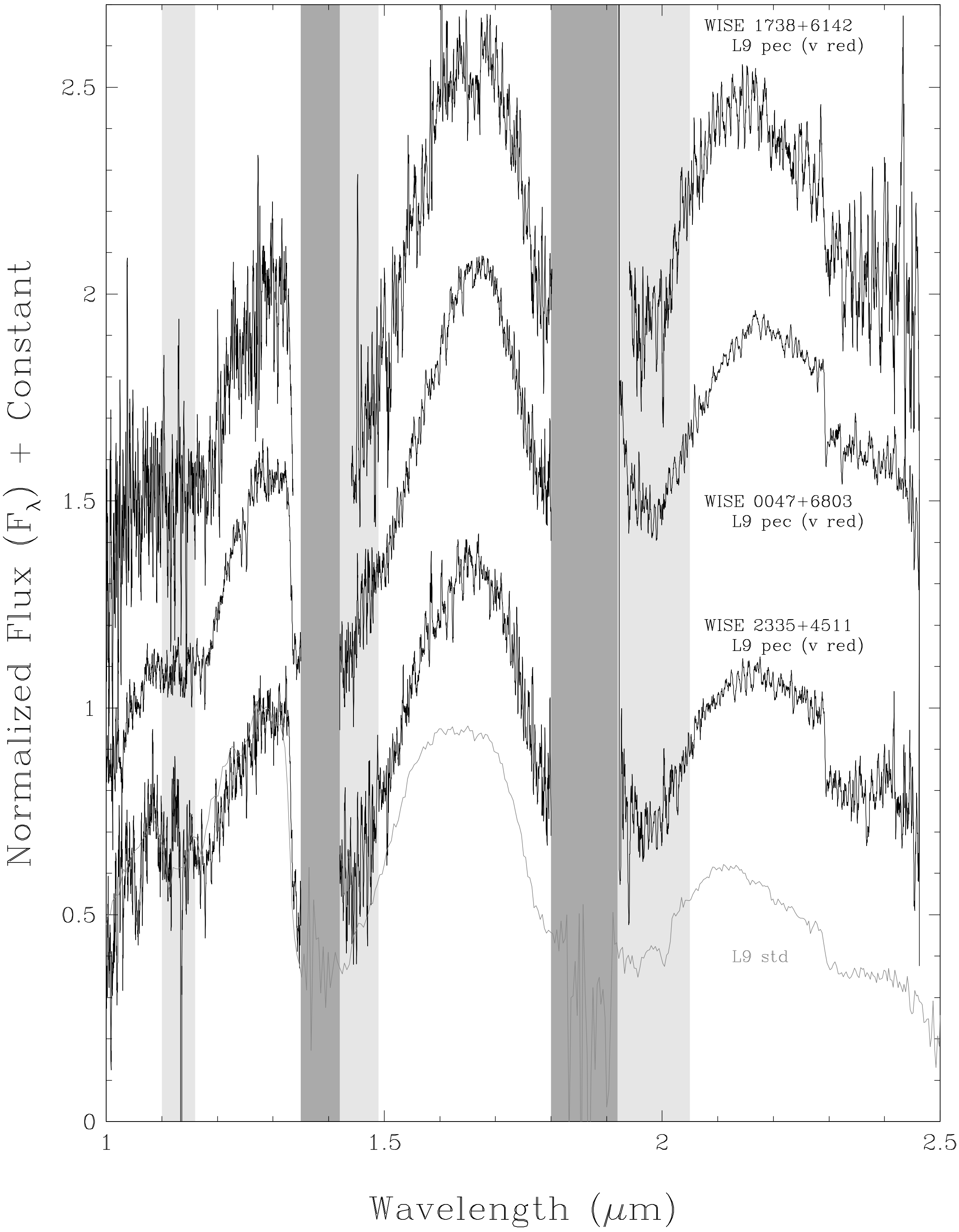

WISE 1738+6142
L9 pec (v red)
WISE 0047+6803
L9 pec (v red)
WISE 2335+4511
L9 pec (v red)
L9 std
Normalized Flux ($F_\lambda$) + Constant
Wavelength ($\mu$m)
0
0.5
1
1.5
2
2.5
1
1.5
2
2.5

*Figure 14:*

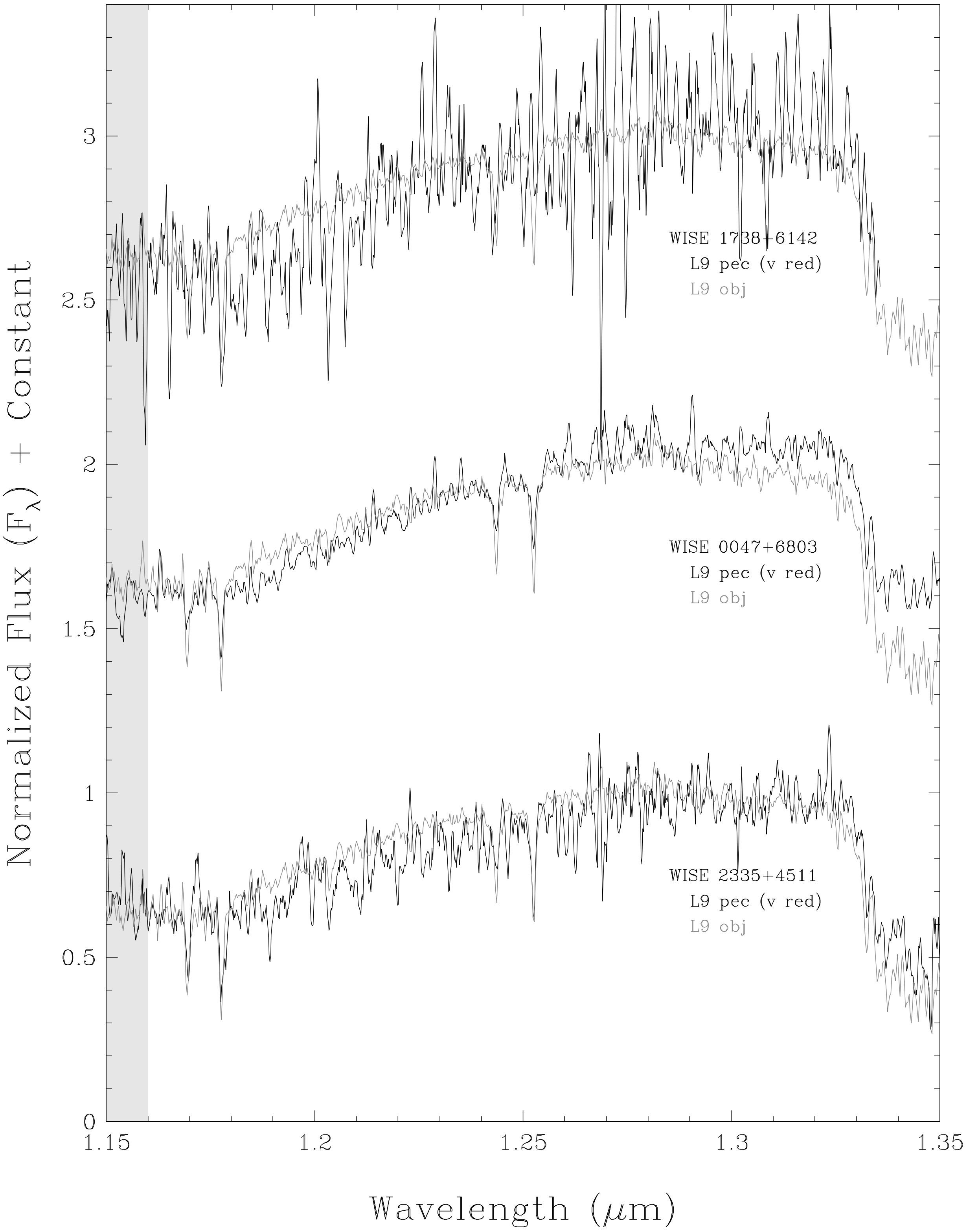

*Figure 15:*

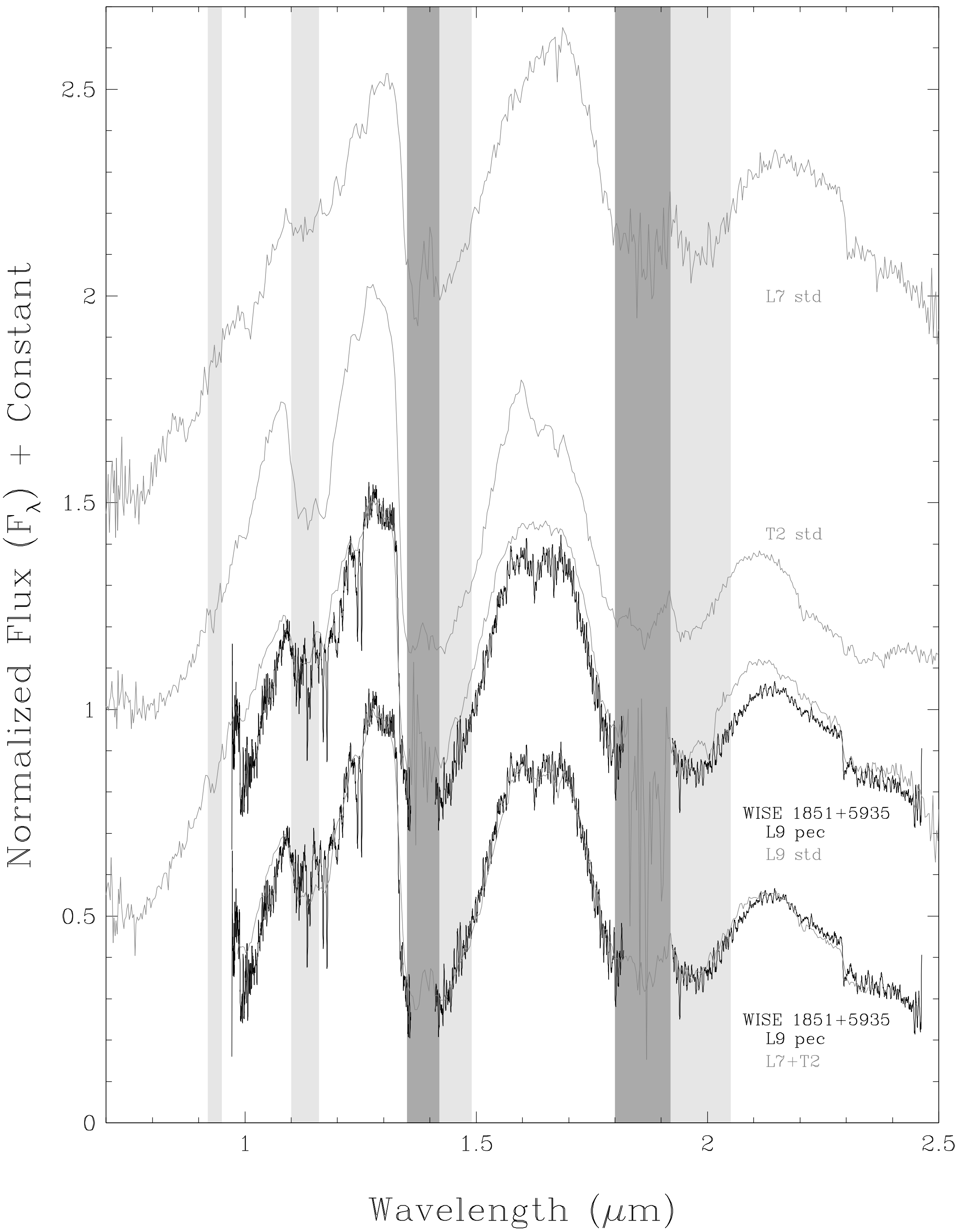

*Figure 16:*

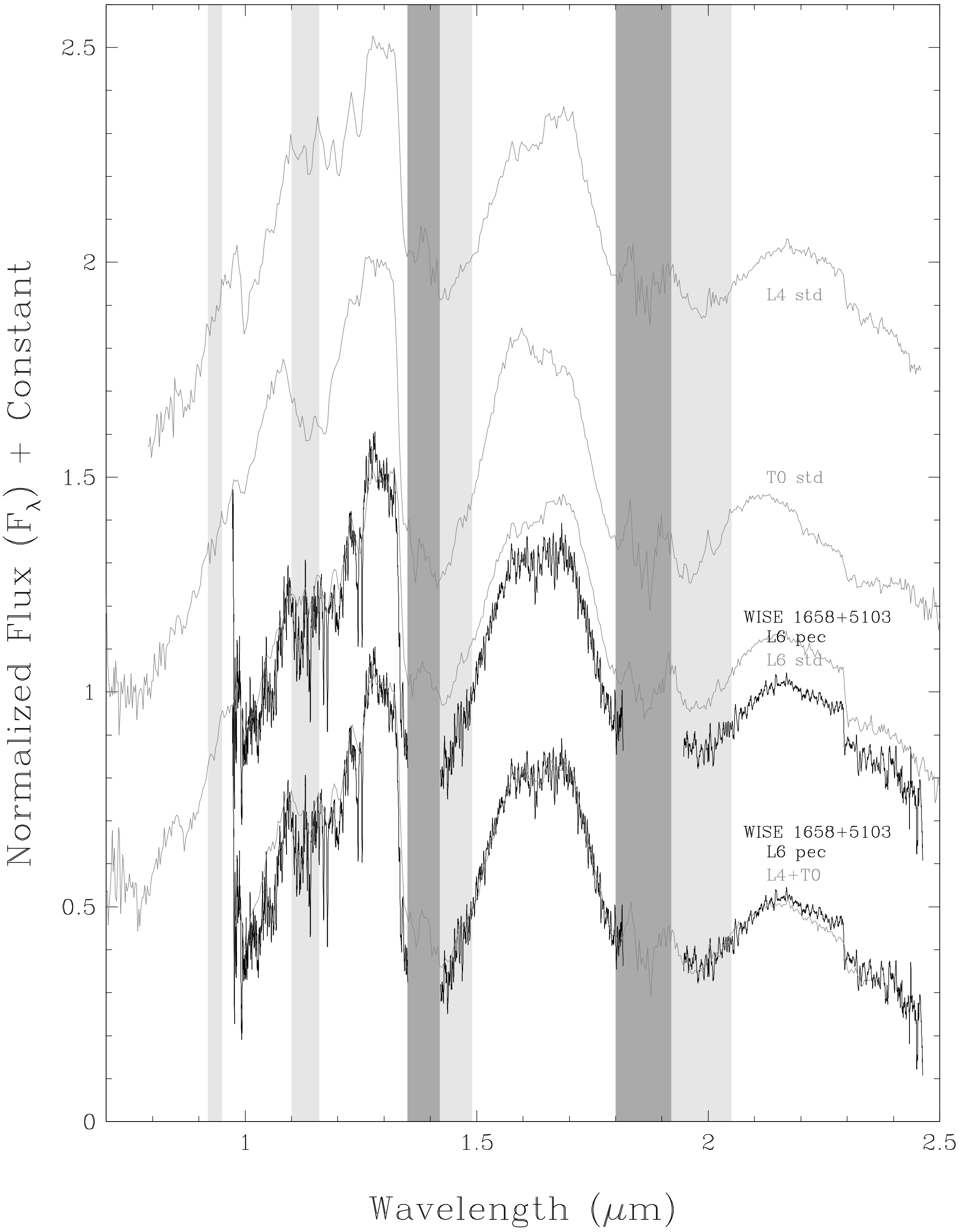

*Figure 17:*

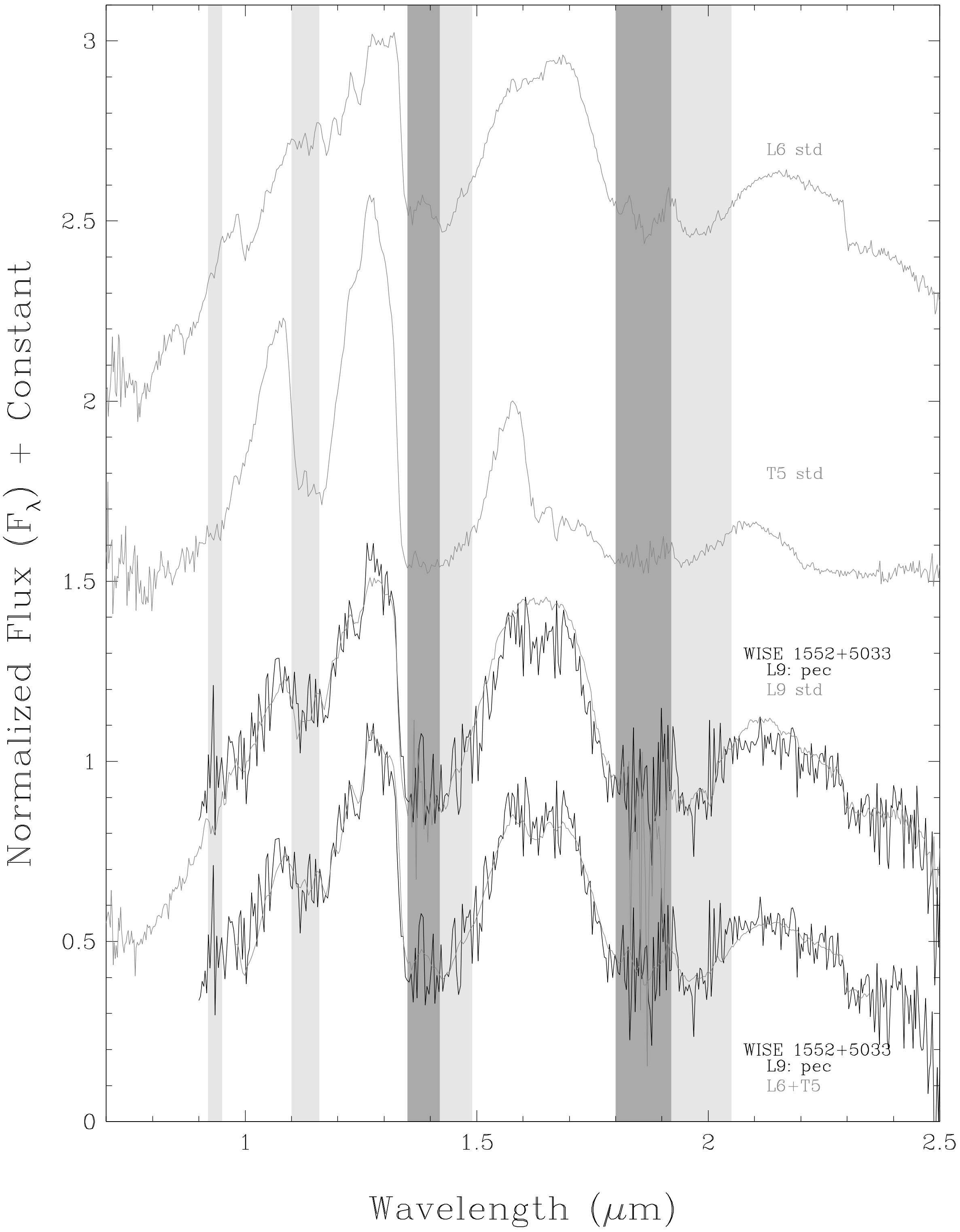

*Figure 18:*

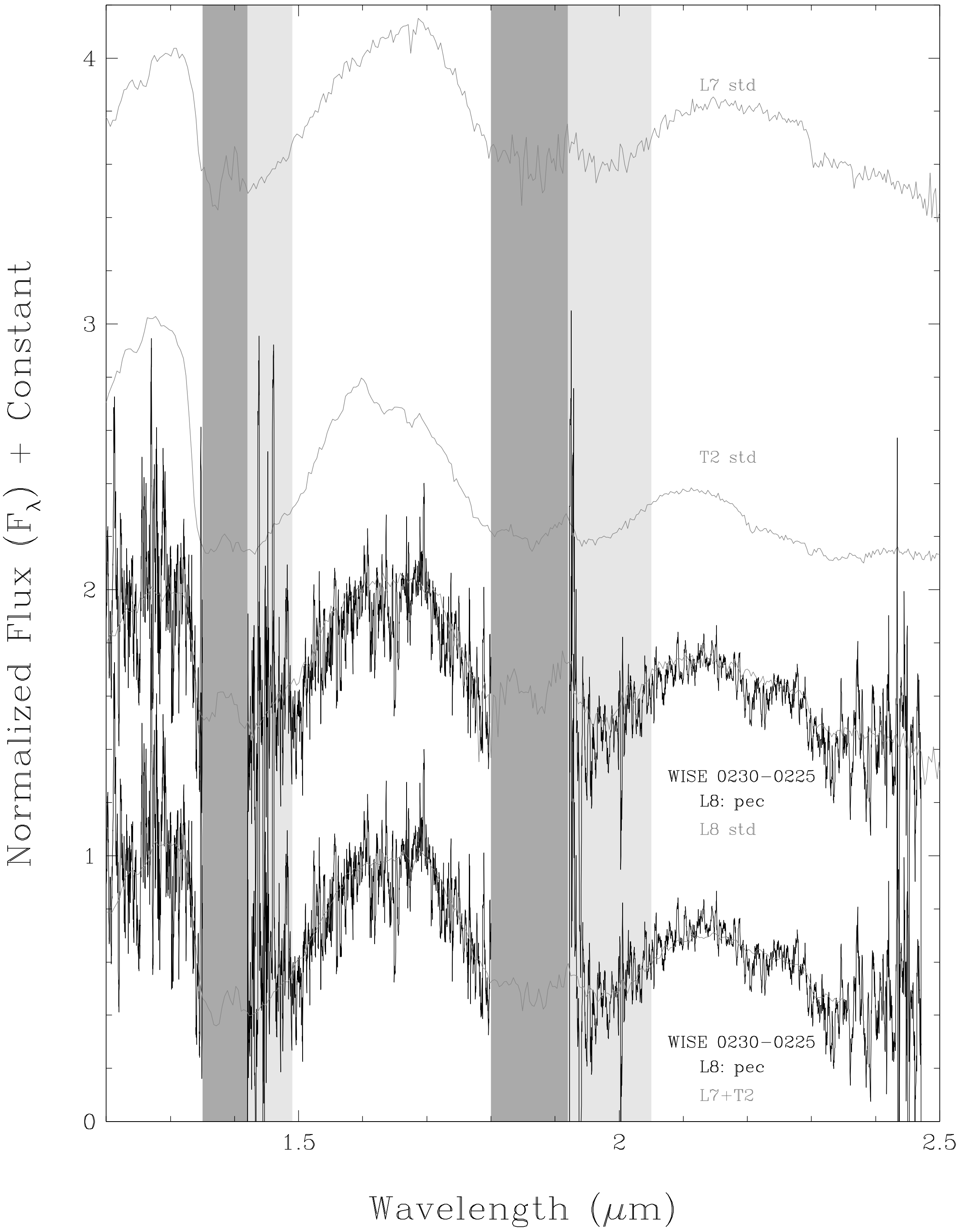

*Figure 19:*

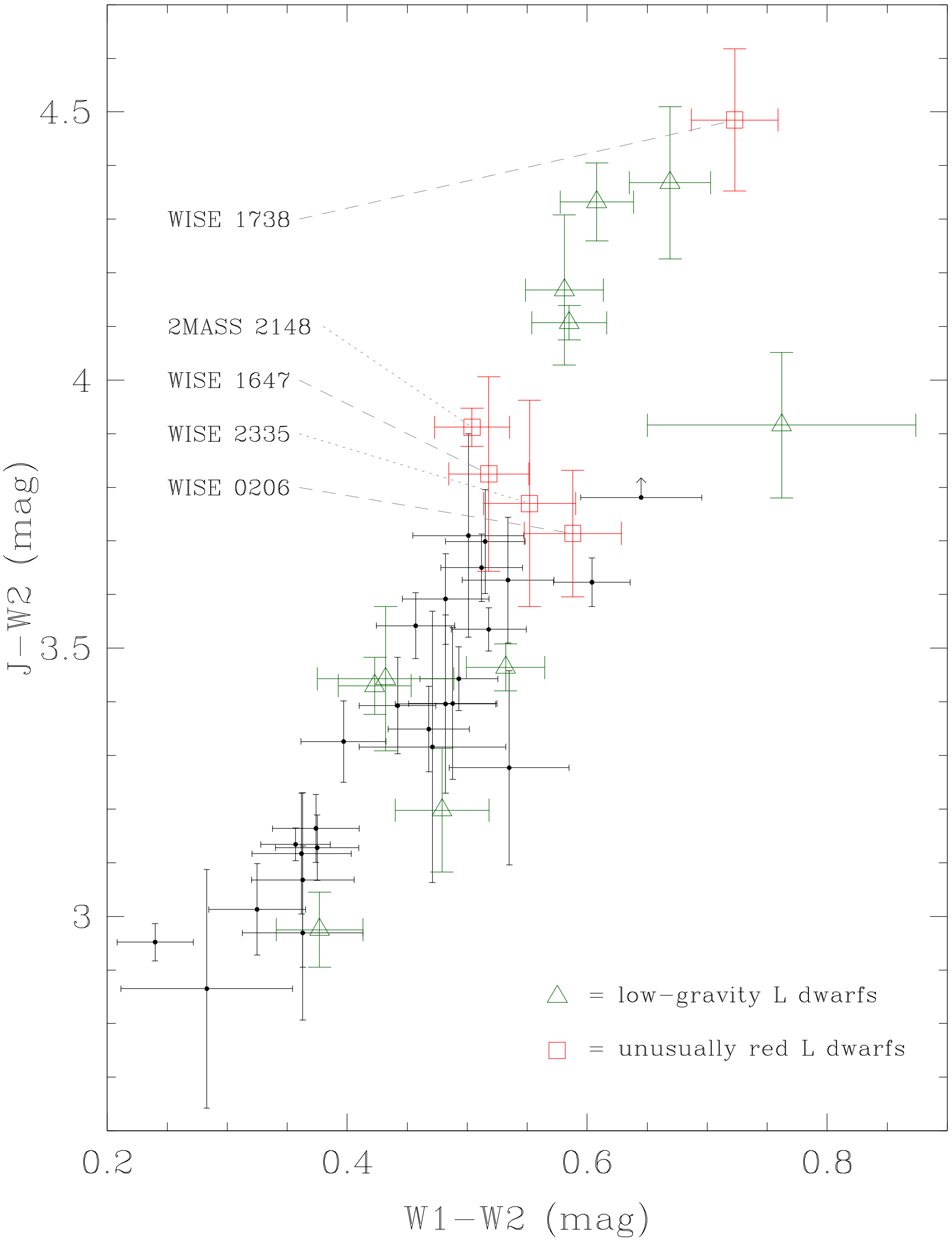

*Figure 20:*

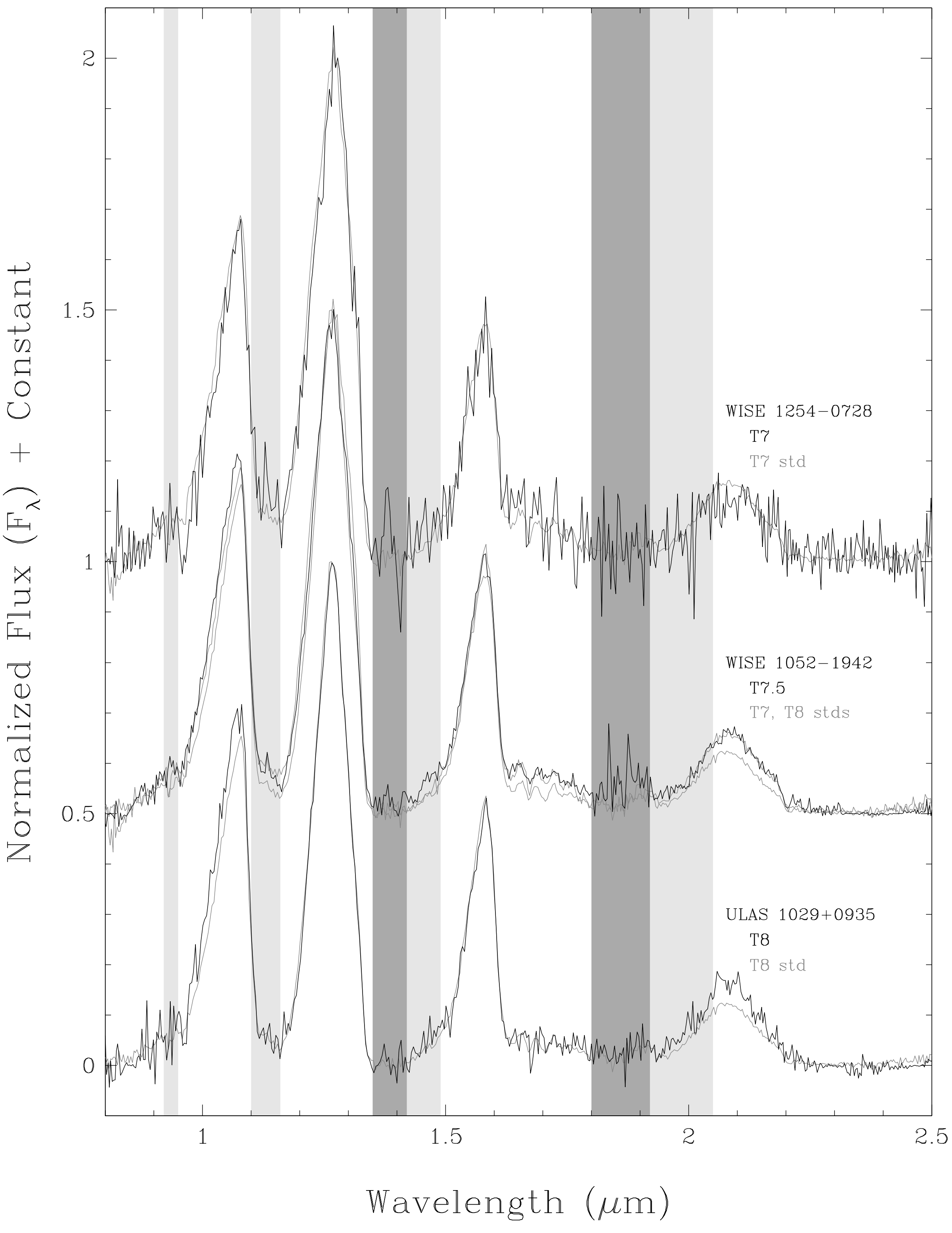

*Figure 21:*

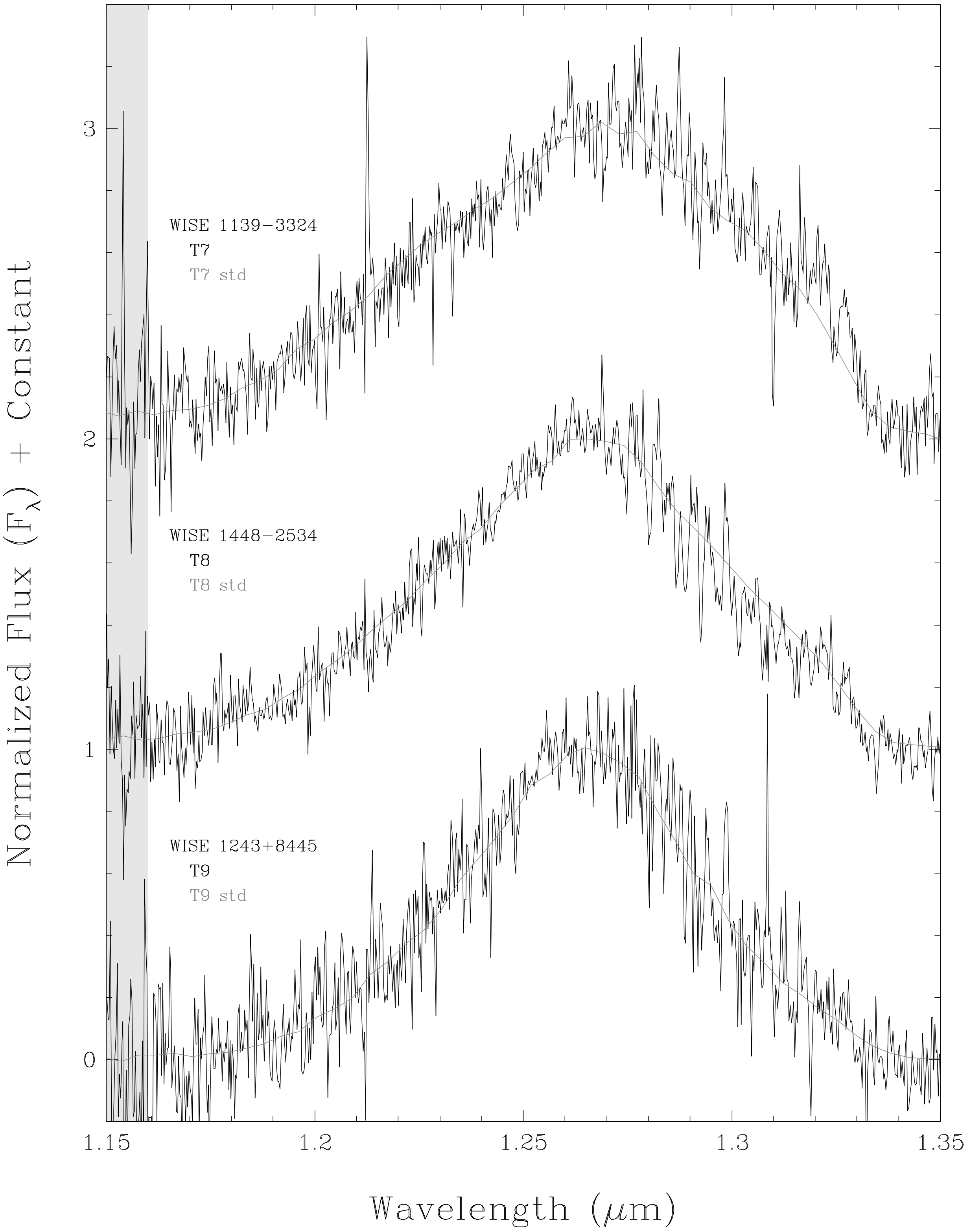

*Figure 22:*

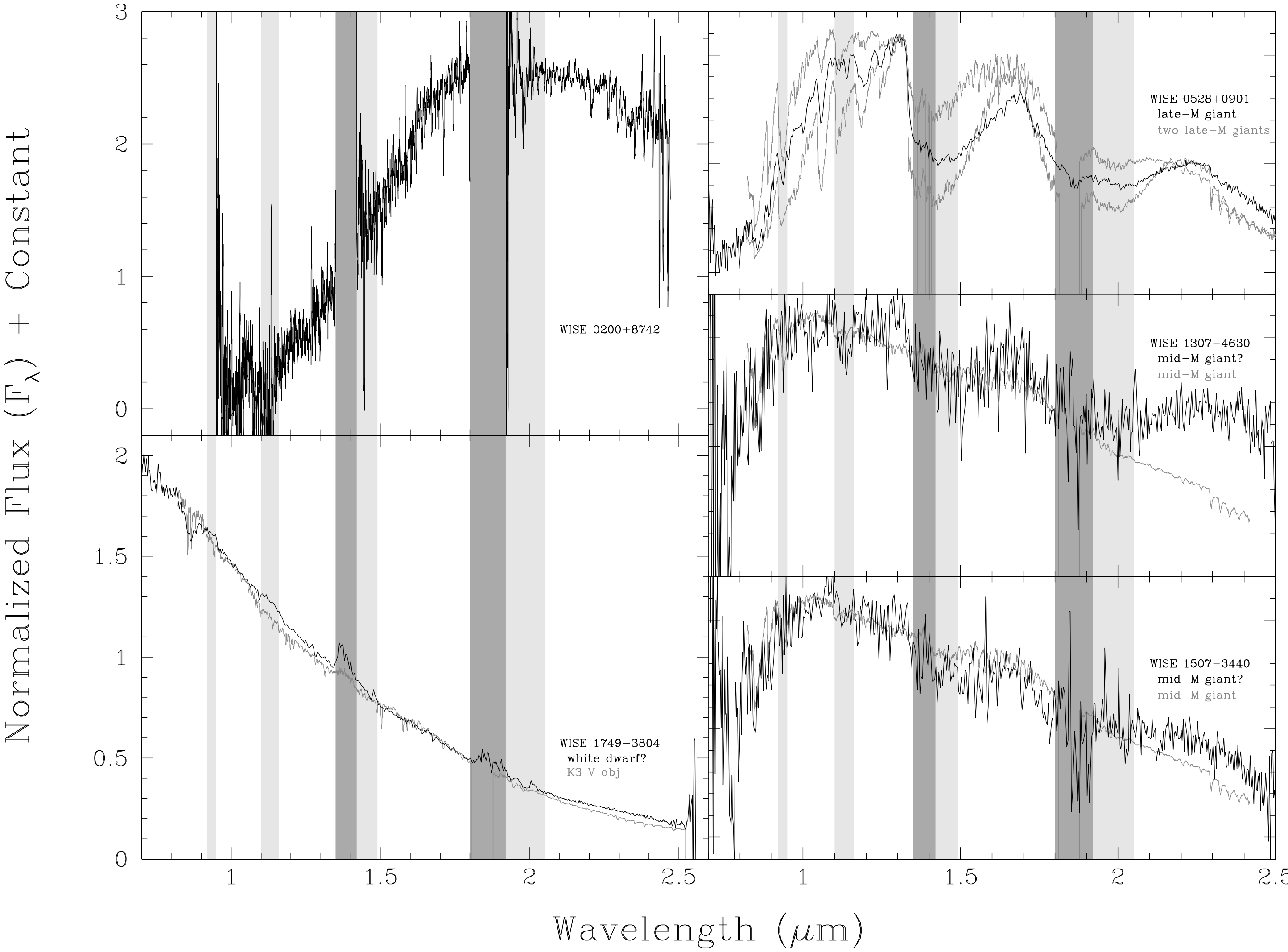